\RequirePackage{fix-cm}
\documentclass[superscriptaddress,secnumarabic,amssymb,amsmath,nobibnotes,aps,prd,showkeys,showpacs,nofootinbib,twocolumn,a4paper,epjc3]{svjour3}
\smartqed  
\RequirePackage{graphicx}
 \RequirePackage{mathptmx}      
\RequirePackage{latexsym}
\RequirePackage[numbers,sort&compress]{natbib}
\RequirePackage[colorlinks,citecolor=blue,urlcolor=blue,linkcolor=blue]{hyperref}
\journalname{Eur. Phys. J. C}
\usepackage{graphicx}
\usepackage{dcolumn}
\usepackage{bm}
\usepackage[
scale=0.7, marginratio={1:1, 2:3}, ignoreall,
text={7in,10in},centering,
margin=1.5in,
total={6.5in,8.75in}, top=1.2in, left=0.9in, includefoot,
height=10in,a5paper,hmargin={3cm,0.8in},
]{geometry}
\RequirePackage{graphicx}
\RequirePackage{mathptmx}      
\RequirePackage{latexsym}
\RequirePackage[colorlinks,citecolor=blue,urlcolor=blue,linkcolor=blue]{hyperref}
\usepackage[english]{babel}
\usepackage{amssymb,amsmath,amsfonts,graphicx, graphics, setspace}
\usepackage{bigints}
\usepackage{anysize}
\usepackage{float}
\usepackage{fancyhdr}
\usepackage{subfigure}
\usepackage{pdflscape}
\usepackage{epsfig}
\usepackage{epsf}
\usepackage{epstopdf}
\usepackage{hyperref} 
\newtheorem{thm}{Theorem}

\newtheorem{lem}[thm]{Lemma}

\usepackage{amsmath}
\usepackage{amsfonts}
\usepackage{amssymb}
\usepackage{subfigure}
\usepackage{epstopdf}
\usepackage{epsf}
\usepackage{url} 
\usepackage{pdflscape} 
\usepackage{amsmath}
\usepackage{amssymb} 
\usepackage{epsfig} 
\usepackage{amsfonts}
\usepackage{graphicx}
\usepackage{subfigure}
\usepackage{times,fancyhdr}
\usepackage{enumitem}
\setlist[enumerate,2]{label=\roman*)}

\def\case#1/#2{\textstyle\frac{#1}{#2}}

\newcommand{\be}{\begin{equation}}
\newcommand{\ee}{\end{equation}}
\newcommand{\ben}{\begin{eqnarray}}
\newcommand{\een}{\end{eqnarray}}

\usepackage{amsmath}
\usepackage{amsfonts}
\usepackage{amssymb}
\usepackage{widetext}
\usepackage{subfigure}
\usepackage{epstopdf}
\usepackage{epsf}
\def\ex{e_1{}^1}
\def\ey{e_2{}^2}

\def\R{{}^3\!R}

\def\sp{\sigma_+}

\def\ex{e_1{}^1}
\def\ey{e_2{}^2}

\def\R{{}^3\!R}

\def\y{\vartheta}
\def\z{\zeta}

\begin{document}
\title{Averaging Generalized Scalar Field Cosmologies II: Locally Rotationally Symmetric Bianchi I and flat Friedmann-Lemaître-Robertson-Walker models}
\author{Genly Leon \thanksref{e1,addr1} \and Sebasti\'an Cu\'ellar \thanksref{e2, addr1}  \and Esteban Gonz\'alez \thanksref{e3,addr2} \and Samuel Lepe \thanksref{e4,addr3}  \and Claudio Michea \thanksref{e5,addr1} \and  Alfredo D. Millano \thanksref{e6,addr1}}
\thankstext{e1}{genly.leon@ucn.cl}
\thankstext{e2}{sebastian.cuellar01@ucn.cl}
\thankstext{e3}{esteban.gonzalezb@usach.cl}
\thankstext{e4}{samuel.lepe@pucv.cl}
\thankstext{e5}{claudio.ramirez@ce.ucn.cl}
\thankstext{e6}{alfredo.millano@alumnos.ucn.cl}
\institute{Departamento  de  Matem\'aticas,  Universidad  Cat\'olica  del  Norte, Avda. Angamos  0610,  Casilla  1280  Antofagasta,  Chile\label{addr1} \and Universidad de Santiago de Chile (USACH), Facultad de Ciencia, Departamento de F\'isica, Chile \label{addr2}
\and Instituto de F\'isica, Facultad de Ciencias, Pontificia Universidad Cat\'olica de Valpara\'iso, 
Av. Brasil 2950, Valpara\'iso, Chile \label{addr3}
}
\date{\today}

\maketitle

\begin{abstract}
Scalar field cosmologies with a generalized harmonic potential and a matter fluid with a barotropic Equation of State (EoS) with barotropic index $\gamma$ for the Locally Rotationally Symmetric (LRS) Bianchi I and flat Friedmann-Lemaître-Robertson-Walker (FLRW) metrics are investigated.  Methods from the theory of averaging of nonlinear dynamical systems are used to prove that  time-dependent systems and their corresponding time-averaged versions have the same late-time dynamics. Therefore, the simplest time-averaged system determines the future asymptotic behavior. Depending on the  values of   $\gamma$,  the late-time attractors  of physical interests are flat quintessence dominated FLRW universe and  Einstein-de Sitter solution. With this approach, the oscillations entering the system through the Klein-Gordon (KG) equation can be controlled and smoothed out as the Hubble parameter $H$ - acting as time-dependent perturbation parameter - tends monotonically to zero. Numerical simulations are presented  as evidence of such behavior.
\end{abstract}

\keywords{Generalized scalar field cosmologies \and Anisotropic models \and Early Universe \and Equilibrium-points \and Harmonic oscillator}

\section{Introduction}
Mathematical methods have been widely used in cosmology. For example, in reference \cite{Paliathanasis:2016rho} the method of Lie symmetries was applied to Wheeler-De Witt equation in Bianchi class A cosmologies for minimally coupled scalar field gravity and hybrid gravity in General Relativity (GR). Several invariant solutions were determined and classified. In reference \cite{Basilakos:2011rx} a model-independent criterion based on first integrals of motion was used; and in reference \cite{Paliathanasis:2014zxa} dynamical symmetries of the field equations were used  to classify Dark Energy (DE) models in the context of scalar field (quintessence or phantom)  FLRW  cosmologies. Using Noether symmetries in \cite{Basilakos:2011rx} the system was simplified and its integrability was determined. For the exponential potential  as well as some types of hyperbolic potentials, extra Noether symmetries apart of the conservation law were found;  suggesting that these potentials should be preferred along the hierarchy of scalar field potentials. In \cite{Paliathanasis:2014zxa} under the requirement that  field equations admit dynamical symmetries resulted in two potentials, one of them is the well known Unified Dark Matter (UDM) potential and another hyperbolic model. In reference \cite{Barrow:2016wiy}  a mathematical approach to reconstruct the EoS and the inflationary potential of the inflaton field from  observed spectral indices for the density perturbations and the tensor-to-scalar ratio (based on the constraints system) was implemented. In reference \cite{Barrow:2016qkh} an algorithm to generate new solutions of the scalar field equations in FLRW universes was used. Solutions  for pure scalar fields with various potentials in absence and in presence of spatial curvature and other perfect fluids were  obtained. A series of generalizations of Chaplygin gas and bulk viscous cosmological solutions for inflationary universes were found. In reference \cite{Paliathanasis:2016vsw} the $f(T)$  cosmological scenario was studied. In particular, analytical solutions for  isotropic, homogeneous universe containing   dust fluid and radiation, and for an empty anisotropic Bianchi I universe were found. The method of movable singularities of differential equations was used. For the isotropic universe, the solutions are expressed in terms of  Laurent expansion, while for anisotropic universe a family of exact Kasner-like solutions in vacuum is found. In reference \cite{Paliathanasis:2014ofa}  the symmetry classification of the KG equation in Bianchi I spacetime was performed. A geometric method which relates the Lie symmetries of the KG equation with the conformal algebra of the underlying geometry was applied. Furthermore, by means of Lie symmetries that follow from the conformal algebra (which are also Noether symmetries for the KG equation) all the potentials in which the KG equation admits Lie and Noether symmetries were determined. The Lie admitted symmetries are useful to determine the corresponding invariant solution of the KG equation for specific potentials. Additionally,  the classification problem of Lie/Noether point symmetries of the wave equation in Bianchi I spacetime  was solved, and invariant solutions of the wave equation were determined. In reference \cite{Tsamparlis:2015qua} a new method to classify  Bianchi I spacetimes which admits Conformal Killing Vectors (CKV) was developed. The method is useful to study the conformal algebra of Kasner spacetime and other Bianchi type I matter solutions of GR. 

Other useful mathematical methods are asymptotic methods and averaging theory \cite{dumortier,fenichel,Fusco,Berglund,holmes,Kevorkian1,Verhulst}. These methods have been applied in cosmology for example in \cite{Rendall:2006cq,Llibre:2012zz,Alho:2015cza,Leon:2019iwj,Leon:2020ovw,Leon:2020pfy,Leon:2020pvt,Leon:2021lct,Leon:2021hxc} with interest in early and late-time dynamics. Some works related to  Einstein-KG, Maxwell,  Yang-Mills and   Einstein-Vlasov systems are \cite{Rendall:2003ks,TchapndaN.:2003bv,Alcubierre:2003sx,Rendall:2006cq,Liebscher:2012xt,Reiris:2015zaa,Lozanov:2017hjm,Wang:2018fay,Klainerman:2018mge,Ionescu:2019spj,Fajman:2019vma,Alho:2019pku,Siemonsen:2020hcg,Chatzikaleas:2020zmq,Chatzikaleas:2020twz,Barzegar:2020pna,Barzegar:2019nue,Barzegar:2020vzk,Alho:2015cza}. In reference \cite{Fajman:2020yjb}  LRS  Bianchi type III cosmologies with a massive scalar field were studied by means of the theory of averaging of nonlinear dynamical systems. In reference \cite{Fajman:2021cli} a theorem about large-time behavior of solutions of a general class Spatially Homogeneous (SH) cosmologies with oscillatory behavior was presented. The results are based on a first order approximation of $H$, when $H$ is non-negative and monotonic decreasing to zero. 

Inspired in \cite{Leon:2020pfy,Leon:2019iwj,Leon:2020ovw,Leon:2020pvt} we have started the  ``Averaging Generalized Scalar Field Cosmologies'' program which consists in using asymptotic methods and averaging theory to obtain relevant information about the solution's space of scalar field cosmologies with generalized harmonic potential in presence of matter (with a barotropic EoS with barotropic index $\gamma$) minimally coupled to a scalar field. This research program has three steps according to the three cases of study: (I) Bianchi III and open FLRW model \cite{Leon:2021lct}, (II) Bianchi I and flat FLRW model (the present case)  and (III) Kantowski-Sachs and closed FLRW \cite{Leon:2021hxc}. In reference \cite{Leon:2020pvt} relevant results for the aforementioned program were presented. In particular,  interacting scalar field cosmologies with generalized harmonic potentials  for flat and negatively curved FLRW, and for Bianchi I metrics were studied. Using asymptotic and averaging methods stability conditions for several solutions of interest as $ H \rightarrow 0$ were obtained. This analysis suggests that the asymptotic behavior of the time-averaged model is independent of the coupling function and the geometry. Following analogous procedures in references \cite{Leon:2021lct} and \cite{Leon:2021hxc} the cases (I) and (III) of the program were studied. 

For LRS Bianchi III metric in paper I \cite{Leon:2021lct}   was proved that the late-time attractors of   full and time-averaged systems are: a matter dominated FLRW universe if $0\leq \gamma \leq \frac{2}{3}$ (mimicking de Sitter, quintessence or zero acceleration solutions), a matter-curvature scaling solution  if  $\frac{2}{3}<\gamma <1$ and Bianchi III flat spacetime  for $1\leq \gamma\leq 2$. For FLRW metric with $k=-1$  late-time attractors are: a matter dominated FLRW universe if  $0\leq \gamma \leq \frac{2}{3}$ (mimicking de Sitter, quintessence or zero acceleration solutions) and the Milne solution if $\frac{2}{3}<\gamma <2$.
For Kantowski-Sachs metric (see references \cite{KS1,KS2,KS3,KS4,Byland:1998gx}) in paper III \cite{Leon:2021hxc}   was proved  that late-time attractors of full and time-averaged systems are: two anisotropic contracting solutions if $0\leq\gamma < 2$, a non-flat LRS Kasner Bianchi I, a Taub (flat LRS Kasner) and a matter dominated FLRW universe if $0\leq\gamma<\frac{2}{3}$ (mimicking de Sitter, quintessence or zero acceleration solutions). For FLRW metric with $k=+1$ late-time attractors are:  Einstein-de Sitter solution if $0<\gamma<1$, the matter dominated FLRW universe for $0\leq\gamma\leq\frac{2}{3}$ (mimicking de Sitter, quintessence or zero acceleration solutions)  and a matter dominated contracting isotropic solution if $1<\gamma<2$. 
In all the metrics, the matter dominated FLRW universe represents quintessence fluid if $0<\gamma<\frac{2}{3}$.

This paper is devoted to case (II). It is organized as follows: in Section \ref{motivation} we motivate our choice of potential and the topic of averaging in the context of differential equations.  In Section \ref{Sect2} we introduce the model under study. In Section \ref{SECT:II} we apply averaging methods to analyze periodic solutions of a scalar field  with self-interacting potentials within the class of generalized harmonic potentials \cite{Leon:2019iwj} where in particular, Section \ref{SECT3.3} is devoted to LRS Bianchi I model and Section \ref{SECT:IIIA} is devoted to flat FLRW metric. In Section \ref{SECT:III} we study the resulting time-averaged systems where,  in particular, Section \ref{LRSBI} is devoted to LRS Bianchi I models and Section \ref{FLRWflatopen} is devoted the flat FLRW metric. Finally, in Section \ref{Conclusions} our main results are discussed. In  \ref{gBILFZ11} the proof of our main theorem is given and in  \ref{numerics}  numerical evidence supporting the results of Section \ref{SECT:II} is presented. 

\section{Motivation}
\label{motivation}

\subsection{The generalized harmonic potential}

Scalar fields are relevant in the physical description of the universe, particularly, in inflationary scenario \cite{Guth:1980zm,Linde:1983gd,Linde:1986fd,Linde:2002ws,Guth:2007ng}. For example, chaotic inflation is a model of cosmic inflation in which  the potential term takes the form of the harmonic potential $V(\phi)= \frac{m_\phi^2 \phi^2}{2}$ \cite{Linde:1983gd,Linde:1986fd,Linde:2002ws,Guth:2007ng}. 
\newline 
In this research we consider the generalized harmonic potential which incorporates cosine-like corrections 
\begin{equation}\label{pot}
 V(\phi)= \mu ^3 \left[b f \left(1-\cos \left(\frac{\phi }{f}\right)\right)+\frac{\phi ^2}{\mu}\right], \;    b> 0,
\end{equation} with $\mu ^3 b f\ll 1$. 
\newline
Introducing a new parameter $\omega$ through the equation $b \mu ^3+2 f \mu ^2-f \omega ^2=0$, potential \eqref{pot}
can be re-expressed as \begin{equation}
\label{pot_v2}
    V(\phi)=\mu ^2 \phi ^2 + f^2 \left(\omega ^2-2 \mu ^2\right) \left(1-\cos \left(\frac{\phi
   }{f}\right)\right). 
\end{equation}
The applicability of this re-parametrization will be discussed at the end of section \ref{section2-3}.  

Potential \eqref{pot_v2}  has the following generic features:
\begin{enumerate} 
    \item$V$ is a real-valued smooth function  $V\in C^{\infty} (\mathbb{R})$  with  $\lim_{\phi \rightarrow \pm \infty} V(\phi)=+\infty$. 
        \item $V$ is an even function  $V(\phi)=V(-\phi)$.
    \item  $V(\phi)$ has always a local minimum at $\phi=0$;  $V(0)=0, V'(0)=0, V''(0)= \omega^2> 0$.
    \item There is a finite number of values $\phi_c \neq 0$ satisfying $2 \mu ^2 \phi_c +f \left(\omega ^2-2 \mu ^2\right) \sin \left(\frac{\phi_c
   }{f}\right)=0$, which are local maximums or local minimums depending on whether  $V''(\phi_c):= 2 \mu ^2+\left(\omega ^2- 2 \mu ^2\right) \cos \left(\frac{\phi_c }{f}\right)<0$ or $V''(\phi_c)>0$. For $\left|\phi_c\right| >\frac{f(\omega^2-2 \mu^2)}{2 \mu^2}= \phi_*$, this set is empty. 
    \item There exist 
    $V_{\max}= \max_{\phi\in [-\phi_*,\phi_*]} V(\phi)$   and $V_{\min}= \min_{\phi\in [-\phi_*,\phi_*]} V(\phi)=0$. The function $V$ has no upper bound  but it has a lower bound equal to zero.
  \end{enumerate}
The asymptotic features of potential \eqref{pot_v2} are the following.  Near the global minimum $\phi=0$, we have 
$V(\phi) \sim \frac{\omega ^2 \phi ^2}{2}+\mathcal{O}\left(\phi ^3\right), \quad \text{as} \; \phi\rightarrow 0$. 
That is, $\omega^2$ can be related to the mass of the scalar field near its global minimum.  As $\phi\rightarrow \pm \infty$ the cosine- correction is bounded, then  $V(\phi) \sim \mu ^2 \phi ^2+\mathcal{O}\left(1\right) \quad \text{as} \; \phi\rightarrow \pm \infty$.  This makes it suitable to describe oscillatory behavior in cosmology. 

Potential \eqref{pot} or \eqref{pot_v2} is related but not equal to the monodromy potential of  \cite{Sharma:2018vnv} used in the context of loop-quantum gravity, which is a particular case of the general monodromy potential  \cite{McAllister:2014mpa}. 
In references  \cite{Leon:2019iwj,Leon:2020ovw,Leon:2020pvt} it was proved that the potential of \cite{Sharma:2018vnv,McAllister:2014mpa} for $p=2$, say 
$V(\phi)= \mu^3 \left[\frac{\phi^2}{\mu} + b f \cos\left(\frac{\phi}{f}\right)\right]$, $b\neq 0$ is not good to describe the late-time FLRW universe driven by a scalar field  because it has two symmetric local negative minimums  which are related to  Anti-de Sitter solutions. 
Therefore, in \cite{Leon:2019iwj,Leon:2020ovw} the following potential  was studied
\begin{equation}
\label{EQ:23}
    V(\phi)= \frac{\phi^2}{2}+ f\left[1- \cos\left(\frac{\phi}{f}\right)\right], 
\end{equation}
that is obtained by setting  $\mu=\frac{\sqrt{2}}{2}$ and $b \mu=2$ in eq. \eqref{pot}.  On the other hand, setting
$\mu=\frac{\sqrt{2}}{2}$ and  $\omega=\sqrt{2}$, we have 
\begin{equation}
\label{pot28}
 V(\phi)= \frac{\phi ^2}{2} + f^2 \left[1-\cos \left(\frac{\phi }{f}\right)\right]. 
\end{equation}
The potentials \eqref{EQ:23} and \eqref{pot28} provide non-negative local minimums  which can be related to a late-time accelerated universe. The generalized harmonic potentials \eqref{pot_v2},  \eqref{EQ:23} and  \eqref{pot28}
belong to the class of potentials studied by  \cite{Rendall:2006cq}.  Additionally, potentials like  
$V(\phi)=\Lambda^4 \left[1- \cos \left(\frac{\phi}{f}\right)\right]$ are of interest in the context of  axion models \cite{DAmico:2016jbm}. In  \cite{Balakin:2020coe} axionic dark matter with  modified periodic potential for the pseudoscalar
field  \newline $V(\phi, \Phi_*)= \frac{m_A^2 {\Phi_*}^2}{2 \pi^2}\left[1- \cos \left(\frac{2 \pi \phi}{\Phi_*}\right)\right]$  has been studied in the framework of the axionic extension of the Einstein-aether theory. This periodic potential has minima at $\phi =n \Phi_*, n \in \mathbb{Z}$, whereas maxima
are found when $n \rightarrow m +
\frac{1}{2}$. Near the minimum, i.e., $\phi =n \Phi_* + \psi$ with $|\psi|$ a small value, $V \rightarrow \frac{m_A^2 \psi^2}{2}$ where $m_A$ the axion rests mass.

\subsection{Simple example}
Given the ordinary differential equation $\dot{\mathbf{x}}= \mathbf{f}(\mathbf{x}, t,\varepsilon)$ with $\varepsilon\geq 0$ and $\mathbf{f}$ periodic in $t$. One approximation scheme which can be used to solve the full problem is the resolution of the unperturbed problem $\dot{\mathbf{x}}= \mathbf{f}(\mathbf{x}, t,0)$ by setting $\varepsilon=0$ at first and then with the use of the approximated unperturbed solution to formulate variational equations in standard form which can be averaged. The term averaging is related to the approximation  of initial value problems involving perturbations (chapter 11,  \cite{Verhulst}). 
\newline
For example, consider the initial value problem: 
 \begin{equation}
 \label{harm-osc}
 \ddot \phi + \omega^2 \phi = \varepsilon (-2 \dot \phi)
 \end{equation}
 with $\phi(0)$  and $\dot\phi(0)$  prescribed. The unperturbed problem 
 $\ddot \phi +\omega^2 \phi = 0$  
admits the solution 
 $\dot\phi(t)= r_0 \omega \cos (\omega t-\Phi_0), \;  \phi(t)= r_0 \sin (\omega t-\Phi_0)$,
 where $r_0$ and $\Phi_0$ are constants depending on initial conditions. 
 
 Let be defined the amplitude-phase transformation (\cite{Verhulst}, chapter 11):
 \begin{small}
 \begin{equation}
 \dot{\phi}(t)= r(t) \omega \cos (\omega t-\Phi(t)), \;  \phi(t)  = r(t) \sin (\omega t-\Phi(t)),
 \end{equation}
 \end{small}
 such that
 \begin{small}
 \begin{equation}
 \label{eqAA25}
 r=\frac{\sqrt{\dot{\phi}^2(t)+\omega ^2 \phi^2(t)}}{\omega }, \;  \Phi =\omega t-\tan ^{-1}\left(\frac{\omega \phi
   (t)}{\dot \phi(t)}\right). 
 \end{equation}
 \end{small}
 Then,  eq. \eqref{harm-osc} 
 becomes, 
 \begin{equation}
 \label{eq4}
 \dot r= -2 r \varepsilon  \cos ^2(t-\Phi), \;  \dot\Phi = - \varepsilon \sin (2 (t-\Phi )).
 \end{equation}
From \eqref{eq4} it follows that $r$ and $\Phi$ are  slowly varying with time,
 and the system takes the form $\dot y= \varepsilon f(y)$.  The idea is consider only nonzero average of the right-hand-sides keeping $r$ and $\Phi$
 fixed  and leaving out the terms with average zero and ignoring the slow-varying dependence of $r$ and $\Phi$ on $t$ through the averaging process \begin{equation}
\label{timeavrg}
      \bar{\mathbf{f}}(\cdot):=\frac{1}{L} \int_{0}^L \mathbf{f}(\cdot, t) dt, \quad L=\frac{2 \pi}{\omega}.  
\end{equation}
Replacing $r$ and $\Phi$ by their averaged approximations $\bar{r}$ and $\bar{\Phi}$ we obtain the system
 \begin{align}
 \label{eq6}
 & \dot {\bar{r}} = - \varepsilon \omega \bar{r}, 
\quad  \dot{\bar{\Phi}} = 0. 
 \end{align}
   Solving \eqref{eq6} with $\bar{r}(0)=r_0$ and $\bar{\Phi}(0)= \Phi_0$, we obtain  $\bar{\phi}= r_0 e^{-\varepsilon \omega t} \sin (\omega t-\Phi_0)$, which is an accurate approximation of the exact solution
   \begin{align*}
     & \phi(t)=  -r_0  e^{-t \varepsilon }  \sin (\Phi_0) \cos
   \left(t \sqrt{\omega ^2-\varepsilon ^2}\right) \nonumber \\
   & -\frac{r_0 e^{-t
   \varepsilon } \sin \left(t \sqrt{\omega ^2-\varepsilon ^2}\right) (\varepsilon 
   \sin (\Phi_0)-\omega  \cos (\Phi_0))}{\sqrt{\omega
   ^2-\varepsilon ^2}},
   \end{align*} due to 
\begin{align*}
\bar{\phi}(t) - \phi(t)  =  \frac{r_0 \varepsilon  e^{-t \varepsilon } \sin (\Phi_0) \sin (t  \omega )}{\omega } + \mathcal{O}\left(\varepsilon  e^{-t \varepsilon }\right)
\end{align*}
as $\varepsilon \rightarrow 0^+$.

\subsection{General class of systems with a time-dependent perturbation parameter}
\label{section2-3}
Let us consider for example the KG system
\begin{align}
\label{KGharmonic}
   & \ddot \phi + \omega^2 \phi = -3 H \dot \phi, \\
   &\dot{H}= -\frac{1}{2}\dot\phi^2. \label{Friedmann}
\end{align} The similarity between \eqref{harm-osc} and \eqref{KGharmonic}  suggests to treat the latter as a perturbed harmonic
oscillator as well, and to apply averaging in an analogous way. However, care has to be taken because in contrast to $\varepsilon$, $H$ is time-dependent and itself is governed by the evolution equation \eqref{Friedmann}. If it is valid, then a surprising feature of such approach is the possibility of exploiting the fact that it is strictly decreasing and goes to zero by promoting  Hubble parameter $H$  to a time-dependent perturbation parameter in \eqref{KGharmonic} controlling the magnitude of the error between solutions of the  full and time-averaged problems. 
Hence, with strictly decreasing $H$ the error should decrease as well. 
Therefore, it is possible to obtain the information about the large-time behavior of the more complicated full system via an analysis of the simpler averaged system equations by means of dynamical systems techniques \cite{Coddington55,Hale69,AP,wiggins,perko,160,Hirsch,165,LaSalle,aulbach,TWE,coleybook,Coley:94,Coley:1999uh,Copeland:1997et,vandenHoogen:1999qq,bassemah,LeBlanc:1994qm,Heinzle:2009zb,Foster:1998sk,Miritzis:2003ym,Giambo:2008ck,Leon:2014rra,Leon:2010ai,Fadragas:2014mra,Dania&Yunelsy,Leon:2008de,Giambo:2009byn,Tzanni:2014eja}. With this in mind, in \cite{Fajman:2021cli} the long-term behavior of solutions of a general class of systems in standard form was studied: 
\begin{equation}
\label{standard51}
\left(\begin{array}{c}
       \dot{H} \\
        \dot{\mathbf{x}}
  \end{array}\right)= H \left(\begin{array}{c}
       0 \\
       \mathbf{f}^1 (\mathbf{x}, t)
  \end{array}\right) + H^2\left(\begin{array}{c}
       f^{[2]} (\mathbf{x}, t) \\
       \mathbf{0}
  \end{array}\right),
  \end{equation}
where $H$ is positive strictly decreasing in $t$ and $\lim_{t\rightarrow \infty}H(t)=0$. 

In this paper we study systems which are not in the standard form \eqref{standard51} but can be expressed as a series with center in $H=0$ according to the equation
\begin{align}
\label{nonstandtard}
 \left(\begin{array}{c}
       \dot{H} \\
        \dot{\mathbf{x}}
  \end{array}\right)= &\left(\begin{array}{c}
    0 \\
       \mathbf{f}^0 (\mathbf{x}, t)
  \end{array}\right)+ H \left(\begin{array}{c}
       0 \\
       \mathbf{f}^1 (\mathbf{x}, t)
  \end{array}\right) \nonumber \\
  & + H^2\left(\begin{array}{c}
       f^{[2]} (\mathbf{x}, t)  \\
       \mathbf{0}
  \end{array}\right)+ \mathcal{O}(H^3),
  \end{align}
depending on a parameter $\omega$ which is a free frequency that can be tuned to make $\mathbf{f}^0 (\mathbf{x}, t)= \mathbf{0}$. Therefore,  systems  can be expressed in the standard form \eqref{standard51}. In particular, assuming $\omega ^2>2 \mu ^2$ and  setting $f=\frac{b \mu ^3}{\omega ^2-2 \mu ^2}$ (which is equivalent to tune the angular frequency $\omega$) the undesired terms evolving as  $\propto H^{0}$ are eliminated  in the series expansion around $H=0$. 
\section{The model}
\label{Sect2}
It is well-known that there is an interesting hierarchy in Bianchi models \cite{WE,Ryan2016,coleybook,Plebanski2006}. In particular,   LRS Bianchi I model naturally appears as a boundary subset of   LRS Bianchi III model. The last one is an invariant boundary of   LRS Bianchi type VIII model as well. Additionally, LRS Bianchi type VIII can be viewed as an invariant boundary of  LRS Bianchi type IX models \cite{BC1,BC2,BC3,BC4,BC5,BC6}. Bianchi spacetimes contain many important cosmological models that have been used to study anisotropies of primordial universe and its evolution towards the observed isotropy of the present epoch \cite{jacobs2,collins,JB1,JB2,PhysRevD.101.044046}. The list includes FLRW model in the limit of the isotropization.

\noindent In GR the Hubble parameter is always monotonic for Bianchi I and anisotropies decay for $H>0$. Therefore, isotropization occurs \cite{nns1,heu}. The exact solutions of  field equations  have been found in some particular Bianchi spacetimes for an exponential potential \cite{b1,b2,b3}. These exact solutions lead to isotropic homogeneous spacetimes as it was found in references \cite{coley1,coley2}. An anisotropic solution of special interest is  Kasner spacetime \cite{kas1,kas2,kas3,kas4,barcl,barcl2,anan01,anan02}, essential for the description of  BKL singularity  \cite{bkl}.

\noindent
The action integral of interest given by 
\begin{small}
\begin{align}&
 {\cal S}=\int  d{ }^4 x \sqrt{|g|}\left[\frac{1}{2} R   -\frac{1}{2} g^{\mu
\nu}\nabla_\mu\phi\nabla_\nu\phi-V(\phi)+ \mathcal{L}_{m}\right]  \label{eq1}
\end{align} 
\end{small}
\noindent
is expressed in a system of units in which $8\pi G=c=\hslash=1$. In eq. \eqref{eq1} $R$ is the scalar curvature of the spacetime, $\mathcal{L}_{m}$ is the Lagrangian density of matter, $\phi$ is the scalar field, $\nabla_\alpha$ is the covariant derivative and $V(\phi)$ is the scalar field potential defined by \eqref{pot}.

\subsection{LRS Bianchi III, Bianchi I and Kantowski-Sachs models}
Considering that
\begin{align}
  & \lim_{k\rightarrow -1}  k^{-1} \sin^2 (\sqrt{k} \vartheta)=\sinh^2 (\vartheta), \\
  & \lim_{k\rightarrow 0}  k^{-1} \sin^2 (\sqrt{k} \vartheta)= \vartheta^2, \label{BianchiIlimit} \\ 
   & \lim_{k\rightarrow +1}  k^{-1} \sin^2 (\sqrt{k} \vartheta)=\sin^2 (\vartheta), 
\end{align}
the metric element for LRS Bianchi III,  Bianchi I  and Kantowski-Sachs models can be written as \cite{Nilsson:1995ah}
\begin{align}
\label{metric}
   &  ds^2= - dt^2 + \left[{e_1}^1(t)\right]^{-2} dr^2 \nonumber \\
   & + \left[{e_2}^2(t)\right]^{-2}  \left[ d \vartheta^2 + k^{-1} \sin^2 (\sqrt{k} \vartheta)d \zeta^2\right],
\end{align}
where ${e_1}^1$, ${e_2}^2$ and ${e_3}^3= \sqrt{k} {e_2}^2/\sin(\sqrt{k} \vartheta)$ are functions of $t$ which are components of the frame vectors \cite{Coley:2008qd}: $\mathbf{e}_0= \partial_t, \quad  \mathbf{e}_1 = {e_1}^1\partial_r, \quad \mathbf{e}_2={e_2}^2 \partial_\vartheta, \quad \mathbf{e}_3={e_3}^3 \partial_\zeta$. Comparing with reference \cite{Nilsson:1995ah} we have settled the parameters $a=f=0$ and ${e_1}^1(t)=D_2(t)^{-1}, \; {e_2}^2(t)=D_2(t)^{-1}$ and we have used the identifications $(\vartheta, \zeta)=(y, z)$.  The line elements for  spatially homogeneous self-similar LRS models have been given by Wu in \cite{Chao:1980ky}. We concentrate only in the spatially homogeneous but anisotropic class  with the exception of spatially homogeneous LRS Bianchi V, that is: LRS Bianchi III ($k=-1$), Bianchi I  ($k=0$) and Kantowski-Sachs  ($k=+1$) \cite{Fadragas:2013ina}. It is useful to define a representative length $\ell(t)$ along worldlines of $\mathbf{u}=\partial_t$  for describing the volume expansion (contraction) behavior of the congruence completely by \cite{vanElst:1996dr} 
\begin{equation}
\frac{\dot\ell (t)}{\ell (t)}= H(t):=     -\frac{1}{3}\frac{d}{d t} \ln\left[ {e_1}^1(t) ( {e_2}^2(t))^2\right],
\label{HubbleGen}
\end{equation}
where dots denote derivatives with respect to  time $t$, $H(t)$ is the Hubble parameter in terms of $\ell (t)$ its time derivative. The anisotropic parameter $\sigma_{+}(t)$ is defined by
\begin{equation}
\sigma_+ = \frac{1}{3}\frac{d}{d t} \ln\left[ {e_1}^1(t) ( {e_2}^2(t))^{-1}\right].
\end{equation} 
The variation of \eqref{eq1}  for the 1-parameter family of metrics \eqref{metric} leads to \cite{Fadragas:2013ina}: 
\begin{align}
    & 3 H^2 + k K = 3 {\sigma_+}^2+ \rho_m + \frac{1}{2}\dot\phi^2 +V(\phi), \label{EQQ32}
\\
    & -3 ({\sigma_+}+H)^2 -2 \dot{\sigma_+}- 2 \dot{H} - k K \nonumber \\
    & = (\gamma-1)\rho_m + \frac{1}{2}\dot\phi^2-V(\phi), \label{EQQ33}
\\
      & -3 {\sigma_+}^2 +3  {\sigma_+} H-3 H^2 + \dot{\sigma_+}- 2 \dot{H} \nonumber \\
      & = (\gamma-1)\rho_m + \frac{1}{2}\dot\phi^2-V(\phi), \label{EQQ34}
\end{align}
where for the matter component we use barotropic EoS $p_m=(\gamma-1)\rho_m$ with $p_{m}$ the pressure of the fluid,  $\rho_{m}$ is the energy density and the barotropic index is a constant $\gamma$ which satisfies $0\leq \gamma\leq 2$. 
\newline
The  Gauss curvature of the spatial 2-space and 3-curvature scalar are \cite{Coley:2008qd}
\begin{equation}
 K= ( {e_2}^2(t))^{2}, \quad   \R= 2 k K. \label{def3R}
\end{equation} 
Furthermore, the evolution equation of the Gauss curvature of the spatial 2-space is
\begin{equation}
\label{Gauss}
    \dot{K}= -2 ({\sigma_+}+ H)K, 
\end{equation}
while the evolution for ${e_1}^1$ is given by \cite{Coley:2008qd}
\begin{equation}
\label{eqe1evol}
    \dot{{e_1}^1}= - (H-2 \sigma_+){e_1}^1. 
\end{equation}
From eqs. \eqref{EQQ33} and \eqref{EQQ34} the shear equation
\begin{equation}
 \dot{\sigma_+}= -3 H {\sigma_+} -\frac{k K}{3}\label{EQQ39}
\end{equation} is obtained.  
Eqs. \eqref{EQQ32}, \eqref{EQQ33}, \eqref{EQQ34} and  \eqref{EQQ39} give the Raychaudhuri equation 
\begin{equation}
\label{Raychaudhuri}
    \dot{H}=- H^2 -2 {\sigma_+}^2 -\frac{1}{6}(3 \gamma -2) \rho_m -\frac{1}{3}\dot{\phi}^2+\frac{1}{3} V(\phi).
\end{equation}
Finally, the matter and KG equations are  
\begin{align}
& \dot{\rho_m}=-3 \gamma H \rho_m, \label{cons1}
\\
& \ddot{\phi}= -3 H \dot{\phi}- \frac{d V(\phi)}{d \phi}. \label{cons2}\end{align}
\noindent
In this paper we will focus our study in LRS Bianchi I model. Therefore, using eq. \eqref{BianchiIlimit}  the metric \eqref{metric} reduces to
\begin{align}
\label{metricLRSBI}
   &  ds^2= - dt^2 + A^2(t) dr^2 
 + B^2(t)  \left(d \vartheta^2 +  \vartheta^2 d \zeta^2\right),
\end{align} 
where the functions $A(t)$ and $B(t)$ are interpreted as the scale factors: $A(t)= {e_1}^1(t)^{-1}$ and $B^2(t)= K(t)^{-1}$.

\subsection{FLRW models}
The general line element for spherically symmetric models can be written as \cite{Coley:2008qd}
\begin{align}
\label{sphsymm}
   &  ds^2= - dt^2 + \left[{e_1}^1(t,r)\right]^{-2} dr^2 \nonumber \\
   & + \left[{e_2}^2(t,r)\right]^{-2} (d\y^2 + \sin^2 \y\, d\z^2).
\end{align}
Spatially homogeneous spherically symmetric models, that are not Kantowski-Sachs,  are the FLRW models  where the metric can be written as 
\begin{align}
\label{metricFLRW}
& ds^2 = - dt^2 + a^2(t) \Big[ dr^2+  f^2(r) (d\y^2 + \sin^2 \y\, d\z^2)\Big],
\\
\label{fx_FLRW}
& \text{with}\; f(r) = \sin r,\ r,\ \sinh r,
\end{align}
for closed, flat and open FLRW models, respectively. 
In comparison with metric \eqref{sphsymm}, the frame coefficients are given by $\ex = a^{-1}(t)$ and $\ey = a^{-1}(t) f^{-1}(r)$ where $a(t)$ is the scale factor. The anisotropic parameter $\sigma_{+}= \frac13\frac{\partial}{\partial t} \ln(\ex/\ey)$ vanishes and the Hubble parameter \eqref{HubbleGen} can be written as $H = \frac{d}{d t} \ln\left[a(t)\right]$. Furthermore, by calculating $\R$  we obtain
\be
    \R = \frac{6k}{a^2},\quad
    k = 1,0,-1,
\ee
for closed, flat and open FLRW, respectively. Therefore, evolution/constraint equations reduce to
\begin{subequations}
	\label{Non_minProb1FLRW0-1}
	\begin{align}
	&\ddot{\phi}= -3 H \dot{\phi}-V'(\phi),
\\
	&\dot{\rho_m}=- 3\gamma H\rho_m,
	\\
	&\dot{a} = a H, 
	\\
	& \dot{H}=-\frac{1}{2}\left(\gamma \rho_m+{\dot \phi}^2\right)+\frac{k}{a^2},
	\\
	& 3H^2=\rho_m+\frac{1}{2}{\dot{\phi}}^2+V(\phi)-\frac{3 k}{a^2}.
	\end{align}
\end{subequations}

\section{Averaging scalar field cosmologies}
\label{SECT:II}

As in reference \cite{Fajman:2020yjb} we construct a time-averaged version of the original system and prove that it shares the same late-time dynamics of the original system. 

\subsection{Bianchi I metric}
\label{SECT3.3}
In this section   averaging methods are applied for Bianchi I metrics for the generalized harmonic potential \eqref{pot} minimally coupled to matter. 

Setting $k=0$ in Eqs. \eqref{EQQ32}, \eqref{EQQ39} and \eqref{Raychaudhuri} we obtain
\begin{align}
    &\dot{H}=- H^2 -2 {\sigma_+}^2 -\frac{1}{6}(3 \gamma -2) \rho_m -\frac{1}{3}\dot{\phi}^2+\frac{1}{3} V(\phi),\\
    &  \dot{\sigma_+}= -3 H {\sigma_+},  \label{EQQ39new} \\
    & 3 H^2 = 3 {\sigma_+}^2+ \rho_m + \frac{1}{2}\dot\phi^2 +V(\phi). \label{NewEQQ32}
\end{align}
Using the characteristic length scale $\ell$ along worldlines of the 4-velocity field  such that  $H=\frac{\dot \ell}{\ell}$, defining $\ell_0$ the current value of $\ell$ such that 
\begin{equation}
\label{characteristic-length}
\frac{\ell(t)}{\ell_0}= \left[ {e_1}^1(t) ( {e_2}^2(t))^2\right]^{-\frac{1}{3}}, \; \tau =  \ln\left(\frac{\ell(t)}{\ell_0}\right),
\end{equation} 
and denoting by convention $t=0$ the current time,  then $ \left(\frac{\ell(0)}{\ell_0}\right)^3= \frac{1}{{e_1}^1(0) ( {e_2}^2(0))^2} =1$ and $\tau(0)=0$.
Using the definition \eqref{HubbleGen} and integrating \eqref{EQQ39new} we obtain $ \sigma_+ = \sigma_0  \ell_0^3/\ell^3$, where $\sigma_0$ is an integration constant, which is the value of $\sigma_+$ when $\ell=\ell_0$. The term $G_0(\ell)=\sigma_0^2 \ell_0^6/\ell^6$, which corresponds to anisotropies in Bianchi I metric, does not correspond to a fluid component in the model. However, it can be interpreted as a stiff-matter fluid for flat FLRW metric with scale factor $a(t)=\ell(t)$. The term $\sigma_+^2$ dilutes very fast with expansion,  isotropizing if $H>0$. 

\noindent The evolution equation for matter and the KG equation do not depend on $k$. Therefore,  the field equations are deduced:
\begin{subequations}
\label{Non_minProb1BI}
\begin{align}
	&\ddot{\phi}= -3 H \dot{\phi}-V'(\phi),  \\
	&\dot{\rho_m}=- 3\gamma H\rho_m, \\
	&\dot \ell = \ell H, \\
	& \dot{H}=-\frac{1}{2}\left(\gamma \rho_m+{\dot \phi}^2\right)-\frac{3 \sigma_0^2 \ell_0^6}{\ell^6}, \\
	& 3H^2=\rho_m+\frac{1}{2}\dot\phi^2+V(\phi)+\frac{3 \sigma_0^2 \ell_0^6}{\ell^6}.
	\end{align}
	\end{subequations}
Now, we define Hubble normalized variables 
\begin{align}
\label{vars}
&\Omega= \frac{\omega r}{\sqrt{6 } H}, \;  \Sigma= \frac{\sp}{H},
\end{align}
along with $r$ and $\Phi$ which are defined in \eqref{eqAA25}  and $\sigma_+ = \sigma_0  \ell_0^3/\ell^3$  is obtained by integrating \eqref{EQQ39new}. Then, we obtain  the system
\begin{subequations}
\label{unperturbed1}
 \begin{align}
   & \dot{\Omega}= -\frac{b \gamma  f \mu ^3 \Omega}{H} \sin \scriptscriptstyle ^2\left(\frac{\sqrt{\frac{3}{2}} H \Omega  \sin (t \omega -\Phi )}{f \omega }\right) \nonumber \\
   & -\frac{b \mu ^3}{\sqrt{6} H} \cos (t \omega -\Phi ) \sin \scriptscriptstyle \left(\frac{\sqrt{6} H \Omega
    \sin (t \omega -\Phi )}{f \omega }\right) \nonumber \\
    & +\frac{3}{2} H \Omega  \left(-(\gamma -2) \Sigma ^2+\gamma -\frac{2 \gamma  \mu ^2 \Omega ^2}{\omega ^2}\right) \nonumber \\
    & +H \cos ^2(t \omega -\Phi )
   \scriptscriptstyle\left(\Omega ^3 \left(\gamma  \left(\frac{3 \mu ^2}{\omega ^2}-\frac{3}{2}\right)+3\right)-3 \Omega \right) \nonumber \\
   & +\frac{\left(\omega ^2-2 \mu ^2\right) \Omega  \sin (2 t \omega -2 \Phi )}{2 \omega },
\end{align}
\begin{align}
   & \dot{\Sigma}=-\frac{b \gamma  f
   \mu ^3 \Sigma}{H} \sin ^2 \scriptscriptstyle\left(\frac{\sqrt{\frac{3}{2}} H \Omega  \sin (t \omega -\Phi )}{f \omega }\right)\nonumber \\
   & +H \Bigg(-\frac{3}{2} (\gamma -2) \Sigma  \left(\Sigma ^2-1\right) \nonumber \\
   & -\frac{3 \gamma  \mu ^2 \Sigma 
   \Omega ^2 \sin ^2(t \omega -\Phi )}{\omega ^2}\Bigg) \nonumber \\
   & -\frac{3}{2} (\gamma -2) \Sigma  H \Omega ^2 \cos ^2(t \omega -\Phi ),
\\
   & \dot{\Phi}= -\frac{b \mu ^3 \sin (t \omega -\Phi )}{\sqrt{6} H \Omega } \sin \scriptscriptstyle\left(\frac{\sqrt{6} H \Omega  \sin
   (t \omega -\Phi )}{f \omega }\right) \nonumber\\
   & -\frac{3}{2} H \sin (2 (t \omega -\Phi)) \nonumber \\
   & +\frac{\left(\omega ^2-2 \mu ^2\right) \sin ^2(t \omega -\Phi )}{\omega }, 
\\
      & \dot H= -(1+q)H^2, \label{RaychBI}
\end{align}
\end{subequations}
where the deceleration parameter $q$ is given by 
\begin{small}
\begin{align}
   & q =-1  -\frac{b \gamma  f \mu ^3}{H^2} \sin ^2 \scriptscriptstyle\left(\frac{\sqrt{\frac{3}{2}} H \Omega \sin (t \omega -\Phi )}{f \omega
   }\right) \nonumber \\
   & +\frac{3}{2} \left(-(\gamma -2) \Sigma ^2+\gamma -\frac{2 \gamma  \mu ^2 \Omega ^2 \sin ^2(t \omega -\Phi )}{\omega ^2}\right) \nonumber \\
   & -\frac{3}{2} (\gamma -2) \Omega ^2 \cos ^2(t \omega -\Phi ). \label{qBI}
\end{align}
\end{small}
Denoting $\mathbf{x}= \left(\Omega, \Sigma, \Phi \right)^T$  system \eqref{unperturbed1} can be symbolically written  in form \eqref{nonstandtard}. Notice that  using the condition $b \mu ^3+2 f \mu ^2-f \omega ^2=0$, the function
\begin{equation}
\mathbf{f}^0 (\mathbf{x}, t) 
= \left(
\begin{array}{c}
 \frac{\Omega  \left(f \omega ^2-\mu ^2 (b \mu +2 f)\right) \sin (2 t \omega -2 \Phi  )}{2 f \omega } \\
 0 \\
 \frac{\left(-\frac{b \mu ^3}{f}-2 \mu ^2+\omega ^2\right) \sin ^2(t \omega -\Phi  )}{\omega } \\
\end{array}
\right),
\end{equation} 
in the eq. \eqref{nonstandtard} becomes trivial. Hence, we obtain: 
\begin{small}
\begin{align}
& \dot{\mathbf{x}}= H \mathbf{f}(\mathbf{x}, t)+  \mathcal{O}(H^2), \label{equx}\\
& \dot{H}= -\frac{3}{2} H^2 \Bigg(\gamma  \left(1-\Sigma ^2-\Omega ^2\right) +2 \Sigma ^2+ 2\Omega ^2 \cos^2(\Phi -t
   \omega )\Bigg) \nonumber \\
   & + \mathcal{O}(H^3), \label{EQ:61b}
\end{align}
\end{small}
\noindent where \eqref{EQ:61b} is Raychaudhuri equation and 
\begingroup\makeatletter\def\f@size{8}\check@mathfonts
\begin{align}
\label{EQ:87}
   & \mathbf{f}(\mathbf{x}, t) \nonumber \\
   & =
   \left(\begin{array}{c}
\frac{3}{2} \Omega  \Bigg(\gamma  \left(1-\Sigma ^2-\Omega ^2\right)+2 \Sigma ^2  
+2 \left(\Omega ^2-1\right) \cos^2(\Phi -t \omega )\Bigg) \\\\
 \frac{3}{2} \Sigma  \Bigg(-\gamma  \left(\Sigma ^2+\Omega
   ^2-1\right)+2 \Sigma ^2 
   +2\Omega ^2 \cos^2  (\Phi -t \omega )-2\Bigg)\\\\
-\frac{3}{2} \sin (2 t \omega -2\Phi)
      \end{array}
   \right).
\end{align}
\endgroup
Replacing $\dot{\mathbf{x}}= H \mathbf{f}(\mathbf{x}, t)$ with $\mathbf{f}(\mathbf{x}, t)$ as defined in \eqref{EQ:87} 
by $\dot{\mathbf{y}}= H  \bar{\mathbf{f}}(\mathbf{y})$ with  $\mathbf{y}= \left(\bar{\Omega}, \bar{\Sigma}, \bar{\Phi} \right)^T$ and $\bar{\mathbf{f}}$ as defined by \eqref{timeavrg},
we obtain the averaged system: 
\begin{align}
    &\dot{\bar{\Omega}}=\frac{3}{2} H \bar{\Omega}  \left(\gamma  \left(1-\bar{\Sigma} ^2-\bar{\Omega }^2\right)+2 \bar{\Sigma} ^2+\bar{\Omega}^2-1\right), \label{Ieq24}
\\
    &\dot{\bar{\Sigma}}=\frac{3}{2} H \bar{\Sigma}  \left(\gamma  \left(1-\bar{\Sigma} ^2-\bar{\Omega}^2\right)+2 \bar{\Sigma}^2+\bar{\Omega}^2-2\right), \label{Ieq25}
\\
    &\dot{\bar{\Phi}}=0, \label{Ieq26}
 \\
    & \dot{H}= -\frac{3}{2} H^2
   \left(\gamma  \left(1-\bar{\Sigma}^2-\bar{\Omega}^2\right)+2 \bar{\Sigma}^2+\bar{\Omega }^2\right). \label{Ieq27}
\end{align}
Proceeding in analogous way as in references \cite{Alho:2015cza,Alho:2019pku} but for 3 dimensional systems  instead of a 1-dimensional one, we implement a local nonlinear transformation
\begin{align}
   & \Omega =\Omega_{0} +H g_1(H , \Omega_{0}, \Sigma_{0}, \Phi_{0}, t), \nonumber\\
   & \Sigma=\Sigma_{0} +H  g_2 (H , \Omega_{0}, \Sigma_{0}, \Phi_{0},  t), \nonumber\\
   & \Phi=\Phi_{0} +H g_3 (H , \Omega_{0}, \Sigma_{0}, \Phi_{0}, t), \label{quasilinear211}
  \end{align}
which in vector form can be written as
\begin{equation}
\label{EQT53}
    \mathbf{x}= \psi(\mathbf{x}_0):=\mathbf{x}_0 + H \mathbf{g}(H, \mathbf{x}_0,t),
\end{equation}
where $\mathbf{x}_0=\left(\Omega_{0}, \Sigma_{0}, \Phi_{0}\right)^T$ and 
\begin{equation}
\label{eqT55}
 \mathbf{g}(H, \mathbf{x}_0,t)= \left(\begin{array}{c}
    g_1(H , \Omega_{0}, \Sigma_{0}, \Phi_{0}, t)\\
    g_2(H , \Omega_{0}, \Sigma_{0}, \Phi_{0}, t)\\
    g_3(H , \Omega_{0}, \Sigma_{0}, \Phi_{0}, t)\\
 \end{array}\right).   
\end{equation}
\newline 
Taking derivative of \eqref{EQT53} with respect to $t,$ we obtain 
\begin{small}
\begin{align}
    & \dot{\mathbf{x}_0}+ \dot{H} \mathbf{g}(H, \mathbf{x}_0,t)  \nonumber \\
    & + H \Bigg(\frac{\partial }{\partial t} \mathbf{g}(H, \mathbf{x}_0,t) + \dot{H} \frac{\partial }{\partial H} \mathbf{g}(H, \mathbf{x}_0,t)    + \mathbb{D}_{\mathbf{x}_0} \mathbf{g}(H, \mathbf{x}_0,t) \cdot \dot{\mathbf{x}_0}\Bigg) \nonumber \\
    & = \dot{\mathbf{x}}, \label{EQT56}
    \end{align}
    \end{small}
    where 
    \begin{equation}
        \mathbb{D}_{\mathbf{x}_0} \mathbf{g}(H, \mathbf{x}_0,t)= \left(\begin{array}{ccc}
             \frac{\partial g_1}{\partial \Omega_0}&  \frac{\partial g_1}{\partial \Sigma_0} &  \frac{\partial g_1}{\partial \Phi_0}\\
             \frac{\partial g_2}{\partial \Omega_0}&  \frac{\partial g_2}{\partial \Sigma_0} &  \frac{\partial g_2}{\partial \Phi_0}\\
                \frac{\partial g_3}{\partial \Omega_0}&  \frac{\partial g_3}{\partial \Sigma_0} &  \frac{\partial g_3}{\partial \Phi_0}\\
        \end{array}
                \right)
    \end{equation}    
is the  Jacobian matrix of $\mathbf{g}(H, \mathbf{x}_0,t)$ for the vector  $\mathbf{x}_0$.  The function $$\mathbf{g}(H, \mathbf{x}_0,t)$$ is conveniently chosen. 
\newline By substituting \eqref{equx} and \eqref{EQT53} in \eqref{EQT56} we obtain 
\begin{small}
\begin{align}
       & \Bigg(\mathbf{I}_3 + H \mathbb{D}_{\mathbf{x}_0} \mathbf{g}(H, \mathbf{x}_0,t)\Bigg) \cdot \dot{\mathbf{x}_0}= H \mathbf{f}(\mathbf{x}_0 + H \mathbf{g}(H, \mathbf{x}_0,t),t) \nonumber \\
       & -H \frac{\partial }{\partial t} \mathbf{g}(H, \mathbf{x}_0,t) -\dot{H} \mathbf{g}(H, \mathbf{x}_0,t) -H \dot{H} \frac{\partial }{\partial H} \mathbf{g}(H, \mathbf{x}_0,t), 
\end{align}
\end{small}
where 
$\mathbf{I}_3= \left(\begin{array}{ccc}
             1 & 0 & 0\\
             0 & 1 &  0\\
             0 & 0 & 1\\
        \end{array}
                \right)$ is the $3\times 3$ identity matrix.
\begin{widetext}
Then we obtain 
  \begin{small}  
  \begin{align}
 & \dot{\mathbf{x}_0} = \Bigg(\mathbf{I}_3 + H \mathbb{D}_{\mathbf{x}_0} \mathbf{g}(H, \mathbf{x}_0,t)\Bigg)^{-1} \cdot \Bigg(H \mathbf{f}(\mathbf{x}_0 + H \mathbf{g}(H, \mathbf{x}_0,t),t)-H \frac{\partial }{\partial t} \mathbf{g}(H, \mathbf{x}_0,t) -\dot{H} \mathbf{g}(H, \mathbf{x}_0,t) -H \dot{H} \frac{\partial }{\partial H} \mathbf{g}(H, \mathbf{x}_0,t)\Bigg).  
\end{align}
\end{small}
Using eq. \eqref{EQ:61b}, we have $ \dot{H}= \mathcal{O}(H^2)$. Hence,
\begin{small}
\begin{align}
    & \dot{\mathbf{x}_0} = \underbrace{\Bigg(\mathbf{I}_3 - H \mathbb{D}_{\mathbf{x}_0} \mathbf{g}(0, \mathbf{x}_0,t) +  \mathcal{O}(H^2)\Bigg)}_{3\times 3 \: \text{matrix}} \cdot \underbrace{\Bigg(H \mathbf{f}(\mathbf{x}_0, t)-H \frac{\partial }{\partial t} \mathbf{g}(0, \mathbf{x}_0,t) +   \mathcal{O}(H^2)\Bigg)}_{3\times 1 \; \text{vector}}= \underbrace{H \mathbf{f}(\mathbf{x}_0, t)-H \frac{\partial }{\partial t} \mathbf{g}(0, \mathbf{x}_0,t) +   \mathcal{O}(H^2)}_{3\times 1 \; \text{vector}}.\label{eqT59}
    \end{align} 
    \end{small}
\end{widetext}
\noindent The strategy is to use eq. \eqref{eqT59} for choosing conveniently $\frac{\partial }{\partial t} \mathbf{g}(0, \mathbf{x}_0,t)$  in order to prove that 
\begin{align}
 & \dot{\Delta\mathbf{x}_0}= -H G(\mathbf{x}_0, \bar{\mathbf{x}}) +   \mathcal{O}(H^2), \label{EqY60}
  \end{align}
where $\bar{\mathbf{x}}=(\bar{\Omega},  \bar{\Sigma}, \bar{\Phi})^T$ and  $\Delta\mathbf{x}_0=\mathbf{x}_0 - \bar{\mathbf{x}}$. The function $G(\mathbf{x}_0, \bar{\mathbf{x}})$ is unknown at this stage. 
\newline 
By construction we neglect dependence of $\partial g_i/ \partial t$ and $g_i$ on $H$, i.e., assume $\mathbf{g}=\mathbf{g}(\mathbf{x}_0,t)$ because dependence of $H$ is dropped out along with higher order terms in eq. \eqref{eqT59}. Next, we solve the differential equation  for $\mathbf{g}(\mathbf{x}_0,t)$:  
\begin{align}
     & \frac{\partial }{\partial t} \mathbf{g}(\mathbf{x}_0,t) = \mathbf{f}(\mathbf{x}_0, t) - \bar{\mathbf{f}}(\bar{\mathbf{x}}) + G(\mathbf{x}_0, \bar{\mathbf{x}}). \label{eqT60}
\end{align}
\noindent  where we have considered $\mathbf{x}_0$ and $t$ as independent variables. 
\newline
The right hand side of \eqref{eqT60} is almost periodic of period $L=\frac{2\pi}{\omega}$ for large times. Then, implementing the average process \eqref{timeavrg} on right hand side of \eqref{eqT60}, where the slow-varying dependence of quantities $\mathbf{x}_0=(\Omega_{0}, \Sigma_{0}, \Phi_0)^T$ and  $\bar{\mathbf{x}}=(\bar{\Omega},  \bar{\Sigma}, \bar{\Phi})^T$  on $t$ are ignored through the averaging process, we obtain \begin{align}
    & \frac{1}{L}\int_0^{L} \Bigg[\mathbf{f}(\mathbf{x}_0, s) - \bar{\mathbf{f}}(\bar{\mathbf{x}}) +G(\mathbf{x}_0, \bar{\mathbf{x}}) \Bigg] ds  \nonumber \\
    & = \bar{\mathbf{f}}( {\mathbf{x}}_0)-\bar{\mathbf{f}}(\bar{\mathbf{x}} )+G(\mathbf{x}_0, \bar{\mathbf{x}}). \label{newaverage}
\end{align}
Defining 
\begin{equation}
  G(\mathbf{x}_0, \bar{\mathbf{x}}):=  -\left(\bar{\mathbf{f}}( {\mathbf{x}}_0)-\bar{\mathbf{f}}(\bar{\mathbf{x}})\right)
\end{equation} the average \eqref{newaverage} is zero so that $\mathbf{g}(\mathbf{x}_0,t)$ is bounded.
\newline 
Finally, eq. \eqref{EqY60} transforms to 
\begin{align}
 & \dot{\Delta\mathbf{x}_0}= H \left(\bar{\mathbf{f}}( {\mathbf{x}}_0)-\bar{\mathbf{f}}(\bar{\mathbf{x}})\right) +   \mathcal{O}(H^2)  \label{EqY602}
  \end{align}
and eq. \eqref{eqT60} 
is simplified to 
\begin{align}
     & \frac{\partial }{\partial t} \mathbf{g}(\mathbf{x}_0,t) = \mathbf{f}(\mathbf{x}_0, t) - \bar{\mathbf{f}}( \mathbf{x}_0). \label{eqT602}
\end{align}
Theorem \ref{LFZ11} establishes the existence of the vector \eqref{eqT55}. 
\begin{thm}
\label{LFZ11} Let  the functions 
$\bar{\Omega}, \bar{\Sigma}, \bar{\Phi}$  and $H$ be defined as solutions of the averaged equations \eqref{Ieq24}, \eqref{Ieq25}, \eqref{Ieq26} and \eqref{Ieq27}. Then, there exist continuously differentiable functions $g_1, g_2$ and $g_3$ such that  $\Omega, \Sigma, \Phi$ are locally given by  \eqref{quasilinear211}  where $\Omega_{0}, \Sigma_{0}, \Phi_{0}$ are zero order approximations of $\Omega, \Sigma, \Phi$ as $H\rightarrow 0$. Then, functions $\Omega_{0}, \Sigma_{0}, \Phi_0$ and averaged solution $\bar{\Omega},  \bar{\Sigma}, \bar{\Phi}$  have the same limit as $t\rightarrow \infty$.
Setting $\Sigma=\Sigma_0=0$  analogous results for flat FLRW model are derived. 
\end{thm}
\textbf{Proof.} 
The proof is  given in \ref{gBILFZ11}. 
\newline 
Theorem \ref{LFZ11} implies that $\Omega , \Sigma$ and $\Phi$  evolve according to the averaged equations \eqref{Ieq24}, \eqref{Ieq25}, \eqref{Ieq26}  as $H\rightarrow 0$ because \eqref{quasilinear211} is a formal near-identity (this means that $\mathbb{D}_{\mathbf{x}_0}\psi|_{H=0}=I$) nonlinear change of coordinates, the first order solutions $\Omega_{0}, \Sigma_{0}, \Phi_0$ and averaged solutions $\bar{\Omega},  \bar{\Sigma}, \bar{\Phi}$  have the same limit when $t\rightarrow \infty$  by monotony and non-negativity of $H$ and the limit $H\rightarrow 0$. 
\subsection{Flat FLRW metric.}
\label{SECT:IIIA}
In this case the field equations are obtained from \eqref{Non_minProb1FLRW0-1} by setting $k=0$. We obtain $\dot{\Omega}$, $\dot \Phi$ and $\dot{H}$ by substituting $\Sigma=0$ in \eqref{unperturbed1} and  \eqref{qBI}. 
Finally, we obtain the Taylor expansion:
\begin{small}
\begin{align}
& \dot{\mathbf{x}}= H \mathbf{f}(t, \mathbf{x}) + \mathcal{O}(H^2), \;   \mathbf{x}= \left(\Omega,  \Phi \right)^T, \nonumber\\
& \dot{H}= -H^2 \left(\frac{3}{2} \gamma\left(1-\Omega^2\right)  +3 \Omega^2 \cos ^2(t \omega -\Phi)\right)+  \mathcal{O}(H^3),
\\
\label{EQ:50}
 &  \mathbf{f}(\mathbf{x}, t)  = 
   \left(\begin{array}{c}
   \frac{3}{2} \gamma\left(1-\Omega^2\right) +3  \Omega \left(\Omega^2-1\right) \cos ^2(t \omega -\Phi) \\\\
    -\frac{3}{2} \sin (2 t \omega -2 \Phi)
      \end{array}
   \right).
\end{align} 
\end{small}
Replacing $\dot{\mathbf{x}}= H \mathbf{f}(\mathbf{x}, t)$ and   $\mathbf{f}(\mathbf{x}, t)$ as defined by \eqref{EQ:50} with $\dot{\mathbf{y}}= H  \bar{f}(\mathbf{y})$  where  $\mathbf{y}= \left(\bar{\Omega},  \bar{\Phi} \right)^T$  with the time averaging \eqref{timeavrg}, we obtain the following time-averaged system: 
\begin{align}
    &\dot{\bar{\Omega}}=-\frac{3}{2} H \; \bar{\Omega}   (\gamma -1) \left(\bar{\Omega} ^2-1\right), \label{eq46}\\
   &\dot{\bar{\Phi}}=0. \label{eq48}
\end{align}
The time-averaged Raychaudhuri equation for flat FLRW metric is obtained by setting $\bar{\Sigma}=0$ in eq.  \eqref{Ieq27}. \newline
Theorem \ref{LFZ11} applies for Bianchi I and the invariant set $\Sigma=0$ corresponds to flat FLRW models.

\section{Qualitative analysis of averaged systems}
\label{SECT:III}
\noindent According to Theorem \ref{LFZ11}    for Bianchi I and flat FLRW, Hubble parameter $H$ plays the role of a time-dependent perturbation parameter controlling the magnitude of the error between solutions of  full and time-averaged problems. Thus, oscillations are viewed as perturbations. In time-averaged system   Raychaudhuri equation \eqref{Ieq27} decouples by using a new time variable $\tau$ through $\frac{d f}{d \tau}= \frac{1}{ {H}}\frac{d f}{d t}$. Therefore, the analysis of the original system  is reduced to study the corresponding averaged equations. 

\subsection{Bianchi I metric}
\label{LRSBI}
The averaged system \eqref{Ieq24}, \eqref{Ieq25}, \eqref{Ieq26} and \eqref{Ieq27} is transformed to 
\begin{subequations}
\label{avrgsyst}
\begin{align}
& \frac{d\bar{\Omega}}{d \tau}=\frac{3}{2} \bar{\Omega} \left(\gamma  \left(1-\bar{\Sigma}^2-\bar{\Omega}^2\right)+2 \bar{\Sigma} ^2+ \bar{\Omega}^2-1\right), \label{guidingC1}\\
& \frac{d\bar{\Sigma}}{d \tau}= \frac{3}{2} \bar{\Sigma}\left(\gamma  \left(1-\bar{\Sigma} ^2-\bar{\Omega}^2\right)+2 \bar{\Sigma} ^2+\bar{\Omega}^2-2\right), \label{guidingC2}\\
&     \frac{d{{\bar{\Phi}}}}{d \tau}=0,   \\
& \frac{d{H}}{d \tau} =   -\frac{3}{2} H
  \left(\gamma  \left(1-\bar{\Sigma}^2-\bar{\Omega}^2\right)+2 \bar{\Sigma}^2+\bar{\Omega }^2\right),
\end{align}
\end{subequations}
by defining the logarithmic time $\tau$ through $\frac{d{ {t}}}{d \tau} = 1/{ {H}}$.

\noindent
We investigate the 2D guiding system \eqref{guidingC1}-\eqref{guidingC2}.
\newline The function $\bar{\Omega}_{m}=1-\bar{\Sigma} ^2-\bar{\Omega} ^2$  is interpreted as an averaged Hubble-normalized density parameter for the matter component.  
Therefore, imposing the energy condition $\bar{\Omega}_{m}\geq 0$ the phase space is  
\begin{equation}
    \left\{(\bar{\Omega}, \bar{\Sigma})\in \mathbb{R}^2: \bar{\Omega} ^2+\bar{\Sigma}^2 \leq  1, \bar{\Omega} \geq 0\right\}. 
\end{equation} 
Before the discussion we introduce the following concept. 
A set of non-isolated singular points is said to be normally hyperbolic if the only eigenvalues with zero real parts are those whose corresponding eigenvectors are tangent to the set.
\newline 
Since by definition any point on a set of non-isolated singular points will have at least
one eigenvalue which is zero, all points in the set are non-hyperbolic. However, a set which is
normally hyperbolic can be completely classified as per its stability by considering the signs of eigenvalues in the remaining directions (i.e., for a curve, in the
remaining $n-1$ directions) (see \cite{aulbach}, pp. 36).
\newline
The resulting 2D guiding system \eqref{guidingC1}-\eqref{guidingC2} has the following equilibrium points:
\begin{enumerate}
\item $T: (\bar{\Omega},\bar{\Sigma})=(0,-1)$ with eigenvalues $\left\{\frac{3}{2},3 (2-\gamma )\right\}$.
\begin{enumerate}
         \item It is a source for $0\leq \gamma <2,$
         \item It is nonhyperbolic for $\gamma=2.$ 
     \end{enumerate}
For $\gamma=2$ the point $T$ is contained in the unstable  normally hyperbolic line of equilibrium points (\cite{aulbach}, pp. 36)  $\mathcal{L}_1: (\bar{\Omega}, \bar{\Sigma})=(0,\bar{\Sigma}^*)$. 

Using a representative length scale $\ell$ which is defined in \eqref{characteristic-length}
and denoting by convention $t=0$, the current time then $ \left(\frac{\ell(0)}{\ell_0}\right)^3= \frac{1}{{e_1}^1(0) ( {e_2}^2(0))^2} =1$ and $\tau(0)=0$.

Starting with  Raychaudhuri equation \eqref{RaychBI} and evaluating it at $T$ we obtain: 
\begin{equation}
\left\{\begin{array}{c}
     \dot{H}= -3 H^2  \\\\
     \dot{\ell}=\ell H
\end{array}\right.
\implies \left\{\begin{array}{c} H(t)= \frac{H_{0}}{3 H_{0} t+1} \\\\
\ell(t)= {\ell_0}
   \sqrt[3]{3 H_{0} t+1}
\end{array}\right..
\end{equation}
$\bar{\Sigma}=-1$ implies $\sigma_{+}=-H=- \frac{H_{0}}{3 H_{0} t+1}$. From eq.
\eqref{Gauss} it follows that $K=( {e_2}^2(t))^{2}=c_2^{-1}$ is a constant. Substituting back $K$ in equation  \eqref{eqe1evol} is obtained:
\begin{align}
   & \dot{{e_1}^1}=-\frac{3H_{0}}{3 H_{0} t+1} {e_1}^1, \; {e_1}^1(0)=c_2. 
\end{align}
Hence, 
\begin{align}
& {{e_1}^1}(t)=\frac{c_2}{3 H_{0} t+1}.
\end{align}
Finally,  line element \eqref{metricLRSBI} becomes
\begin{align}
   &  ds^2= - dt^2 + \frac{\left(3 H_{0} t+1\right)^2}{c_2^2} dr^2 \nonumber \\
   & + c_2  \left[ d \vartheta^2 +  \vartheta^2 d \zeta^2\right].
\end{align}
Therefore, the corresponding solution can be expressed as  Taub-Kasner solution ($p_1=1, p_2= 0, p_3= 0$)  where the scale factors of Kasner solution are $t^{p_i}, i=1,2,3$ with $p_1+p_2+p_3=1, p_1^2+p_2^2+p_3^2=1$  \cite{WE} (Sect 6.2.2 and p. 193, Eq. (9.6)).

 \item $Q: (\bar{\Omega},\bar{\Sigma})=(0,1)$ with eigenvalues $\left\{\frac{3}{2},3 (2-\gamma )\right\}$. 
     \begin{enumerate}
         \item It is a source for $0\leq \gamma <2.$
         \item It is nonhyperbolic for $\gamma=2.$ In this case, $Q$ is included in a normally hyperbolic line of equilibrium points $\mathcal{L}_1: (\bar{\Omega}, \bar{\Sigma})=(0,\bar{\Sigma}^*).$
     \end{enumerate}
     Evaluating Raychaudhuri equation \eqref{RaychBI} at $Q$ and integrating $H,$ we obtain 
\begin{equation}
 H(t)= \frac{H_{0}}{3 H_{0} t+1}.
\end{equation}
 $\bar{\Sigma}=1$ implies $\sigma_{+}=H=\frac{H_{0}}{3 H_{0} t+1}$. 
 Hence, eqs.  \eqref{Gauss} and   \eqref{eqe1evol} become 
 \begin{equation}
    \dot{K}= -\frac{4 H_{0} K}{3 H_{0}
   t+1}, \; K(0)=c_1^{-1},
\end{equation}
and 
\begin{equation}
    \dot{{e_1}^1}=\frac{H_{0}}{3 H_{0} t+1}{e_1}^1, \; {e_1}^1(0)=c_1. 
\end{equation}
Then, by integration 
\begin{align}
    & {e_1}^1(t)= c_1 \sqrt[3]{3 H_{0} t+1},\\
    & K(t)= \frac{1}{c_1\left(3 H_{0} t+1\right)^{4/3}}.
\end{align}
Then, line element \eqref{metricLRSBI} becomes
\begin{align}
   &  ds^2= - dt^2 + c_1^{-2}\left( {3 H_{0} t+1}\right)^{-\frac{2}{3}} dr^2 \nonumber \\
   & +  {c_1^{-1}}{\left(3 H_{0} t+1\right)^{4/3}} \left[ d \vartheta^2 +  \vartheta^2 d \zeta^2\right].
\end{align}
Therefore, the corresponding solution can be expressed as non-flat LRS Kasner ($p_1=-\frac{1}{3}, p_2= \frac{2}{3}, p_3= \frac{2}{3}$) Bianchi I solution (\cite{WE} Sect. 6.2.2 and Sect. 9.1.1 (2)). 
    \item $F_0: (\bar{\Omega}, \bar{\Sigma})=(0,0)$ with eigenvalues \newline $\left\{\frac{3 (\gamma -1)}{2},\frac{3 (\gamma -2)}{2}\right\}$.
   \begin{enumerate}
       \item It is a sink for $0\leq\gamma <1$. 
       \item It is a saddle for $1<\gamma <2$.
       \item It is nonhyperbolic for $\gamma=1,2.$
   \end{enumerate} 
If $\gamma=1,$ $F_0$ is included in a normally hyperbolic line of equilibrium points $\mathcal{L}_2:  (\bar{\Omega}, \bar{\Sigma})=(\bar{\Omega}^*,0).$ If $\gamma=2,$ $F_0$ is included in normally hyperbolic line of equilibrium points $\mathcal{L}_1: (\bar{\Omega}, \bar{\Sigma})=(0,\bar{\Sigma}^*).$
   
Evaluating eq. \eqref{RaychBI} at $F_0$ we obtain 
\begin{equation}
\label{scaling1}
\left\{\begin{array}{c}
      \dot{H}= -\frac{3}{2} \gamma H^2\\\\ 
      \dot{\ell}=\ell H  
       \end{array} \right.\implies 
       \left\{\begin{array}{c}
       H(t)= \frac{2 H_{0}}{3 \gamma  H_{0} t+2} \\\\
       \ell(t)=
   \ell_{0} \left(\frac{3 \gamma  H_{0} t}{2}+1\right)^{\frac{2}{3
   \gamma }}
\end{array}\right..
\end{equation}
          That is, the line element \eqref{metricLRSBI} becomes
    \begin{align}
   &  ds^2= - dt^2 + \ell_{0}^2 \left(\frac{3 \gamma  H_{0} t}{2}+1\right)^{\frac{4}{3
   \gamma }} dr^2 \nonumber \\
   & + \ell_{0}^2 \left(\frac{3 \gamma  H_{0} t}{2}+1\right)^{\frac{4}{3
   \gamma }}  \left[ d \vartheta^2 +  \vartheta^2 d \zeta^2\right]. \label{eq60}
\end{align}
The corresponding solution is a matter dominated FLRW Universe    with $\bar{\Omega}_m=1$ (mimicking de Sitter, quintessence or zero acceleration solutions).  
   
    \item $F: (\bar{\Omega}, \bar{\Sigma})=(1,0)$,  with eigenvalues \newline $\left\{-\frac{3}{2},-3 (\gamma -1)\right\}$. 
   \begin{enumerate}
       \item It is a saddle $0\leq \gamma <1$.
       \item It is a sink for $1<\gamma\leq 2$.
       \item It is nonhyperbolic for $\gamma=1$.  
   \end{enumerate}
For $\gamma=1$ point $F$ is contained in a stable normally hyperbolic line (\cite{aulbach}, pp. 36) of equilibrium points $\mathcal{L}_2:  (\bar{\Omega}, \bar{\Sigma})=(\bar{\Omega}^*,0)$.

   Evaluating  Raychaudhuri equation \eqref{RaychBI} at equilibrium point $F$ we have
   \begin{align}
& \dot{H} =\frac{b^2 \gamma  \mu ^6 }{\omega ^2-2 \mu ^2} \sin ^2 \scriptscriptstyle\left(\frac{\sqrt{\frac{3}{2}} H
   \left(\omega ^2-2 \mu ^2\right) \sin (\Phi -t \omega )}{b \mu ^3 \omega
   }\right) \nonumber \\
   & +\frac{1}{2} H^2 \Big(-3 \gamma +\frac{6
   \gamma  \mu ^2 \sin ^2(\Phi -t \omega )}{\omega ^2} \nonumber \\
   & +3 (\gamma -2) \cos
   ^2(\Phi -t \omega )\Big).
\end{align}
Therefore, 
     \begin{equation}
     \dot{H}   \sim -3 H^2 \cos ^2(t \omega -\Phi),
     \end{equation} for large $t$. 
     In average, $\Phi$ is a constant, setting $\Phi=0$ for simplicity and integrating we obtain 
     \begin{equation}
         H(t)=\frac{4 H_{0} \omega }{6 H_{0} t \omega +3 H_{0} \sin (2 t
   \omega )+4 \omega }, 
     \end{equation}
     where $H_{0}$ is the current value of $H(t)$. 
     Finally, $H(t)\sim \frac{2}{3 t}$ for large $t$. 
 Eqs.  \eqref{Gauss} and  \eqref{eqe1evol} become
     \begin{equation}
         \dot{e_1^1}=-\frac{2 {e_1^1}}{3 t}, \; \dot{K}=-\frac{4 K}{3    t},
     \end{equation}
     with general solution
     \begin{equation}
     \dot{e_1^1}(t)= \frac{c_1}{t^{2/3}}, \; K(t)= 
   \frac{c_2}{t^{4/3}}. 
     \end{equation}
Then, line element \eqref{metricLRSBI} becomes
\begin{align}
   &  ds^2= - dt^2 + c_1^{-2} {t^{4/3}} dr^2 \nonumber \\
   & +  {c_2^{-1}}{t^{4/3}} \left[ d \vartheta^2 +  \vartheta^2 d \zeta^2\right]. \label{eq61}
\end{align}  
For large $t$,  $F$  can be associated with Einstein- de Sitter solution (\cite{WE}, Sec 9.1.1 (1)) with $\gamma= 1$).
\end{enumerate}
\begin{figure*}[t!]
    \centering
    \subfigure[]{\includegraphics[width=0.4\textwidth]{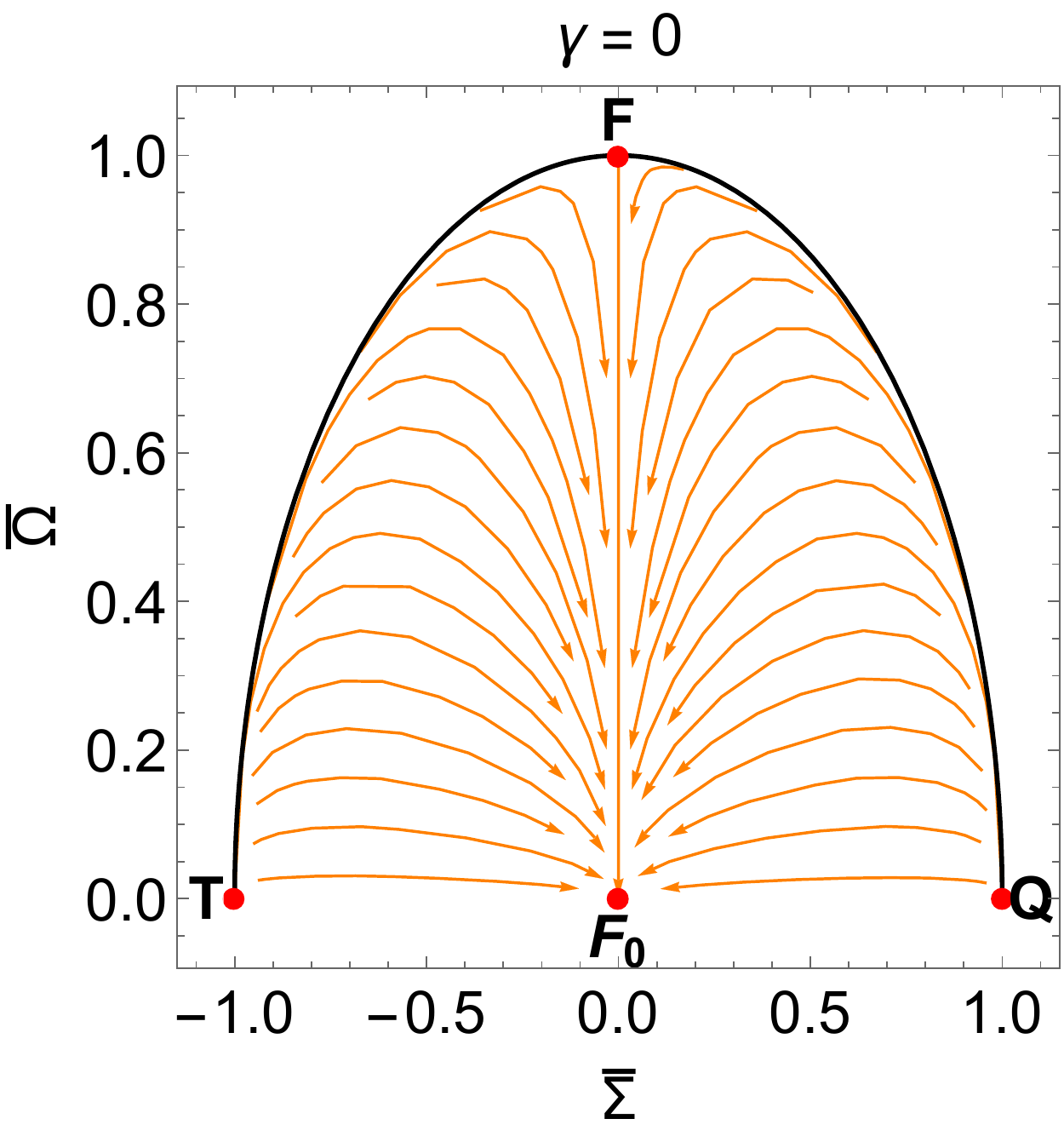}}
    \subfigure[]{\includegraphics[width=0.4\textwidth]{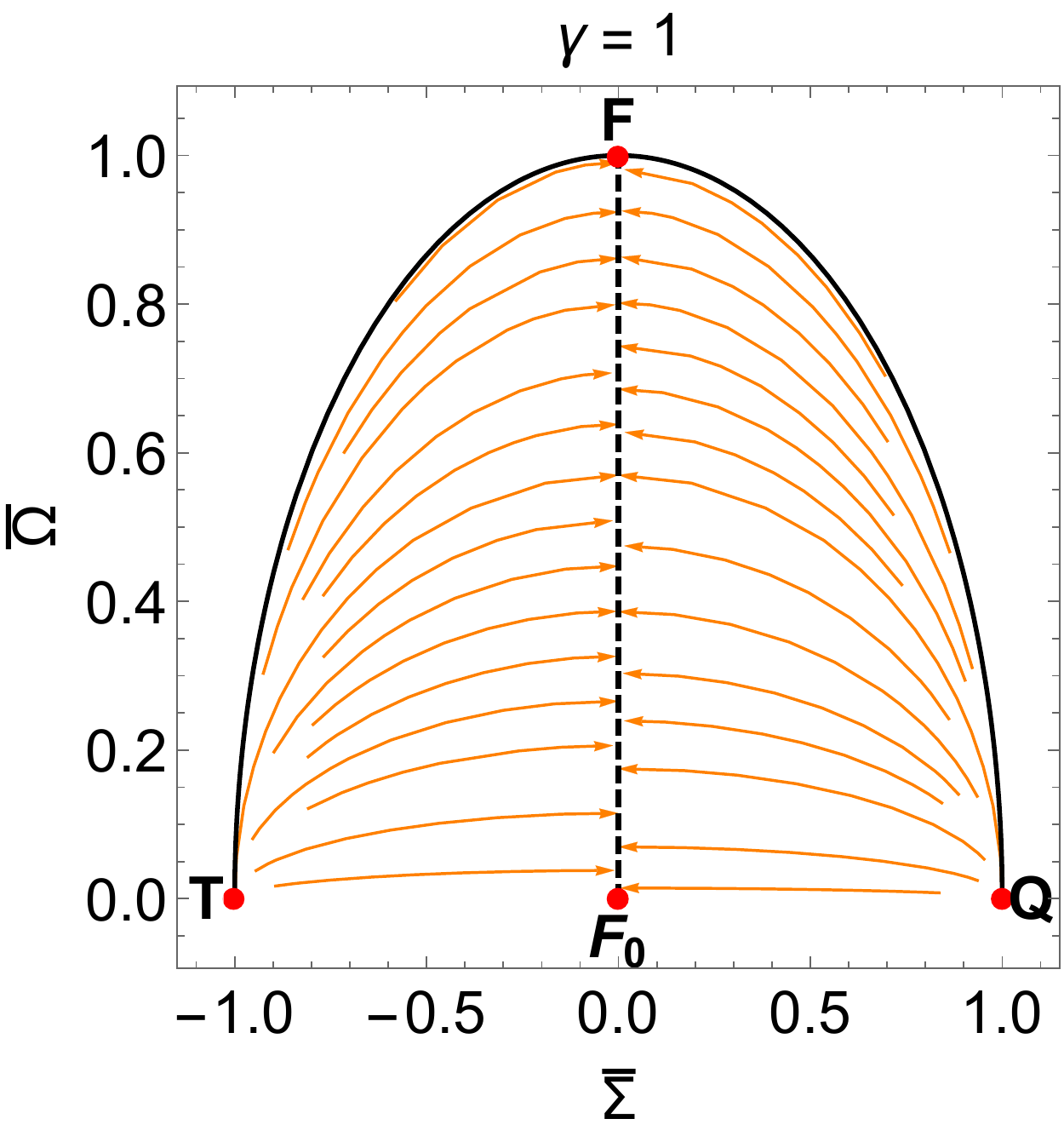}}
    \subfigure[]{\includegraphics[width=0.4\textwidth]{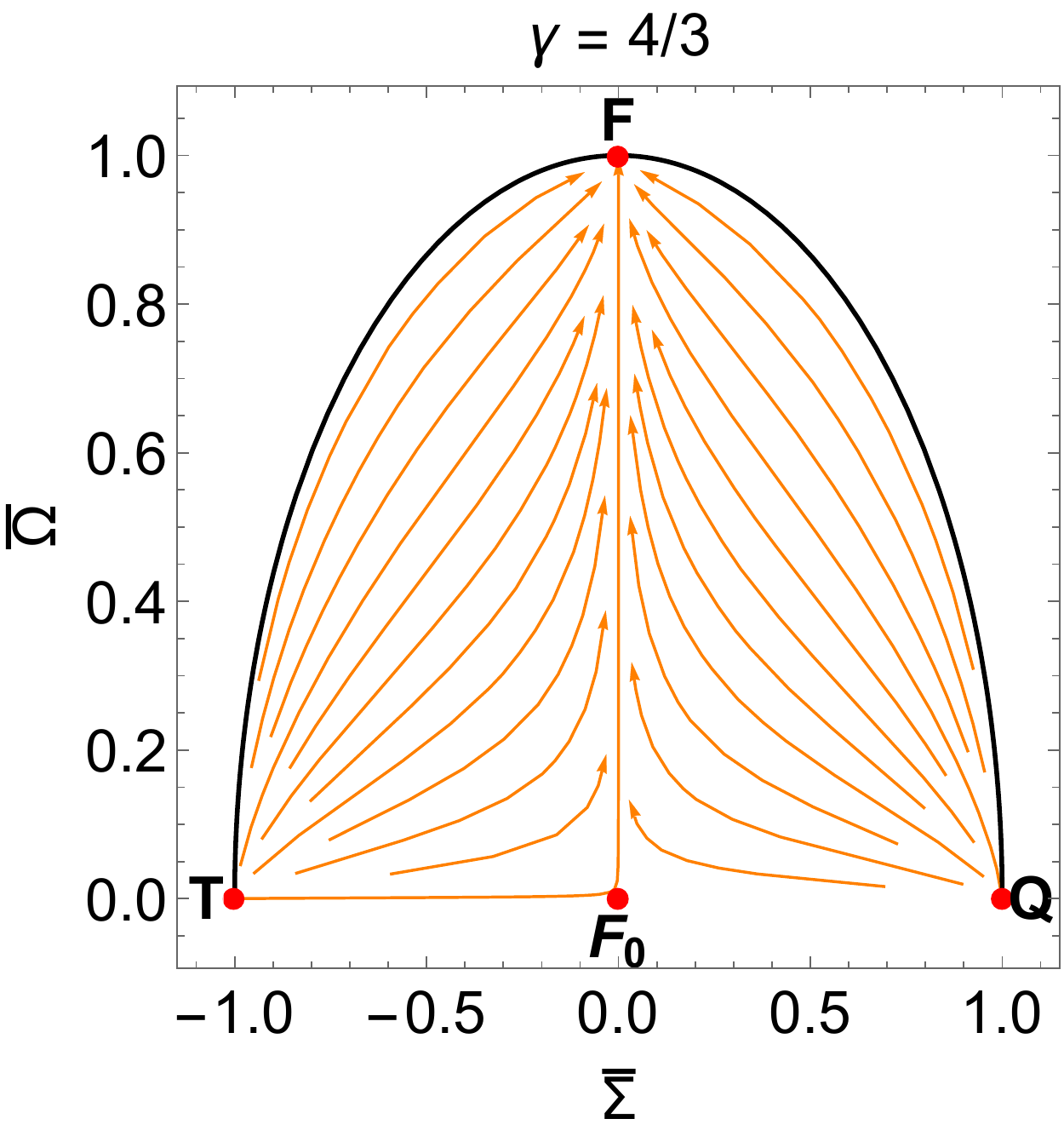}}
    \subfigure[]{\includegraphics[width=0.4\textwidth]{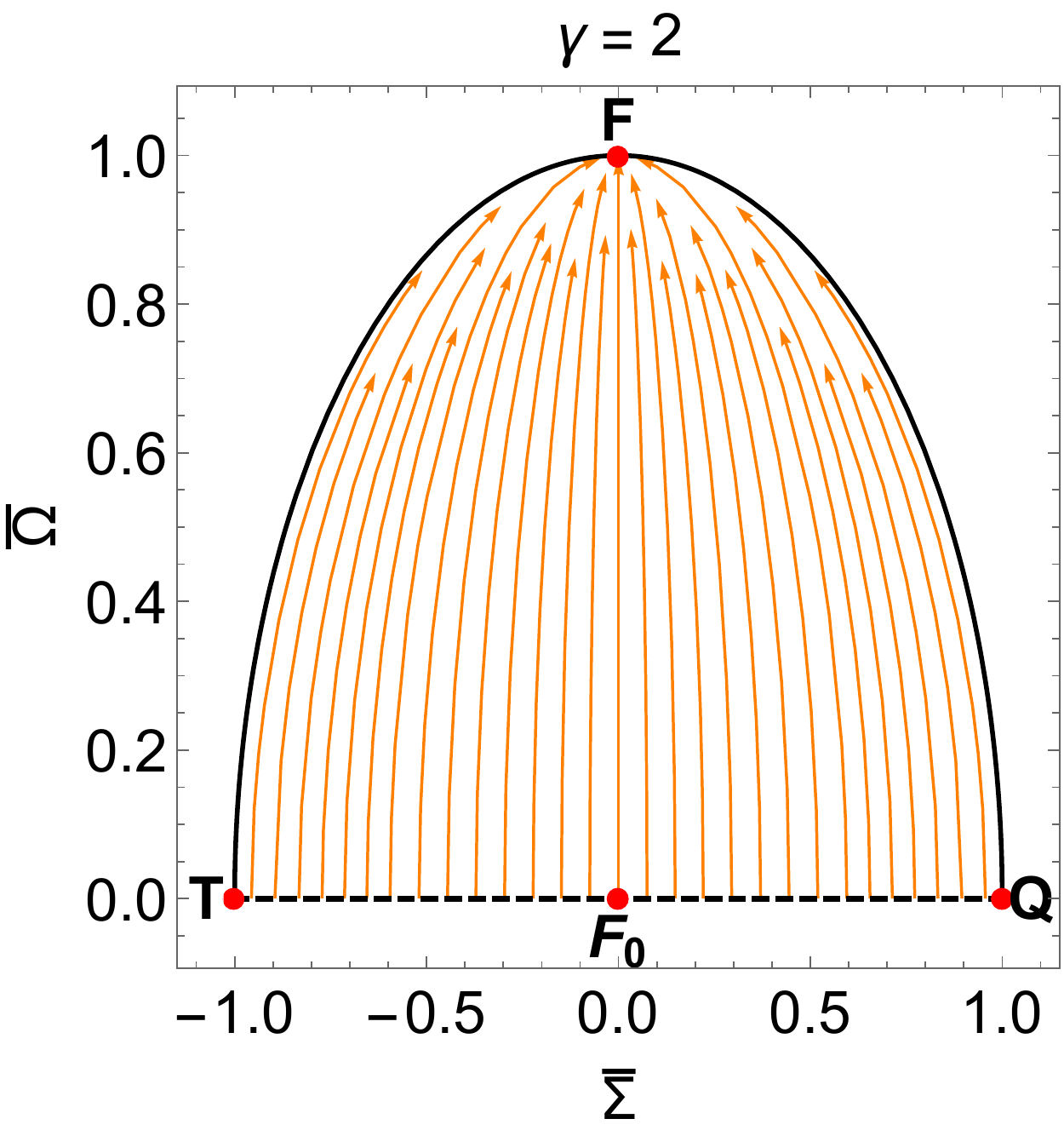}}
    \caption{Phase plane for system \eqref{guidingC1}, \eqref{guidingC2} for different choices of $\gamma.$ For $\gamma=1$ the dashed vertical line $\mathcal{L}_2:  (\bar{\Omega}, \bar{\Sigma})=(\bar{\Omega}^*,0)$ is stable.  For $\gamma=2$ the dashed horizontal unstable line $\mathcal{L}_1: (\bar{\Omega}, \bar{\Sigma})=(0,\bar{\Sigma}^*)$.}
    \label{Bianchi1plots2d}
\end{figure*}
\noindent
In Figure \ref{Bianchi1plots2d}  a phase plane for system \eqref{guidingC1}, \eqref{guidingC2} for different choices of $\gamma$ is presented. 

\begin{table*}[t]
    \centering    
    \caption{Exact solutions associated with  equilibrium points of the reduced
averaged system \eqref{guidingC1}, \eqref{guidingC2}. $A(t)$ and $B(t)$ denote the scale factors of  metric \eqref{metricLRSBI} and $c_1, c_2, \ell_0 \in \mathbb{R}^+$. }
    \resizebox{\textwidth}{!} {\begin{tabular}{lccc}\hline
         Point & $A(t)$ & $B(t)$ & Solution  \\\hline 
        $T$ & $ \frac{\left(3 H_{0} t+1\right)}{c_2}$ &  $\sqrt{c_2}$ & Taub-Kasner solution ($p_1=1, p_2= 0, p_3= 0$) \\ \hline 
        $Q$ & $c_1^{-2}\left( {3 H_{0} t+1}\right)^{-1/3}$  & $ {c_1^{-1}}{\left(3 H_{0} t+1\right)^{2/3}}$ & Non-flat LRS Kasner ($p_1=-\frac{1}{3}, p_2= \frac{2}{3}, p_3= \frac{2}{3}$) Bianchi I solution \\ \hline 
       $F$ & $c_1^{-1} {t^{2/3}}$   & ${c_2^{-1/2}}{t^{2/3}}$ & Einstein-de Sitter solution \\\hline
        $F_0$ &  $a_{0} \left(\frac{3 \gamma  H_{0} t}{2}+1\right)^{\frac{2}{3
   \gamma }}$ & $ a_{0} \left(\frac{3 \gamma  H_{0} t}{2}+1\right)^{\frac{2}{3
   \gamma }}$ & Flat matter dominated FLRW universe \\\hline 
     \end{tabular}}
    \label{Metrics}
\end{table*}

\noindent System (\ref{guidingC1}),  (\ref{guidingC2}) when $\gamma=1$ reduces to 
\begin{align}
      &\frac{d \bar{\Omega}}{d \tau}=\frac{3}{2} \bar{\Sigma}^2 \bar{\Omega},  \quad \frac{d\bar{\Sigma}}{d \tau}=-\frac{3}{2} \bar{\Sigma }\left(1-\bar{\Sigma}^2\right).  \label{141}
\end{align}
From the above system we obtain 
\begin{equation}
    \frac{d}{d \tau} \ln \left(\frac{\bar{\Sigma}^2}{\bar{\Omega}^2}\right)=-3 \implies \frac{\bar{\Sigma}^2}{{\bar{\Sigma}_0}^2}=\frac{\bar{\Omega}^2}{{\bar{\Omega}_0}^2} e^{-3 \tau}.
\end{equation}
We have assumed that the orbit passes by $(\bar{\Omega}, \bar{\Sigma})= (\bar{\Omega}_0, \bar{\Sigma}_0)$ at  time $\tau_0=0$. These values are identified with the current epoch.
Using equations \eqref{141} and the chain rule we have 
\begin{equation}
    \frac{d \bar{\Sigma}^2}{d \bar{\Omega}}=2 \bar{\Sigma} { \frac{d \bar{\Sigma}}{d \tau}}\Big{/}{\frac{d \bar{\Omega}}{d \tau}}= -2\frac{(1-\bar{\Sigma}^2)}{\bar{\Omega}}.
\end{equation}
Then, 
\begin{equation}
\label{eqT95}
    \frac{d (1- \bar{\Sigma}^2)}{d \bar{\Omega}}=2\frac{(1-\bar{\Sigma}^2)}{\bar{\Omega}}.
\end{equation}
Due to  $(\bar{\Omega}, \bar{\Sigma})= (\bar{\Omega}_0, \bar{\Sigma}_0)$ at the time $\tau_0=0$ by solving \eqref{eqT95} with the variables separation method  it follows that 
\begin{equation}
\frac{1- \Sigma^2}{1- \Sigma_0^2} = \frac{\Omega^2}{\Omega_0^2}.   
\end{equation}
Finally, the orbits of system \eqref{141} 
are given by 
\begin{equation}
    \bar{\Sigma}^2= 1- \frac{\left(1-\bar{\Sigma}_0^2\right) \bar{\Omega} ^2}{\bar{\Omega}_0^2}.
\end{equation}

\noindent In table \ref{Metrics} exact solutions are associated with  equilibrium points of the reduced
averaged system  \eqref{guidingC1}, \eqref{guidingC2} are summarized, where $A(t)$ and $B(t)$ denote the scale factors of the metric \eqref{metricLRSBI} and $c_1, c_2, \ell_0 \in \mathbb{R}^+$ are integration constants. 

 \subsubsection{Late-time behavior}
 Results from the linear stability analysis which are combined with Theorem \ref{LFZ11} lead to: 
\begin{thm}
\label{thm9}
The late-time attractors of full system \eqref{unperturbed1} and time-averaged system  \eqref{avrgsyst} for LRS Bianchi I line element are:

     \begin{table}[t]
    \centering    
    \caption{Exact solutions associated with  equilibrium points of the reduced
averaged equation \eqref{EquationOmega}. $a(t)$ denotes the scale factor of  metric \eqref{metricFLRW} in  flat case and $a_0 \in \mathbb{R}^+$. }
\footnotesize\setlength{\tabcolsep}{4pt}
    \begin{tabular}{lcc}\hline
         Point & $a(t)$ & Solution  \\\hline 
          $F$ & $ a_{0}  \left(\frac{3    H_{0} t}{2}+1\right)^{\frac{2}{3
     }}$ & Einstein-de Sitter solution \\\hline
        $F_0$ &  $ a_{0} \left(\frac{3 \gamma  H_{0} t}{2}+1\right)^{\frac{2}{3
   \gamma }}$ & Flat matter dominated FLRW universe \\\hline 
    \end{tabular}
    \label{MetricsflatFLRW}
\end{table}
\begin{enumerate}
    \item[(i)]  The flat matter dominated FLRW Universe $F_0$  with  the line element \eqref{eq60}
   if  $0< \gamma < 1$. $F_0$ represents a  quintessence fluid if $0<\gamma<\frac{2}{3}$ or a zero-acceleration model if $\gamma=\frac{2}{3}$. 
Taking limit $\gamma=0$ we have $\ell(t)=\ell_{0} \left(\frac{3 \gamma  H_{0} t}{2}+1\right)^{\frac{2}{3
   \gamma }}\rightarrow \ell_0  e^{H_0 t}$, i.e., a de Sitter solution.
  \item[(ii)] The scalar field dominated solution $F$   with  line element \eqref{eq61} if $1<\gamma\leq 2$. 
For large $t$  the equilibrium point  can be associated with Einstein-de Sitter solution.
\end{enumerate}
\end{thm}
For $\gamma=1$, $F_0$ and $F$ are stable because they belong to stable normally hyperbolic line of equilibrium points. For $\gamma=2$, $F$ is asymptotically stable (see figures \ref{Bianchi1plots2d} (b) and \ref{Bianchi1plots2d} (d)).
\subsection{Flat FLRW metric.}
\label{FLRWflatopen}

For flat FLRW Universe ($k=0$) and for $\gamma\neq 1$, we obtain the following time-averaged system in the new logarithmic time variable $\tau$: 
\begin{subequations}
\label{avrgsystFLRW}
\begin{align}
     & \frac{d\bar{\Omega}}{d \tau}=-\frac{3}{2}  \bar{\Omega}  (\gamma -1) \left(\bar{\Omega} ^2-1\right), \label{EquationOmega}\\ 
    &  \frac{d{{\bar{\Phi}}}}{d \tau}=0,   
   \\
    & \frac{d {H}}{d \tau}=  -\frac{3}{2} H \left[  \gamma (1-  \bar{\Omega}^2) + \bar{\Omega}^2  \right]. \label{EquationH}
\end{align}
\end{subequations}
Equation \eqref{EquationOmega} has solution
\begin{equation}
      \bar{\Omega}(\tau)=\frac{ \Omega_{0}e^{\frac{3 \gamma  \tau}{2}}}{\sqrt{\Omega_{0}^2 e^{3 \gamma 
   \tau}+e^{3 \tau} \left(1-\Omega_{0}^2\right)}},
   \end{equation}
   where $\bar{\Omega}(0)=\Omega_0$. 
   
\noindent   Equation \eqref{EquationOmega} has the following equilibrium points
   \begin{enumerate}
       \item $F_0:  \bar{\Omega}=0$ with eigenvalue $-\frac{3}{2} (1-\gamma )$. It is a sink for $0<\gamma<1$ or a source for $1<\gamma \leq 2$.
       \item $F: \bar{\Omega}=1$ with eigenvalue $3 (1-\gamma )$. It is a source for $0<\gamma<1$ or a sink for $1<\gamma \leq 2.$
   \end{enumerate}
 Evaluating averaged Raychaudhuri equation \eqref{EquationH}
  at $F_0$ we obtain equations  \eqref{scaling1}. Then,  metric \eqref{metricFLRW} in flat case becomes 
     \begin{align}
   &  ds^2= - dt^2 + a_{0}^2 \left(\frac{3 \gamma  H_{0} t}{2}+1\right)^{\frac{4}{3
   \gamma }} dr^2 \nonumber \\
   & + a_{0}^2 \left(\frac{3 \gamma  H_{0} t}{2}+1\right)^{\frac{4}{3
   \gamma }} r^2  \left[ d \vartheta^2 + \vartheta^2 d \zeta^2\right]. \label{eq69}
\end{align} 
Evaluating averaged Raychaudhuri equation \eqref{EquationH}
  at $F_1$ we obtain 
\begin{equation}
\label{scaling2}
\left\{\begin{array}{c}
      \dot{H}= -\frac{3}{2}  H^2\\\\ 
      \dot{a}=a H  
       \end{array} \right.\implies 
       \left\{\begin{array}{c}
       H(t)= \frac{2 H_{0}}{3    H_{0} t+2} \\\\
       a(t)=
   a_{0} \left(\frac{3    H_{0} t}{2}+1\right)^{\frac{2}{3
     }}
\end{array}\right..
\end{equation}
Then,  metric \eqref{metricFLRW} in   flat FLRW case becomes 
   \begin{align}
   &  ds^2= - dt^2 + a_{0}^2 \left(\frac{3    H_{0} t}{2}+1\right)^{\frac{4}{3
     }} dr^2 \nonumber \\
   &  a_{0}^2 \left(\frac{3    H_{0} t}{2}+1\right)^{\frac{4}{3
     }} r^2  \left[ d \vartheta^2 + \vartheta^2 d \zeta^2\right]. \label{eq70}
\end{align}

\noindent In table \ref{MetricsflatFLRW} exact solutions which are associated with equilibrium points of the reduced averaged equation \eqref{EquationOmega} are presented.

 For $\gamma=1$, the time-averaged system truncated at order $\mathcal{O}(H^4)$ is given by 
\begin{subequations}
\label{avrgflatFLRWgamma1}
\begin{small}
\begin{align}
 & \frac{d\bar{\Omega}}{d \tau}=\frac{9 H^2 \bar{\Omega}^5 \left(\omega ^2-2 \mu ^2\right)^3}{32 b^2 \mu ^6 \omega ^4}\\
& \frac{d\bar{\Phi}}{d \tau}= \frac{3 H \bar{\Omega}^2 \left(\omega ^2-2 \mu ^2\right)^3}{8 b^2 \mu ^6 \omega ^3}-\frac{3 H^3 \bar{\Omega}^4 \left(\omega ^2-2 \mu ^2\right)^5}{32
   b^4 \mu ^{12} \omega ^5},\\
&    \frac{d H}{d\tau}=-\frac{3 H}{2}-\frac{9 H^3 \bar{\Omega}^4 \left(\omega ^2-2 \mu ^2\right)^3}{32 b^2 \mu ^6 \omega ^4}. 
\end{align}
\end{small}
\end{subequations}
Assuming that $H$ is an explicit function of $\Omega$ and using the chain rule, we obtain $H'(\bar{\Omega})= { \frac{d H}{d\tau}}\Big{/}{\frac{d\bar{\Omega}}{d \tau}}$. 
From the first and third equations of \eqref{avrgflatFLRWgamma1} we obtain 
\begin{small}
\begin{equation}
H'(\bar{\Omega})=\frac{16 b^2 \mu ^6 \omega ^4}{3 \bar{\Omega} ^5 H(\bar{\Omega}) \left(2 \mu ^2-\omega ^2\right)^3}-\frac{H(\bar{\Omega})}{\bar{\Omega}}. 
\end{equation}
\end{small}
Given $H_0$ and $\bar{\Omega}_0$, the initial values of $H$ and $\bar{\Omega}$ when $\tau=0$, i.e., $H(\bar{\Omega}_0)=H_0$ we obtain the solution
\begin{small}
\begin{align}
 &  H(\bar{\Omega})= \frac{\sqrt{\frac{16 b^2 \mu ^6 \omega ^4}{3 \left(\omega ^2-2 \mu ^2\right)^3} \left(1-\frac{\bar{\Omega} ^2}{\bar{\Omega}_{0}^2}\right)+H_0^2 \Omega ^2 \bar{\Omega}_{0}^2}}{\bar{\Omega} ^2}. 
\end{align}
\end{small}
Then,  equations \eqref{avrgflatFLRWgamma1} can be expressed as 
\begin{subequations}
\begin{small}
\begin{align}
    & \frac{d \bar{\Omega }}{d\tau}= \frac{3 \bar{\Omega }}{2}  -\frac{3 \bar{\Omega }^3 \left(16 b^2 \mu ^6 \omega ^4+3 H_0^2 \bar{\Omega }_0^4 \left(2 \mu ^2-\omega ^2\right)^3\right)}{32 b^2 \mu ^6 \omega ^4 \bar{\Omega }_0^2}, 
   \\
    &\frac{d\bar{\Phi}}{d \tau}= \frac{3 \left(\omega ^2-2 \mu ^2\right)^3 \sqrt{\frac{16 b^2 \mu ^6 \omega ^4 \left(1-\frac{\bar{\Omega }^2}{\bar{\Omega }_0^2}\right)}{3 \left(\omega ^2-2 \mu ^2\right)^3}+H_0^2 \bar{\Omega }_0^2 \bar{\Omega
   }^2}}{8 b^2 \mu ^6 \omega ^3}  \nonumber \\
   & -\frac{3 \left(\omega ^2-2 \mu ^2\right)^5 \left(\frac{16 b^2 \mu ^6 \omega ^4 \left(1-\frac{\bar{\Omega }^2}{\bar{\Omega }_0^2}\right)}{3 \left(\omega ^2-2 \mu ^2\right)^3}+H_0^2
   \bar{\Omega }_0^2 \bar{\Omega }^2\right)^{3/2}}{32 b^4 \mu ^{12} \omega ^5 \bar{\Omega }^2}.
\end{align}
\end{small}
\end{subequations}
By integration we obtain
\begin{small}
\begin{equation*}
    \bar{\Omega}(\tau)=\frac{4 b \mu ^3 e^{3 \tau/2} \omega ^2 \bar{\Omega}_0}{\sqrt{16 b^2 \mu ^6 e^{3 \tau} \omega ^4+3 H_0^2 \left(1-e^{3 \tau}\right) \bar{\Omega}_0^4 \left(\omega ^2-2 \mu ^2\right)^3}},
\end{equation*}
\end{small}
which satisfies 
$\lim_{\tau\rightarrow -\infty}\bar{\Omega}(\tau)=0$  
and 
$\lim_{\tau\rightarrow +\infty}\bar{\Omega}(\tau)= \frac{4 b \mu ^3 \omega ^2 \bar{\Omega}_0}{\sqrt{16 b^2 \mu ^6 \omega ^4+3 H_0^2 \bar{\Omega}_0^4 \left(2 \mu ^2-\omega ^2\right)^3}}$.
Furthermore, 
\begin{small}
\begin{equation*}
    H(\tau)= \frac{H_0 e^{-3
   \tau } \sqrt{16 b^2 \mu ^6 e^{3 \tau } \omega ^4+3 H_0^2 \left(e^{3 \tau }-1\right) \bar{\Omega}_0^4 \left(2 \mu ^2-\omega ^2\right)^3}}{4 b \mu ^3 \omega ^2}, 
\end{equation*}
\end{small}
satisfies 
 $\lim_{\tau\rightarrow -\infty}H(\tau)=\infty$ and 
 $\lim_{\tau\rightarrow +\infty}H(\tau)=0$.

\noindent Finally, 
\begingroup\makeatletter\def\f@size{7}\check@mathfonts
\begin{equation*}
 \frac{d\bar{\Phi}}{d \tau}=-\frac{3 H_0 e^{-3 \tau } 
 \bar{\Omega}_0^2 \left(\omega ^2-2 \mu ^2\right)^3 \left(H_0^2 \bar{\Omega}_0^2 \left(\omega ^2-2 \mu ^2\right)^2-4 b^2 \mu ^6 e^{3 \tau } \omega ^2\right)}{8 b^3 \mu ^9 \omega ^3 \sqrt{{16 b^2 \mu ^6 e^{3 \tau } \omega ^4+3 H_0^2 \left(e^{3 \tau }-1\right) \bar{\Omega}_0^4 \left(2 \mu ^2-\omega ^2\right)^3}}},
   \end{equation*}
   \endgroup
  is integrable leading to $\bar{\Phi}(\tau)$. 

    \begin{figure}[t!]
       \centering
       \subfigure[]{\includegraphics[width=0.45\textwidth]{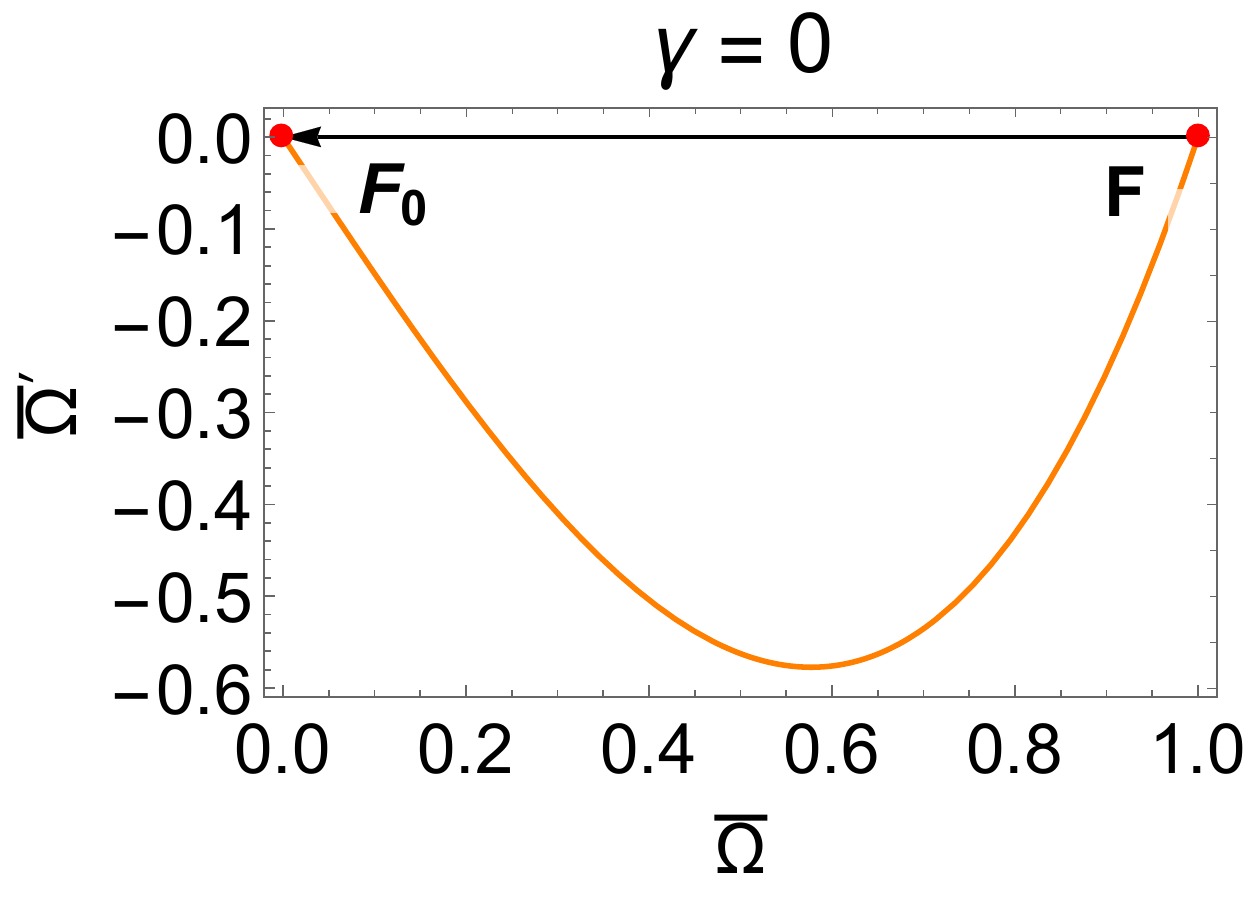}}
       \subfigure[]{\includegraphics[width=0.45\textwidth]{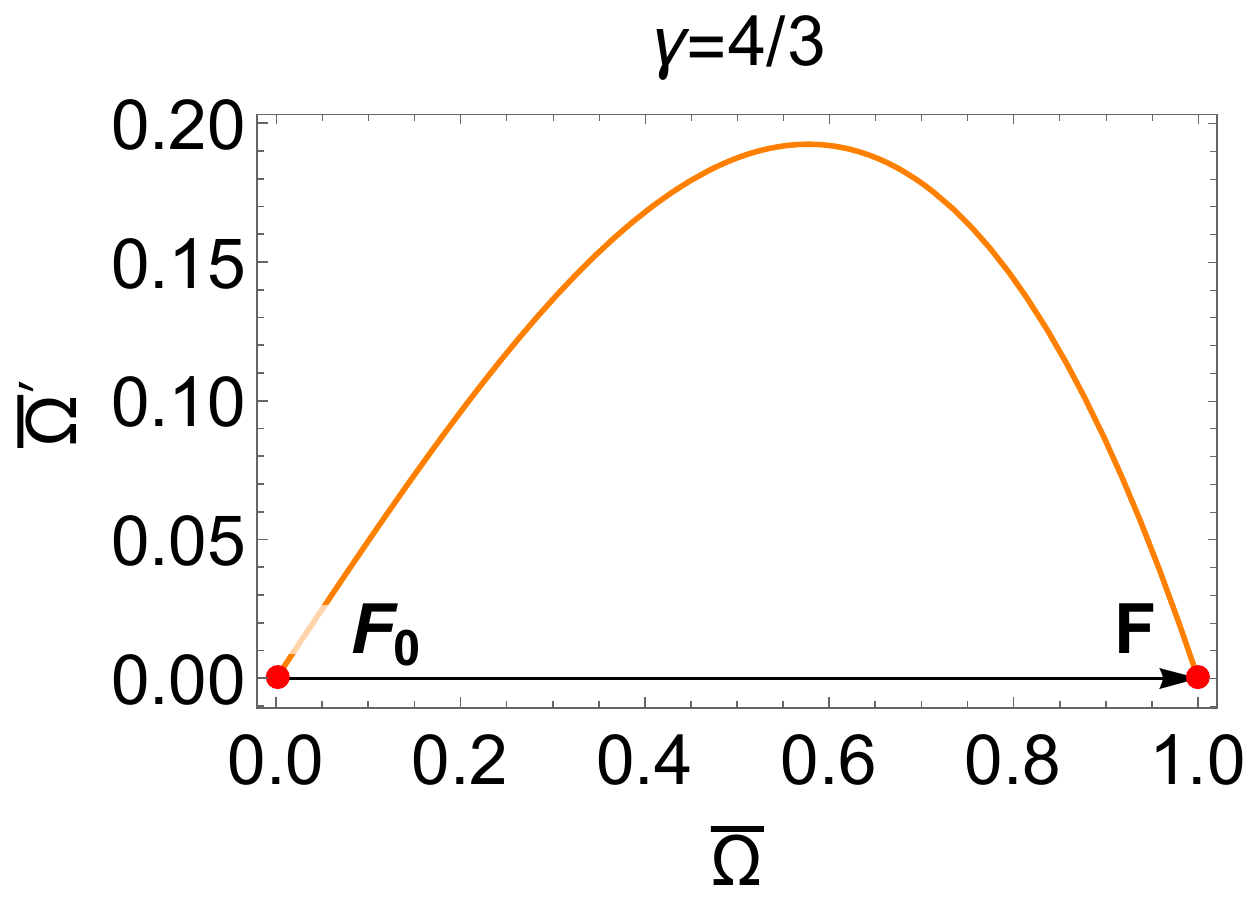}}
        \caption{Phase plot of equation \eqref{EquationOmega} for different choices of $\gamma.$}
       \label{fig:my_label}
   \end{figure}
   
\noindent In Figure \ref{fig:my_label}  the phase plot of equation \eqref{EquationOmega} for different choices of $\gamma$ is presented. The arrows indicate the direction of the flow for a 1-dimensional dynamical system (see, e.g. \cite{STROGATZ} pages 16-17, Figure 2.2.1).

\noindent Analyzing the dynamics on the extended phase space $(\bar{\Omega}, H)$ for $\gamma=1$ (see phase plot \ref{fig:FlatFLRWDust}), the solutions tend to finite $\Omega$ and $H=0$. The equilibrium point $F: (\bar{\Omega},H)=(1,0)$ is the attractor of the horizontal line with $\bar{\Omega}=1$. $F_0:(\bar{\Omega},H)=(0,0)$ is the attractor of the horizontal line with $\bar{\Omega}=0$. The solution of the full system with same initial conditions is affected by boundary effects at $\bar{\Omega}=1$ and has large oscillations for large values of $H$, but they are damped as $H\rightarrow 0$. For $\gamma=2$ (see phase plot \ref{fig:FlatFLRWStiff}) the attractor is $F$. Only the solution with initial value $\bar{\Omega}_0=0$ tends to $F_0$.

\subsubsection{Late-time behavior}
The results from the linear stability analysis are combined with Theorem \ref{LFZ11} (for $\Sigma=0$) and lead to: 
\begin{thm}
\label{thm10}
The late-time attractors of   full system  \eqref{unperturbed1} with $\Sigma=0$  and  averaged system \eqref{EquationOmega}  are: 
\begin{enumerate}
    \item[(i)] The flat matter dominated FLRW Universe $F_0$
   with line element \eqref{eq69} 
if $0\leq \gamma < 1$. $F_0$ represents a  quintessence fluid if $1<\gamma<\frac{2}{3}$ or a zero-acceleration model if $\gamma=\frac{2}{3}$. 
We have $a(t)=a_{0} \left(\frac{3 \gamma  H_{0} t}{2}+1\right)^{\frac{2}{3
   \gamma }}\rightarrow a_0  e^{H_0 t}$ as $\gamma\rightarrow 0$, i.e.,  a de Sitter solution is recovered.

\item[(ii)]    The scalar field dominated solution $F$   with  line element \eqref{eq70}
if $1<\gamma\leq 2$.   
For large $t$  the equilibrium point  can be associated with Einstein-de Sitter solution.
    \end{enumerate}
\end{thm}
Observe in figure \ref{fig:FlatFLRWDust} that as $H\rightarrow 0$ the values of $\bar{\Omega}$ lying on the orange line (solution of time-averaged system at fourth order) give an upper bound to the value $\Omega$ of the original system. Therefore, by controlling the error of averaged higher order system, one can also control the error in the original one.

\section{Conclusions}
\label{Conclusions}
This is the second paper of the  ``Averaging Generalized Scalar Field Cosmologies'' program that was initiated in reference \cite{Leon:2021lct}. This program consists in using asymptotic methods and averaging theory to obtain relevant information about solution's space of scalar field cosmologies in presence of a matter fluid with  EoS with barotropic index $\gamma$ minimally coupled to a scalar field with generalized harmonic potential \eqref{pot}. 

\noindent According to this research program, in paper I \cite{Leon:2021lct} was proved that late-time attractors in LRS Bianchi III model are: a  matter dominated flat FLRW universe  if  $0\leq \gamma \leq \frac{2}{3}$ (mimicking de Sitter, quintessence or zero acceleration solutions), a matter-curvature scaling solution  if  $\frac{2}{3}<\gamma <1$ and Bianchi III flat spacetime  if $1\leq \gamma\leq 2$. Late-time attractors in FLRW metric with $k=-1$ are: a matter dominated FLRW universe  if  $0\leq \gamma \leq \frac{2}{3}$ (mimicking de Sitter, quintessence or zero acceleration solutions) and  Milne solution if $\frac{2}{3}<\gamma <2$.  In all metrics, the matter dominated flat  FLRW universe represents quintessence fluid if $0< \gamma < \frac{2}{3}$.
\noindent 
For LRS Bianchi I and  flat FLRW metrics as well as for LRS Bianchi III and open FLRW, we can use Taylor expansion with respect to $H$ near $H=0$. Hence, the  resulting system can be expressed in standard form \eqref{standard51} after selecting a convenient angular frequency $\omega$ in the transformation \eqref{eqAA25}.  
\noindent Next, we have taken the time-averaged of  previous system  obtaining a system that can be easily studied using dynamical system's tools. 

\noindent In particular, we have proved in Theorem \ref{LFZ11}   that late-time attractors of full and time-averaged systems are the same for some homogeneous metrics. Theorem \ref{LFZ11} implies that $\Omega, \Sigma$ and $\Phi$  evolve according to time-averaged equations \eqref{Ieq24}, \eqref{Ieq25} and \eqref{Ieq26} 
as $H\rightarrow 0$. Therefore, we can establish the stability of a periodic solution as it matches  exactly the stability of a stationary solution of  averaged equation. 

\noindent We have given a rigorous demonstration of Theorem   \ref{LFZ11} in \ref{gBILFZ11} based on the construction of a smooth local near-identity nonlinear transformation, well-defined as $H$ tends to zero. We have used properties of the sup norm and the theorem of the mean values  for a vector function $\bar{\mathbf{f}}: \mathbb{R}^2\longrightarrow \mathbb{R}^2$. We have explained preliminaries of  the method of proof in Section \ref{SECT3.3}. As in paper \cite{Fajman:2020yjb}, our analytical results were strongly supported by numerics in  \ref{numerics} as well.

\noindent More specific, according to Theorem   \ref{LFZ11}   for Bianchi I and flat FLRW metrics,  Hubble parameter $H$ plays the role of a  time-dependent perturbation parameter controlling the magnitude of error between solutions of  full and  time-averaged systems.  Therefore,  analysis of system  is reduced to study time-averaged equations.  In this regard, we have formulated theorems  \ref{thm9} and \ref{thm10} concerning to the late-time behavior of our model. 

\noindent For LRS Bianchi I  late-time attractors of full system \eqref{unperturbed1} and averaged system  \eqref{avrgsyst} are:
\begin{enumerate}
   \item[(i)]  The matter dominated FLRW Universe $F_0$    with line element
\eqref{eq60} if  $0< \gamma < 1$. $F_0$ represents a  quintessence fluid or a zero-acceleration model for $\gamma=\frac{2}{3}$. 
In the limit $\gamma=0$ we have a de Sitter solution.
  \item[(ii)] The scalar field dominated solution $F$   with  line element 
\eqref{eq61} 
if $1<\gamma\leq 2$.   
For large $t$  the equilibrium point  can be associated with  Einstein-de Sitter solution.
\end{enumerate}

\noindent For flat FLRW metric late-time attractors of  full system  \eqref{unperturbed1} with $\Sigma=0$  and averaged system  are: 
\begin{enumerate}
   \item[(i)] The matter dominated FLRW Universe $F_0$
   with line element \eqref{eq69}
if $0\leq \gamma < 1$. $F_0$ represents a  quintessence fluid or a zero-acceleration model for $\gamma=\frac{2}{3}$. 
In the limit $\gamma=0$ we have a de Sitter solution.

\item[(ii)]    The scalar field dominated solution $F$   with  line element  \eqref{eq70}
if $1<\gamma \leq 2$.   
For large $t$  the equilibrium point  can be associated with  Einstein-de Sitter solution.
    \end{enumerate}
It is interesting to note that for LRS Bianchi I and  flat FLRW cases when matter fluid is a cosmological constant,  $H$ tends asymptotically to constant values depending on initial conditions which is consistent to de Sitter expansion (see figures \ref{fig:BICC3D} and \ref{fig:FlatFLRWCC}). 
For dust $\gamma=1$ in flat FLRW metric, we have from  qualitative analysis in Section \ref{FLRWflatopen} that $\bar{\Omega}(\tau)$ tends to a constant and $H(\tau)$ tends to zero as ${\tau\rightarrow +\infty}$. Observe in figure \ref{fig:FlatFLRWDust} that as $H\rightarrow 0$ the values of $\bar{\Omega}$ lying on the orange line (solution of time-averaged system at fourth order) give an upper bound to  values of $\Omega$ in the original system. Therefore, by controlling the error of averaged higher order system, one can also control the error of the original one. 

\noindent We have illustrated that asymptotic methods and averaging theory are powerful tools to investigate scalar field cosmologies with generalized harmonic potential. One evident advantage is that to determine  stability of full oscillation it is not needed to analyze the full dynamics, but only the late-time behavior of time-averaged (simpler) system has to be analyzed.  Interestingly,  we have examined in detail sub-case $\Omega_k=0$ of LRS Bianchi III and open FLRW, i.e., flat limits  LRS Bianchi I and zero-curvature FLRW. We have obtained for LRS Bianchi I that late-time attractors of  full and time-averaged systems are: a flat matter dominated FLRW Universe if  $0\leq \gamma <1$  or  an equilibrium solution if $1<\gamma\leq 2$ which for large $t$ can be associated with  Einstein-de Sitter solution.
For flat FLRW metric, late-time attractors of  full system  and time-averaged system are: a flat  matter dominated  FLRW (mimicking de Sitter, quintessence or zero acceleration solutions) if $0<\gamma<1$ and Einstein-de Sitter solution if $1<\gamma<2$. In all metrics, the matter dominated flat  FLRW universe represents quintessence fluid if $0< \gamma <\frac{2}{3}$.

\section*{Acknowledgements}

This research was funded by  Agencia Nacional de Investigaci\'on y Desarrollo- ANID  through the program FONDECYT Iniciaci\'on grant no.
11180126 and by Vicerrector\'{\i}a de Investigaci\'on y Desarrollo Tecnol\'ogico at
Universidad Cat\'olica del Norte. Ellen de los Milagros Fern\'andez Flores  is acknowledged for proofreading this manuscript and for improving the English. 
We thank anonymous referee for his/her valuable comments which have helped us to improve our work.

\appendix

\section{Proof of Theorem \ref{LFZ11}}
\label{gBILFZ11}

\begin{lem}[\textbf{Gronwall's Lemma (integral form)}]
\label{Gronwall}
 Let be $\xi(t)$ a nonnegative function, summable over  $[0,T]$ which satisfies almost everywhere the integral inequality
        $$
            \xi(t)\leq C_1 \int_0^t \xi(s)ds +C_2, \;  C_1, C_2\geq 0.
        $$
       Then, 
       $$
            \xi(t)\leq C_2   e^{C_1 t},
        $$
        almost everywhere for $t$ in $0\leq t\leq T$.
     In particular, if    $$
            \xi(t)\leq C_1 \int_0^t \xi(s)ds, \;  C_1\geq 0,
        $$
        almost everywhere for $t$ in $0\leq t\leq T$, then  $\xi \equiv 0$,
        almost everywhere for $t$ in $0\leq t\leq T$.
\end{lem}

\begin{lem}
\label{lemma6}
 Let $U \subset \mathbb{R}^n$ be open, $\mathbf{f}: U \rightarrow \mathbb{R}^m$ continuously differentiable and $\mathbf{x}\in U$, $\mathbf{h}\in \mathbb{R}^m$ vectors such that the line segment $\mathbf{x}+z \; \mathbf{h}$,  $0 \leq z \leq 1$ remains in $U$. Then, we have:
\begin{equation}
    \mathbf{f}(\mathbf{x}+\mathbf{h})-\mathbf{f}(\mathbf{x}) = \left (\int_0^1 \mathbb{D}\mathbf{f}(\mathbf{x}+z \; \mathbf{h})\,dz\right)\cdot \mathbf{h},
\end{equation} where  $\mathbb{D} \mathbf{f}$ denotes the Jacobian matrix  of $\mathbf{f}$ and the integral of a matrix is understood as componentwise.
\end{lem}

\textbf{Proof of Theorem \ref{LFZ11}}. 

\textbf{Step 1}: 
From equation   \eqref{EQ:61b}
it follows that $H$ is a  monotonic decreasing function of  $t$  if $0<\Omega^2+ \Sigma^2 <1$. 
This allows to define recursively the bootstrapping sequences 
\begin{align}
    & \left\{\begin{array}{c}
       t_0=t_{*}   \\ \\
        H_0=H(t_{*}) 
    \end{array}\right., \quad  \left\{\begin{array}{c}
       {t_{n+1}}= {t_{n}} +\frac{1}{H_n}   \\ \\
       H_{n+1}= H(t_{n+1})  
    \end{array}\right.,
\end{align}
such that  $\lim_{n\rightarrow \infty}H_n=0$ y $\lim_{n\rightarrow \infty} t_n=\infty$.

\begin{widetext}
\noindent Given  expansions \eqref{quasilinear211} from \eqref{eqT59} we have 
\begin{align*}
   & \dot{\Omega_0}= \frac{3}{2} H {\Omega_0} \Bigg(\gamma  \left(1-{\Sigma_0}^2-{\Omega_0}^2\right)   +2 {\Sigma_0}^2   +2\left({\Omega_0}^2-1\right) \cos^2 ({\Phi_0}-t \omega )\Bigg)   -\frac{\partial g_1}{\partial t}
H +\mathcal{O}\left(H^2\right), 
\\
   & \dot{\Sigma_0}= \frac{3}{2} H {\Sigma_0} \Bigg(-(\gamma -2) \left({\Sigma_0}^2-1\right)-(\gamma -1) \Omega_{0}^2    +{\Omega_0}^2 \cos (2 ({\Phi_0}-t \omega ))\Bigg) -\frac{\partial g_2}{\partial t}
H +\mathcal{O}\left(H^2\right), 
\\
   & \dot{\Phi_0}= - { \left(3
     \cos ({\Phi_0}-t \omega ) \sin ({\Phi_0}-t \omega)\right) H} -\frac{\partial g_3}{\partial t}
H +\mathcal{O}\left(H^2\right). 
\end{align*}
Let  define  
$\Delta \Omega_0= \Omega_0 - \bar{\Omega}, \; \Delta \Sigma_0= \Sigma_0 - \bar{\Sigma}, \; \Delta \Phi_0= \Phi_0 - \bar{\Phi}$ and take same initial conditions at $t=t_n$, such that 
$\Omega_0(t_n)=\bar{\Omega}(t_n)= {\Omega_{n}}, \;   \Sigma_0(t_n)=\bar{\Sigma}(t_n)= {\Sigma}_{n}, $ $ \Phi_0(t_n)=\bar{\Phi}(t_n)= {\Phi}_{n}, \; 0 <  {\Omega_{n}} <1, \; -1 < {\Sigma}_{n}<1.$ 
The system \eqref{eqT602} becomes
\begin{align}\label{EDOgLFZ11}
 & \frac{\partial g_1}{\partial t}=\frac{3}{2} \Omega_0 \left(\Omega_0^2-1\right) \cos (2 (\Phi_0-t \omega )),
\quad   \frac{\partial g_2}{\partial t}= \frac{3}{2} \Sigma_0 \Omega_0^2 \cos (2 (\Phi_0-t \omega )),
\quad  \frac{\partial g_3}{\partial t}=\frac{3}{2} \sin (2 ({\Phi_0}-t \omega )).
\end{align}
Explicit expressions for  $g_i$ are obtained straightforwardly by integration of \eqref{EDOgLFZ11}:
\begin{align}
& g_1 (\Omega_{0}, \Sigma_{0}, \Phi_{0}, t)= \frac{3 \Omega_0 \left(1-\Omega_0^2\right) \sin (2 (\Phi_0-t
   \omega ))}{4 \omega },\\ 
& g_2 (\Omega_{0}, \Sigma_{0}, \Phi_{0}, t)= -\frac{3 \Sigma_0 \Omega_0^2 \sin (2 (\Phi_0-t \omega
   ))}{4 \omega }, \\ 
& g_3 (\Omega_{0}, \Sigma_{0}, \Phi_{0}, t)=  \frac{3 \cos (2 (\Phi_0-t \omega ))}{4 \omega },
\end{align}
where we set three integration functions $C_i(\Omega_{0}, \Sigma_{0}, \Phi_{0}), i=1, 2, 3$ to zero. 
\noindent
The   $g_i, i=1, 2, 3$ are continuously differentiable, such that their partial derivatives are bounded on $t\in [t_n, t_{n+1}]$. 

\noindent The second order expansion around $H=0$ of system \eqref{EqY602} is written as: 
\begin{subequations}
\label{newEqY602}
\begin{align}
& \dot{\Delta\Omega_0}= \frac{1}{2} H \left(3 \bar{\Omega } \left((\gamma -2) \bar{\Sigma }^2+(\gamma -1) \left(\bar{\Omega }^2-1\right)\right)-3 \Omega_0 \left(\gamma  \left(\Sigma_0^2+\Omega_0^2-1\right)-2
   \Sigma_0^2-\Omega_0^2+1\right)\right) \nonumber\\
   & +\frac{H^2}{8 b^2 \mu ^6 \omega ^3} \Bigg[8 \Omega_0^3 \left(2 \mu ^2-\omega ^2\right)^3 \sin ^3(\Phi_0-t \omega ) \cos (\Phi_0-t \omega ) \nonumber\\
   & -9 b^2 \mu ^6 \omega ^2
   \Omega_0 \left(\Omega_0^2-1\right) \sin (2 (\Phi_0-t \omega )) \left( \gamma  \left(1-\Sigma_0^2-\Omega_0^2\right)+2 \Sigma_0^2+\left(4 \Omega_0^2+1\right) \cos (2
   (\Phi_0-t \omega ))+\Omega_0^2\right)\Bigg], \label{firsteq}\\
   & \dot{\Delta\Sigma_0}=\frac{3}{2} H \left(\bar{\Sigma } \left((\gamma -2) \left(\bar{\Sigma }^2-1\right)+(\gamma -1) \bar{\Omega }^2\right)-(\gamma -2)
   \Sigma_0 \left(\Sigma_0^2-1\right)-(\gamma -1) \Sigma_0 \Omega_0^2\right) \nonumber \\
   & -\frac{9 H^2 \Sigma_0 \Omega_0^2 \sin (2 (\Phi_0-t \omega )) \left(\gamma 
   \left(1-\Sigma_0^2-\Omega_0^2\right)+2 \Sigma_0^2+4 \Omega_0^2 \cos (2 (\Phi_0-t \omega ))+\Omega_0^2\right)}{8 \omega }, \label{secondeq}\\
   & \dot{\Delta\Phi_0}=\frac{3 H^2}{8 \omega ^3}  \Bigg[\frac{\left(2 \mu ^2-\omega ^2\right)^3 \bar{\Omega }^2}{b^2 \mu ^6}  -\frac{8 \Omega_0^2 \left(2 \mu ^2-\omega ^2\right)^3 \sin ^4(\Phi_0-t \omega )}{3 b^2 \mu ^6} \nonumber \\
   & +3 \omega ^2 \cos (2 (\Phi_0-t \omega )) \Bigg(\gamma  \left(1-\Sigma_0^2-\Omega_0^2\right) +2 \Sigma_0^2+\left(\Omega_0^2+2\right) \cos (2 (\Phi_0-t \omega ))+\Omega_0^2\Bigg)\Bigg]. \label{thirdeq}
\end{align}
\end{subequations}
\noindent
Denoting $\mathbf{x}_0=(\Omega_0, \Sigma_0)^T$, $\bar{\mathbf{x}}=(\bar{\Omega}, \bar{\Sigma})^T$  equations \eqref{firsteq} and \eqref{secondeq} are reduced to: 
\begin{align*}
 & \dot{\Delta\mathbf{x}_0}= H \left(\bar{\mathbf{f}}( {\mathbf{x}}_0)-\bar{\mathbf{f}}(\bar{\mathbf{x}})\right) +   \mathcal{O}(H^2), 
  \end{align*}
where the vector function $\bar{\mathbf{f}}$ 
is explicitly given  (the last row corresponding to  eq. \eqref{thirdeq} was omitted) by: 
\begin{align*}
    \bar{\mathbf{f}}(y_1, y_2)= \left(
\begin{array}{c}
 -\frac{3}{2}  y_1 \left(-\gamma +(\gamma -1)  y_1^2+(\gamma -2)  y_2^2+1\right) \\
 -\frac{3}{2}  y_2 \left((\gamma -1)  y_1^2+(\gamma -2) \left( y_2^2-1\right)\right) \\
\end{array}
\right).
\end{align*}
It is a vector function with polynomial components in variables $(y_1, y_2)$. Therefore, it is continuously differentiable in all its components. 

\noindent Let be $\Delta\mathbf{x}_0(t)= (\Omega_0-\bar{\Omega},{\Sigma_0}-  \bar{\Sigma})^T$ with $0\leq |\Delta\mathbf{x}_0|:=\max \left\{|\Omega_0-\bar{\Omega}|, |{\Sigma_0}-  \bar{\Sigma}| \right\}< \infty$ in the closed interval $[t_n,t_{n+1}]$. Using same initial conditions for $\mathbf{x}_0$ and $\bar{\mathbf{x}}$ we obtain by integration: 
\begin{align}
 \Delta\mathbf{x}_0(t) = \int_{t_n}^t \dot{\Delta\mathbf{x}_0} d s =  \int_{t_n}^t \left(H \left(\bar{\mathbf{f}}( {\mathbf{x}}_0)-\bar{\mathbf{f}}(\bar{\mathbf{x}})\right) +   \mathcal{O}(H^2)\right) ds. \label{int}
\end{align}
The terms of order $\mathcal{O}(H^2)$ under the integral sign in eq. \eqref{int} come from the second order terms in the series expansion centered in $H=0$ of $\dot \Delta \Omega_0$ and  $\dot \Delta \Sigma_0$ in \eqref{firsteq} and \eqref{secondeq}. These terms are bounded in the interval $[t_n, t_{n+1}]$ by $M_1 H_n^2$, where 
\begin{align*}
& M_1= \max_{t\in[t_{n},t_{n+1}]}  \Bigg\{  \Bigg{|} \frac{1}{8 b^2 \mu ^6 \omega ^3} \Bigg[8 \Omega_0^3 \left(2 \mu ^2-\omega ^2\right)^3 \sin ^3(\Phi_0-t \omega ) \cos (\Phi_0-t \omega ) \nonumber\\
   & -9 b^2 \mu ^6 \omega ^2
   \Omega_0 \left(\Omega_0^2-1\right) \sin (2 (\Phi_0-t \omega )) \left( \gamma  \left(1-\Sigma_0^2-\Omega_0^2\right)+2 \Sigma_0^2+\left(4 \Omega_0^2+1\right) \cos (2
   (\Phi_0-t \omega ))+\Omega_0^2\right)\Bigg]\Bigg{|}, \nonumber \\
   & \Bigg{|}\frac{9  \Sigma_0 \Omega_0^2 \sin (2 (\Phi_0-t \omega )) \left( \gamma 
   \left(1-\Sigma_0^2-\Omega_0^2\right)+2 \Sigma_0^2+4 \Omega_0^2 \cos (2 (\Phi_0-t \omega ))+\Omega_0^2\right)}{8 \omega }\Bigg{|}\Bigg\}
\end{align*}
is finite by continuity of $\bar{\Omega}, \Omega_0, \bar{\Sigma}, \Sigma_0,  \Phi_0$ in the closed interval $[t_{n},t_{n+1}]$. 

\noindent 
Using Lemma \ref{lemma6} we have 
\begin{equation}
   \bar{\mathbf{f}}( {\mathbf{x}}_0(s))-\bar{\mathbf{f}}(\bar{\mathbf{x}}(s)) = \underbrace{\left (\int_0^1 \mathbb{D}   \bar{\mathbf{f}}\left(\bar{\mathbf{x}}(s)+ z \; \left({\mathbf{x}}_0(s) - \bar{\mathbf{x}}(s)\right)\right)\,d z\right)}_{\mathbb{A}(s)}\cdot \left({\mathbf{x}}_0(s) - \bar{\mathbf{x}}(s)\right),
\end{equation} where  $\mathbb{D}\bar{\mathbf{f}}$ denotes the Jacobian matrix  of $\bar{\mathbf{f}}$ and the integral of a matrix is understood as componentwise.

\noindent 
Omitting the dependence on $s$ we calculate the matrix elements of $\mathbb{A}$ as
\begin{equation}
    \mathbb{A} = \left(
\begin{array}{ccc}
 a  & b \\
 c  & d \\ 
\end{array}
\right),
\end{equation}
where 
\begin{align}
    & a= -\frac{3}{2} \left(\frac{1}{3} (\gamma -2) \left(\Sigma_0
   \bar{\Sigma }+\bar{\Sigma }^2+\Sigma_0^2\right)+(\gamma -1)
   \left(\Omega_0 \bar{\Omega }+\bar{\Omega }^2+\Omega_0^2-1\right)\right),\\
   & b= -\frac{1}{2} (\gamma -2)
   \left(\bar{\Sigma } \left(2 \bar{\Omega }+\Omega_0\right)+\Sigma_0 \left(\bar{\Omega }+2 \Omega_0\right)\right), \\
   & c= -\frac{1}{2} (\gamma -1) \left(\bar{\Sigma } \left(2 \bar{\Omega
   }+\Omega_0\right)+\Sigma_0 \left(\bar{\Omega }+2
   \Omega_0\right)\right), \\
   & d=  -\frac{3}{2} \left((\gamma -2)
   \left(\Sigma_0 \bar{\Sigma }+\bar{\Sigma }^2+\Sigma_0^2-1\right)+\frac{1}{3} (\gamma -1) \left(\Omega_0
   \bar{\Omega }+\bar{\Omega }^2+\Omega_0^2\right)\right).
\end{align}
Taking the sup norm
$ |\Delta\mathbf{x}_0|=\max \left\{|\Omega_0-\bar{\Omega}|, |{\Sigma_0}-  \bar{\Sigma}| \right\}$ we have for all $t\in[t_n, t_{n+1}]$: 
\begin{align*}
 & \Big{|}\Delta\mathbf{x}_0(t) \Big{|} =  \Big{|}\int_{t_n}^t \dot{\Delta\mathbf{x}_0} d s \Big{|}= \Bigg{|} \int_{t_n}^t \Big(H \left(\bar{\mathbf{f}}( {\mathbf{x}}_0(s))-\bar{\mathbf{f}}(\bar{\mathbf{x}}(s))\right) + \mathcal{O}(H^2) \Big) ds  \Bigg{|}  \nonumber\\
 & \leq \Bigg{|} \int_{t_n}^t  H \left(\bar{\mathbf{f}}( {\mathbf{x}}_0(s))-\bar{\mathbf{f}}(\bar{\mathbf{x}}(s))\right)   ds  \Bigg{|} + M_1 H_n^2 (t-t_n) = \Bigg{|} \int_{t_n}^t  H \mathbb{A}(s)\cdot \Delta\mathbf{x}_0(s)   ds  \Bigg{|} + M_1 H_n^2 (t-t_n)
\nonumber \\
  &
  \leq H_n \int_{t_n}^t  \Bigg{|} \left(
\begin{array}{cc}
 a  & b  \\
 c & d \\
\end{array}
\right) \cdot \Delta\mathbf{x}_0(s) \Bigg{|} ds + M_1 H_n^2 (t-t_n)
\end{align*}
On the other hand 
\begin{equation*}
    \Bigg{|} \left(
\begin{array}{cc}
 a & b \\
 c & d \\
\end{array}
\right) \cdot \Delta\mathbf{x}_0(s) \Bigg{|}\leq 2  \Bigg{|} \left(
\begin{array}{cc}
 a & b \\
 c & d \\
\end{array}
\right)\Bigg{|} \Big{|}\Delta\mathbf{x}_0(s)\Big{|}
\end{equation*}
where the sup norm of a matrix
$\Bigg{|} \left(
\begin{array}{cc}
 a & b \\
 c & d \\
\end{array}
\right)\Bigg{|}$ is defined by $\max\{|a|, |c|, |b|, |d|\}$. 
\newline
By continuity of $\bar{\Omega}, \Omega_0, \bar{\Sigma}, \Sigma_0$  in $[t_n, t_{n+1}]$ 
\begin{equation*}
    L_1= 2 \max_{s\in[t_n,t_{n+1}]} \Bigg{|} \left(
\begin{array}{cc}
 a(s) & b(s) \\
 c(s) & d(s) \\
\end{array}
\right)\Bigg{|}
\end{equation*}
is finite. Hence, for all $t\in[t_n, t_{n+1}]$ we have: 
\begin{align*}
 & \Big{|}\Delta\mathbf{x}_0(t) \Big{|}  \leq L_1 H_n \int_{t_n}^t  \Big{|}
 \Delta\mathbf{x}_0(s) \Big{|} ds +   M_1 H_n^2 (t-t_n) \leq  L_1 H_n \int_{t_n}^t  \Big{|}  \Delta\mathbf{x}_0(s) \Big{|} ds +   M_1 H_n
\end{align*}
due to $t-t_n\leq {t_{n+1}}- {t_{n}} =\frac{1}{H_n}$.

\noindent Using   Gronwall's Lemma \ref{Gronwall}, we have for $t \in[t_n, t_{n+1}]$: 
\begin{align*}
 & \Big{|} \Delta \mathbf{x}_0(t)  \Big{|} \leq  M_1  H_n   e^{L_1  H_n(t-t_n)} \leq    M_1  {H_n}e^{L_1}.
 \end{align*}
  Then, 
 \begin{align*}
& \Big{|} \Delta \Omega_0(t) \Big{|} \leq    M_1 e^{L_1} {H_n}, \;  \Big{|} \Delta \Sigma_0(t) \Big{|} \leq     M_1 e^{L_1} {H_n}.
\end{align*}
Furthermore, defining
  \begin{align}
      M_2= & \max_{t\in[t_{n},t_{n+1}]}  \Bigg{|} \frac{3 }{8 \omega ^3}  \Bigg[\frac{\left(2 \mu ^2-\omega ^2\right)^3 \bar{\Omega }^2}{b^2 \mu ^6} -\frac{8 \Omega_0^2 \left(2 \mu ^2-\omega ^2\right)^3 \sin ^4(\Phi_0-t \omega )}{3 b^2 \mu ^6} \nonumber \\
& +3 \omega ^2 \cos (2({\Phi_0}-s \omega)) \Big(\gamma  \left(1-{\Sigma_0}^2-{\Omega_0}^2\right)  +2 {\Sigma_0}^2 +\left({\Omega_0}^2+2\right) \cos (2 ({\Phi_0}-s \omega ))+{\Omega_0}^2\Big)\Bigg] \Bigg{|}
  \end{align}
  which is finite by continuity of $\bar{\Omega}, \Omega_0, \bar{\Sigma}, \Sigma_0,  \Phi_0$ in the closed interval $[t_{n},t_{n+1}]$, we obtain from eq. \eqref{thirdeq} that 
\begin{align*}
& |\Delta \Phi_0(t)|
= \Bigg{|} \int_{t_n}^{t} \dot{\Delta\Phi_0}(s)  d s\Bigg{|}  \leq  M_2 H_n^2 (t-t_n) + \Big{|}\mathcal{O}({ {H_n}}^3)\Big{|} \leq 
    |\Delta \Phi_0(t)| \leq  M_2 H_n +\Big{|}\mathcal{O}({ {H_n}}^3)\Big{|},
\end{align*}
due to $t-t_n\leq {t_{n+1}}- {t_{n}} =\frac{1}{H_n}$.
Finally, taking limit $n\rightarrow \infty$ we obtain $H_n\rightarrow  0$. Then,
as $H_n \rightarrow 0$  functions $\Omega_{0}, \Sigma_{0},  \Phi_0$ and  $\bar{\Omega},  \bar{\Sigma},  \bar{\Phi}$  have same limit as $t\rightarrow \infty$.

\textbf{Step 2}: For FLRW the second order expansion around $H=0$ of system \eqref{EqY602} is written as: 
\begin{subequations}
\begin{align}
  &  \dot{\Delta\Omega_0}= -\frac{3}{2}(\gamma-1)H \Delta\Omega_0 \left(1-\Omega_0^2-\Omega_0 \bar{\Omega}-\bar{\Omega}^2\right) \nonumber \\
    & -\frac{9 H^2 \Omega_0 \left(\Omega_0^2-1\right) \sin (2 (\Phi_0-t \omega )) \left(-\gamma  \Omega_0^2+\gamma +\left(4 \Omega_0^2+1\right) \cos (2 (\Phi_0-t \omega
   ))+\Omega_0^2\right)}{8 \omega } \nonumber \\
   & - \frac{ H^2\Omega_0^3 \left(\omega ^2 -2 \mu ^2\right)^3 \sin ^3(\Phi_0-t \omega ) \cos (\Phi_0-t \omega )}{b^2 \mu ^6 \omega ^3}. \label{eq}
\\
  &  \dot{\Delta\Phi_0}= \frac{3 H^2}{8 \omega ^3}  \Bigg[\frac{\left(2 \mu ^2-\omega ^2\right)^3 \bar{\Omega }^2}{b^2 \mu ^6} -\frac{8 \Omega_0^2 \left(2 \mu ^2-\omega ^2\right)^3 \sin ^4(\Phi_0-t \omega )}{3 b^2 \mu ^6} \nonumber \\
& +3 \omega ^2 \cos (2({\Phi_0}-t \omega)) \Big(\gamma  \left(1-{\Omega_0}^2\right)     +\left({\Omega_0}^2+2\right) \cos (2 ({\Phi_0}-t \omega ))+{\Omega_0}^2\Big)\Bigg]. \label{eq2}
\end{align}
\end{subequations}
Then, 
\begin{align*}
& |\Delta \Omega_0(t)| = \Bigg{|} \int_{t_n}^{t} \left( \dot{\Omega_0}(s)- \dot{\bar{\Omega}}(s) \right) d s\Bigg{|}   = \Bigg{|} \int_{t_n}^{t} \left[-\frac{3}{2}(\gamma-1)H \Delta\Omega_0 \left(1-\Omega_0^2-\Omega_0 \bar{\Omega}-\bar{\Omega}^2\right) + \mathcal{O}(H^2) \right]  d s\Bigg{|}  
\end{align*}
and 
\begin{align*}
    & |\Delta \Phi_0(t)|
= \Bigg{|} \int_{t_n}^{t} \left( \dot{\Phi_0}(s)- \dot{\bar{\Phi}}(s)\right) d s\Bigg{|} \nonumber \\
& = \Bigg{|} \bigints_{t_n}^{t} \Bigg\{\frac{3 H^2}{8 \omega ^3}  \Bigg[\frac{\left(2 \mu ^2-\omega ^2\right)^3 \bar{\Omega }^2}{b^2 \mu ^6} -\frac{8 \Omega_0^2 \left(2 \mu ^2-\omega ^2\right)^3 \sin ^4(\Phi_0-t \omega )}{3 b^2 \mu ^6} \nonumber \\
& +3 \omega ^2 \cos (2({\Phi_0}-s \omega)) \Big(\gamma  \left(1-{\Omega_0}^2\right)     +\left({\Omega_0}^2+2\right) \cos (2 ({\Phi_0}-s \omega ))+{\Omega_0}^2\Big)\Bigg] + \mathcal{O}({ {H}}^3) \Bigg\} d s\Bigg{|}.
\end{align*}
By continuity of $\bar{\Omega}, \Omega_0$ and  $\Phi_0$  in $[t_n, t_{n+1}]$ the following finite constants are found:
\begin{equation*}
    L_2= \max_{t\in[t_n,t_{n+1}]} \Bigg{|} \frac{3}{2}  (1-\gamma)  \left(1-\Omega_0^2-\Omega_0 \bar{\Omega}-\bar{\Omega}^2\right)\Bigg{|},
\end{equation*}
\begin{align*}
    M_3= \max_{t\in[t_n,t_{n+1}]}\Bigg{|} & \frac{9 \Omega_0 \left(\Omega_0^2-1\right) \sin (2 (\Phi_0-t \omega )) \left(-\gamma  \Omega_0^2+\gamma +\left(4 \Omega_0^2+1\right) \cos (2 (\Phi_0-t \omega
   ))+\Omega_0^2\right)}{8 \omega } \nonumber \\
   & +  \frac{\Omega_0^3 \left(\omega ^2 -2 \mu ^2\right)^3 \sin ^3(\Phi_0-t \omega ) \cos (\Phi_0-t \omega )}{b^2 \mu ^6 \omega ^3}\Bigg{|} 
\end{align*}
and 
\begin{align*}
 &   M_4= \max_{t\in[t_n,t_{n+1}]} \Bigg{|}  \frac{3}{8 \omega ^3}  \Bigg[\frac{\left(2 \mu ^2-\omega ^2\right)^3 \bar{\Omega }^2}{b^2 \mu ^6} -\frac{8 \Omega_0^2 \left(2 \mu ^2-\omega ^2\right)^3 \sin ^4(\Phi_0-t \omega )}{3 b^2 \mu ^6}\nonumber\\
 & +3 \omega ^2 \cos (2({\Phi_0}-s \omega)) \Big(\gamma  \left(1-{\Omega_0}^2\right)    +\left({\Omega_0}^2+2\right) \cos (2 ({\Phi_0}-s \omega ))+{\Omega_0}^2\Big)\Bigg]\Bigg{|},
\end{align*}
such that the terms proportional to $H^2$ in eq. \eqref{eq} are bounded in absolute value  by $M_3 H_n^2$ and the terms proportional to $H^2$ in eq. \eqref{eq2} are bounded in absolute value  by $M_4 H_n^2$ in the interval $[t_n, t_{n+1}]$. Then, 
\begin{align*}
& |\Delta \Omega_0(t)| 
\leq  \int_{t_n}^{t} \underbrace{\Bigg{|} \frac{3}{2}  (1-\gamma)  \left(1-\Omega_0^2-\Omega_0 \bar{\Omega}-\bar{\Omega}^2\right)\Bigg{|}}_{\leq L_2}\underbrace{H }_{\leq H_n}  \;  |\Delta \Omega _{0}(s)  | d s   + M_3 {H_n}^2 (t-t_n) \nonumber \\
 & \leq  L_2  H_n \int_{t_n}^t |\Delta\Omega _{0}(s)|  ds  + M_3 H_n^2 (t-t_n)  \leq  L_2  H_n \int_{t_n}^t |\Delta\Omega _{0}(s)|  ds  + M_3 H_n
\end{align*}
\noindent due to $t-t_n\leq {t_{n+1}}- {t_{n}} =\frac{1}{H_n}$.
 \end{widetext} 
\noindent 
Using   Gronwall's Lemma \ref{Gronwall}, we have for $t \in[t_n, t_{n+1}]$: 
\begin{align*}
 & \Big{|} \Delta \Omega_0(t) \Big{|} \leq  M_3 H_n   e^{L_2  H_n(t-t_n)} \leq    M_3 {H_n}e^{L_2}.
 \end{align*}
Furthermore, from eq. \eqref{eq2}
we have
\begin{align*}
& |\Delta \Phi_0(t)|
= \Bigg{|} \int_{t_n}^{t}  \dot{\Delta\Phi_0}(s) d s\Bigg{|}   \leq  M_4 H_n^2 (t-t_n) + \Big{|}\mathcal{O}({ {H_n}}^3)\Big{|} \nonumber \\
& \leq 
   M_4 H_n +\Big{|}\mathcal{O}({ {H_n}}^3)\Big{|},
\end{align*}
\noindent due to $t-t_n\leq {t_{n+1}}- {t_{n}} =\frac{1}{H_n}$.
Finally, taking limit $n\rightarrow \infty$ we obtain $H_n\rightarrow  0$. Then,
as $H_n \rightarrow 0$  functions $\Omega_{0},  \Phi_0$ and  $\bar{\Omega},    \bar{\Phi}$  have same limit as $t\rightarrow \infty$. $\square$

Summarizing, according to Theorem   \ref{LFZ11} for Bianchi I and flat FLRW metrics,  Hubble parameter $H$ plays the role of a  time-dependent perturbation parameter controlling the magnitude of the error between solutions of  full and  time-averaged systems. Therefore, the analysis is reduced to study time-averaged equations.  

\section{Numerical simulation}
\label{numerics}

In this Section we present numerical evidence that supports   the main results in Section \ref{SECT:II} by solving numerically  full and averaged systems for each metric, namely LRS Bianchi I and flat  FLRW. 

\noindent
For this purpose, an algorithm in the programming language \textit{Python} was elaborated where systems of differential equations were solved using the \textit{solve\_ivp} code that is provided by the \textit{SciPy} open-source \textit{Python}-based ecosystem. The integration method is an implicit Runge-Kutta method of the Radau IIa family of order $5$ with  relative and absolute tolerances of $10^{-4}$ and $10^{-7}$, respectively. All systems of differential equations were integrated with respect  to  $\tau$ in an integration range of $-40\leq\tau\leq 10$   for the original systems and in an integration range of $-40\leq\tau\leq 100$  for the averaged systems. All of them are partitioned in $10000$ data points. Furthermore, each full and time-averaged systems were solved considering only one matter component. These are cosmological constant ($\gamma=0$), non relativistic matter or dust ($\gamma=1$), radiation ($\gamma=4/3$) and stiff fluid ($\gamma=2$). Vacuum solutions correspond to   $\Omega=\Omega_m\equiv 0$ and  solutions without matter fluid correspond to $\Omega_m\equiv 0$. Finally, we have considered these constants: $\mu=\sqrt{2}/2$, $b=\sqrt{2}/5$ and $\omega=\sqrt{2}$, that lead to the value of $f=\frac{b \mu ^3}{\omega ^2-2 \mu ^2}=1/10$ which fulfills the condition $f\geq 0$. These values are translated into a generalized harmonic potential: 
\begin{equation}
    V(\phi)=\frac{\phi ^2}{2}+\frac{1}{100}(1-\cos(10\phi)).
\end{equation}

\subsection{LRS Bianchi I model}

For LRS Bianchi I metric we integrate:
\begin{enumerate}
    \item The full system \eqref{unperturbed1}.
\item The time-averaged system \eqref{avrgsyst}.
\end{enumerate}
As  initial conditions we use seven data sets which are presented in Table \ref{Tab2}.

    \begin{table}[H]
  \centering    
   \caption{\label{Tab2} Seven initial data sets for the simulation of   full system \eqref{unperturbed1} and averaged system \eqref{avrgsyst} for   Bianchi I metric are displayed. All initial conditions satisfy $\bar{\Sigma}^{2}(0)+\bar{\Omega}^{2}(0)+\bar{\Omega}_{m}(0)$=1.}
\setlength{\tabcolsep}{5pt}
    \begin{tabular}{lcccccc}\hline
   Sol.  &  {$H(0)$} &  {${\bar{\Sigma}(0)}$} &  {$\bar{\Omega}^2(0)$} &  {$\bar{\Omega}_{m}(0)$}    &  {$\bar{\Phi}(0)$}  &  {$t(0)$}  \\\hline
              i &  $0.1$ & $0.1$ & $0.9$ & $0.09$  & $0$ & $0$\\
        ii &  $0.1$ & $0.4$ & $0.1$ & $0.74$  & $0$ & $0$\\
        iii &  $0.1$ & $0.6$ & $0.1$ & $0.54$  & $0$ & $0$\\
        iv &  $0.02$ & $0.48$ & $0.02$ & $0.7496$ & $0$ & $0$\\
        v &  $0.1$ & $0.48$ & $0.02$ & $0.7496$  & $0$ & $0$\\
        vi &  $0.1$ & $0.5$ & $0.01$ & $0.74$ & $0$ & $0$\\
        vii &  $0.1$ & $0$ & $0.685$ & $0.315$ & $0$ & $0$\\\hline
    \end{tabular}
\end{table}

\begin{figure*}
    \centering
    \subfigure[\label{fig:BICC3D} Projections in the space $(\Sigma, H, \Omega^2)$. The surface is given by the constraint $\Omega^{2}=1-\Sigma^{2}$.]{\includegraphics[scale = 0.40]{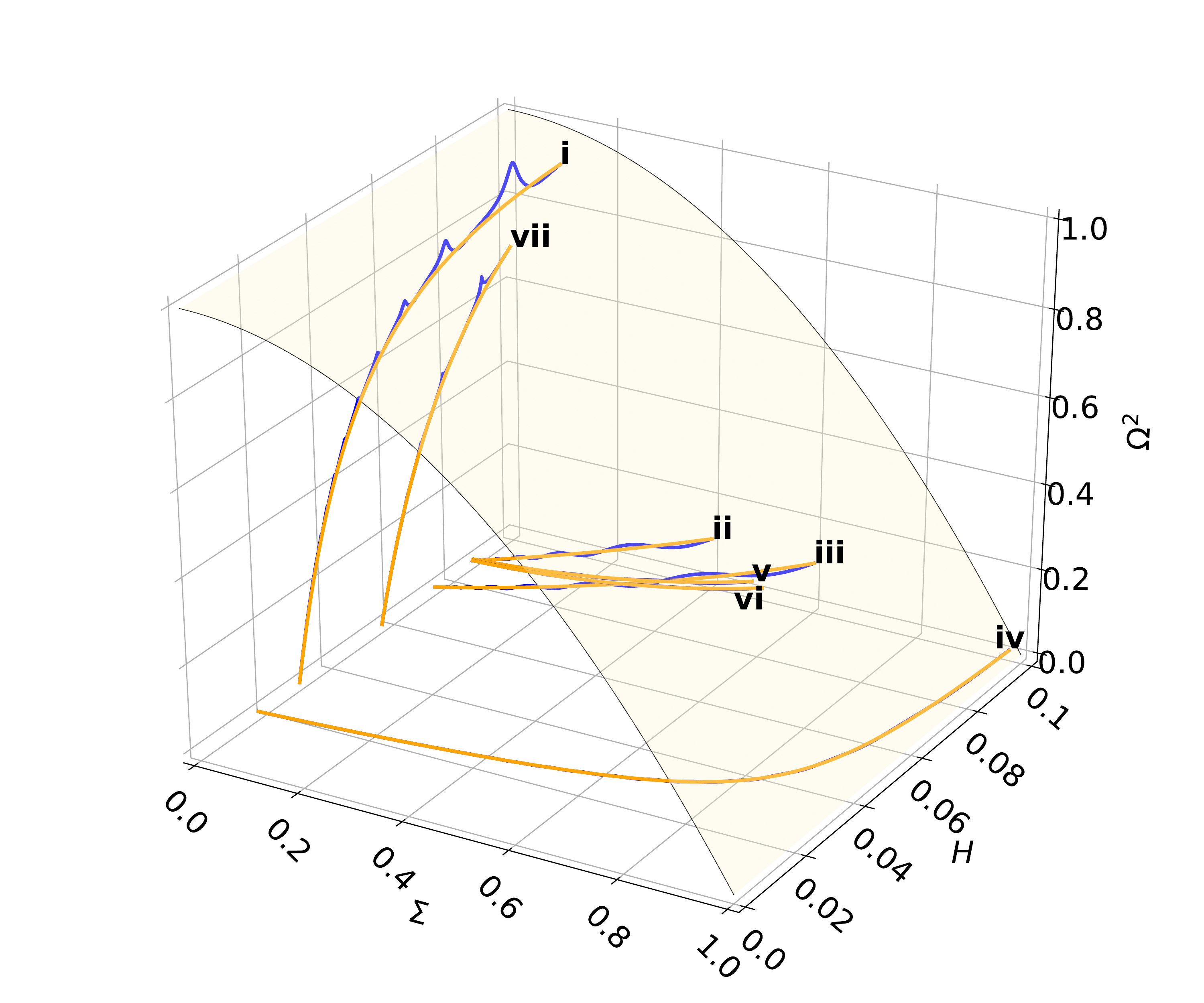}}
    \subfigure[\label{fig:BICC2D} Projection in the space $(\Sigma, \Omega^2)$. The black line represents the constraint $\Omega^{2}=1-\Sigma^{2}$.]{\includegraphics[scale = 0.51]{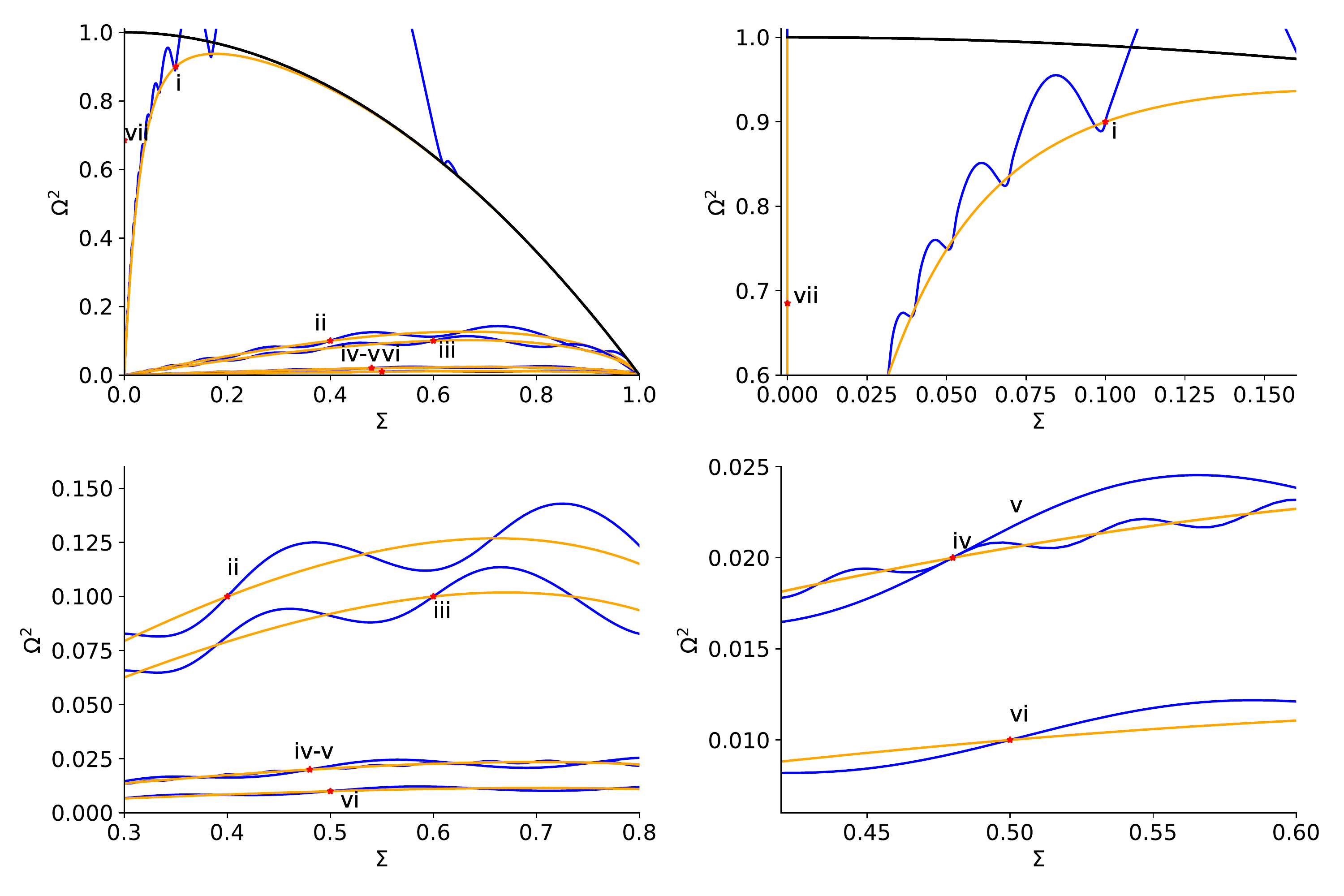}}
    \caption{Some solutions of the full system \eqref{unperturbed1} (blue) and time-averaged system \eqref{avrgsyst} (orange) for  Bianchi I metric when $\gamma=0$. We have used for both systems the initial data sets that are  presented in Table \ref{Tab2}.}
\end{figure*}

\begin{figure*}
    \centering
    \subfigure[\label{fig:BIDust3D} Projections in the space $(\Sigma, H, \Omega^2)$. The surface is given by the constraint $\Omega^{2}=1-\Sigma^{2}$.]{\includegraphics[scale = 0.39]{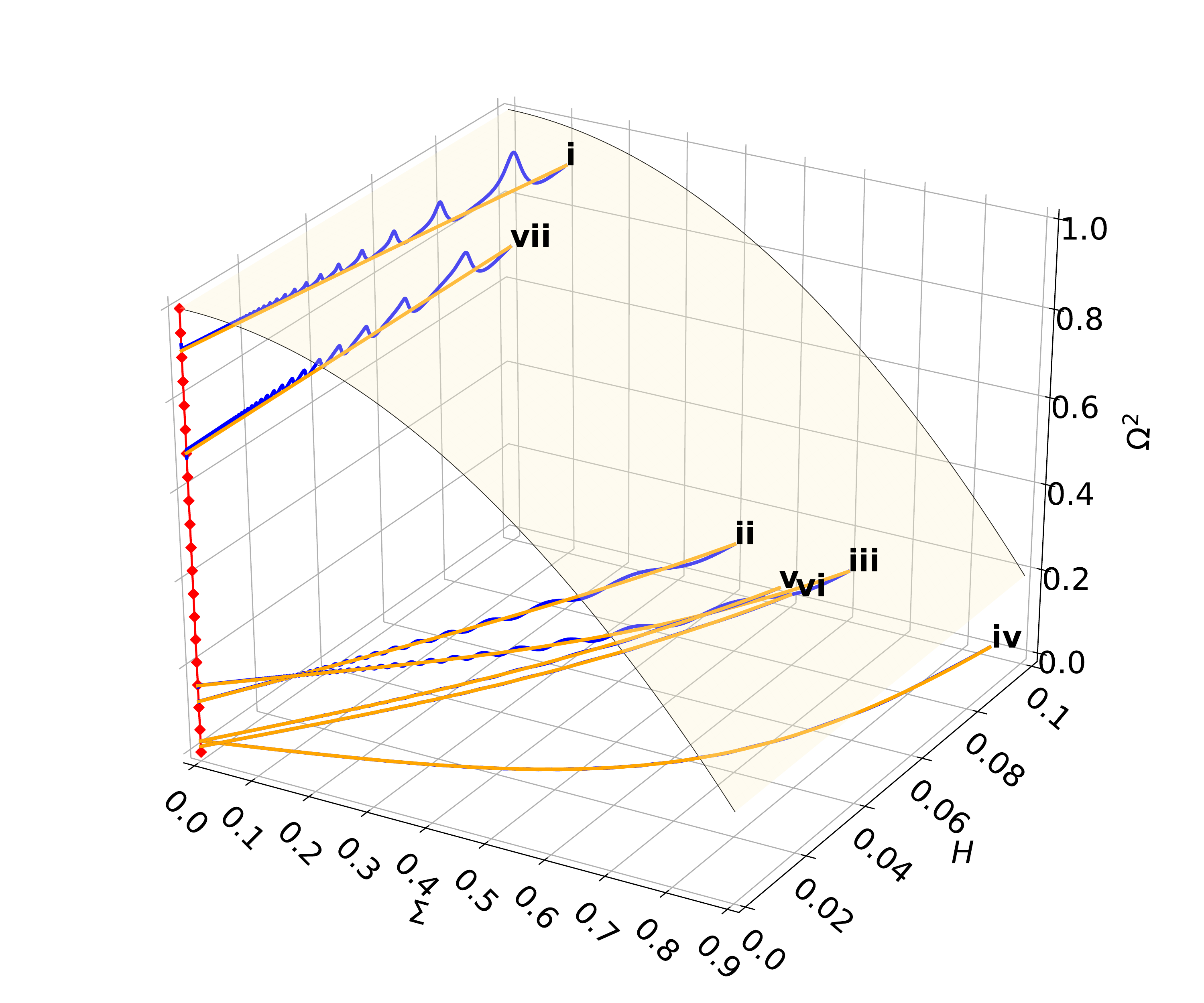}}
    \subfigure[\label{fig:BIDust2D} Projection in the space $(\Sigma, \Omega^2)$. The black line represents the constraint $\Omega^{2}=1-\Sigma^{2}$.]{\includegraphics[scale = 0.50]{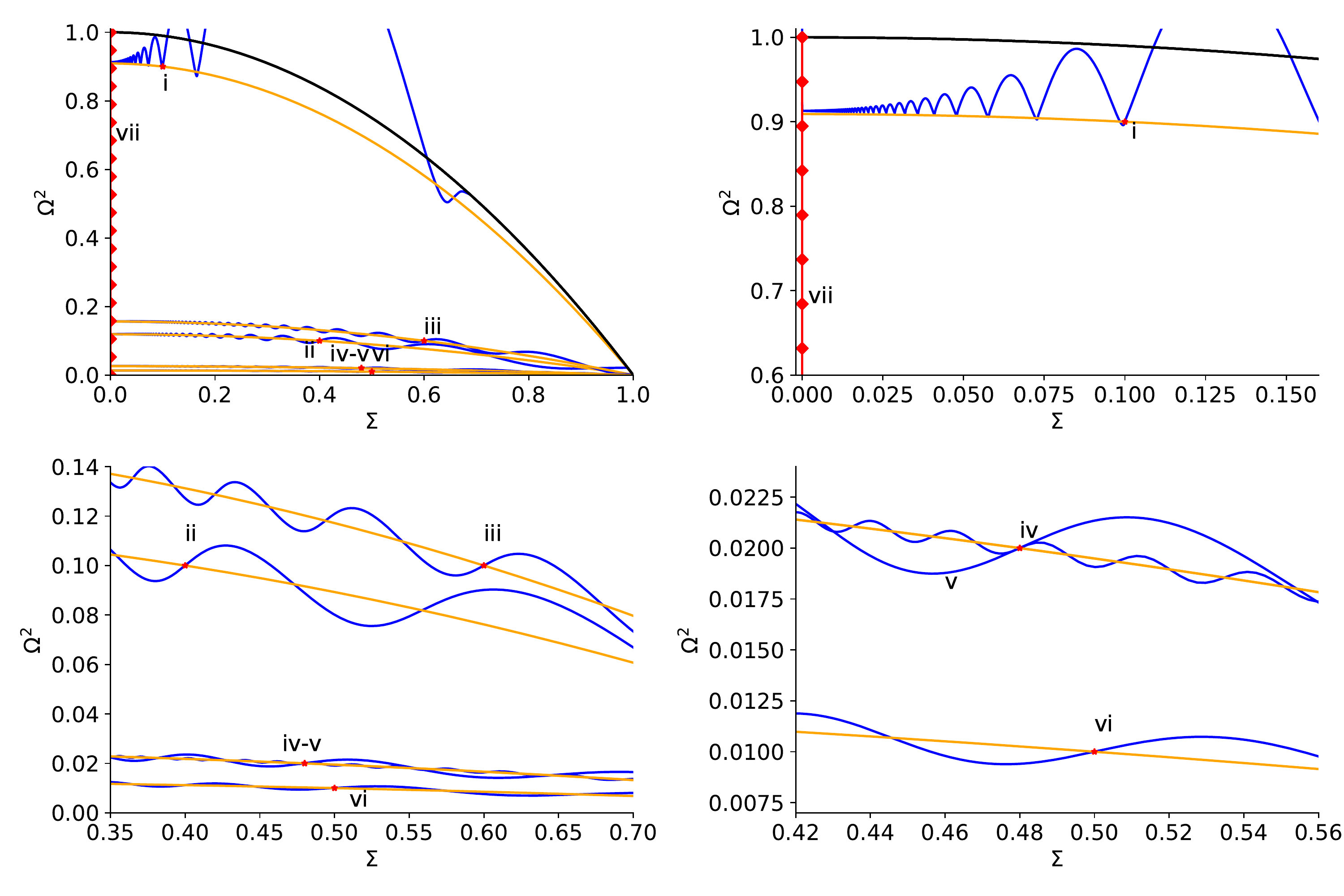}}
    \caption{Some solutions of the full system \eqref{unperturbed1} (blue) and time-averaged system \eqref{avrgsyst} (orange) for  Bianchi I metric when $\gamma=1$. The line denoted by red diamonds corresponds to the attracting line of equilibrium points. We have used for both systems the initial data sets that are presented in Table \ref{Tab2}.}
\end{figure*}

\begin{figure*}
    \centering
    \subfigure[\label{fig:BIRad3D} Projections in the space $(\Sigma, H, \Omega^2)$. The surface is given by the constraint $\Omega^{2}=1-\Sigma^{2}$.]{\includegraphics[scale = 0.40]{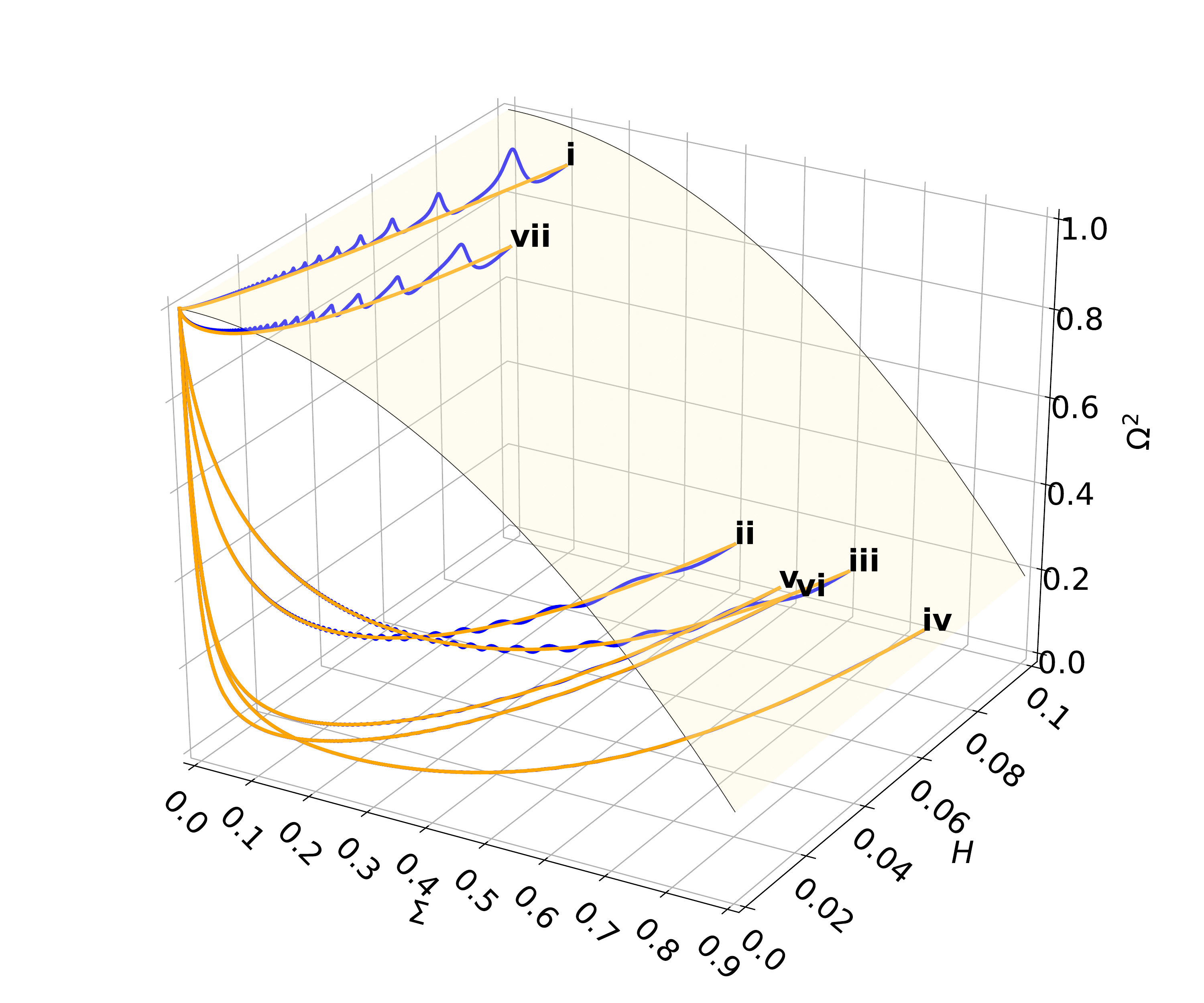}}
    \subfigure[\label{fig:BIRad2D} Projection in the space $(\Sigma, \Omega^2)$. The black line represents the constraint $\Omega^{2}=1-\Sigma^{2}$.]{\includegraphics[scale = 0.51]{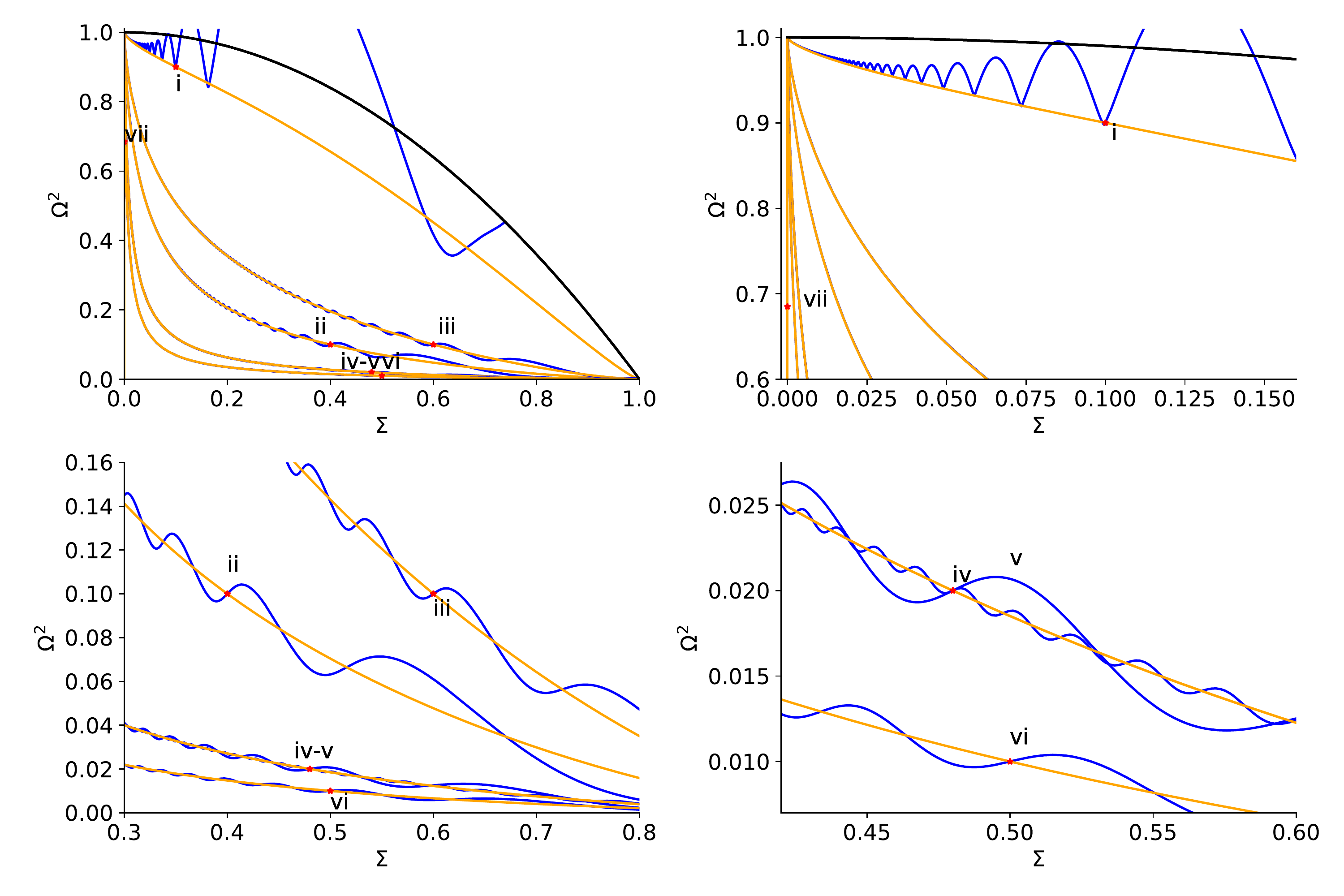}}
    \caption{Some solutions of the full system \eqref{unperturbed1} (blue) and time-averaged system \eqref{avrgsyst} (orange) for  Bianchi I metric when $\gamma=4/3$. We have used for both systems the initial data sets that are  presented in Table \ref{Tab2}.}
\end{figure*}

\begin{figure*}
    \centering
    \subfigure[\label{fig:BIStiff3D} Projections in the space $(\Sigma, H, \Omega^2)$. The surface is given by the constraint $\Omega^{2}=1-\Sigma^{2}$.]{\includegraphics[scale = 0.40]{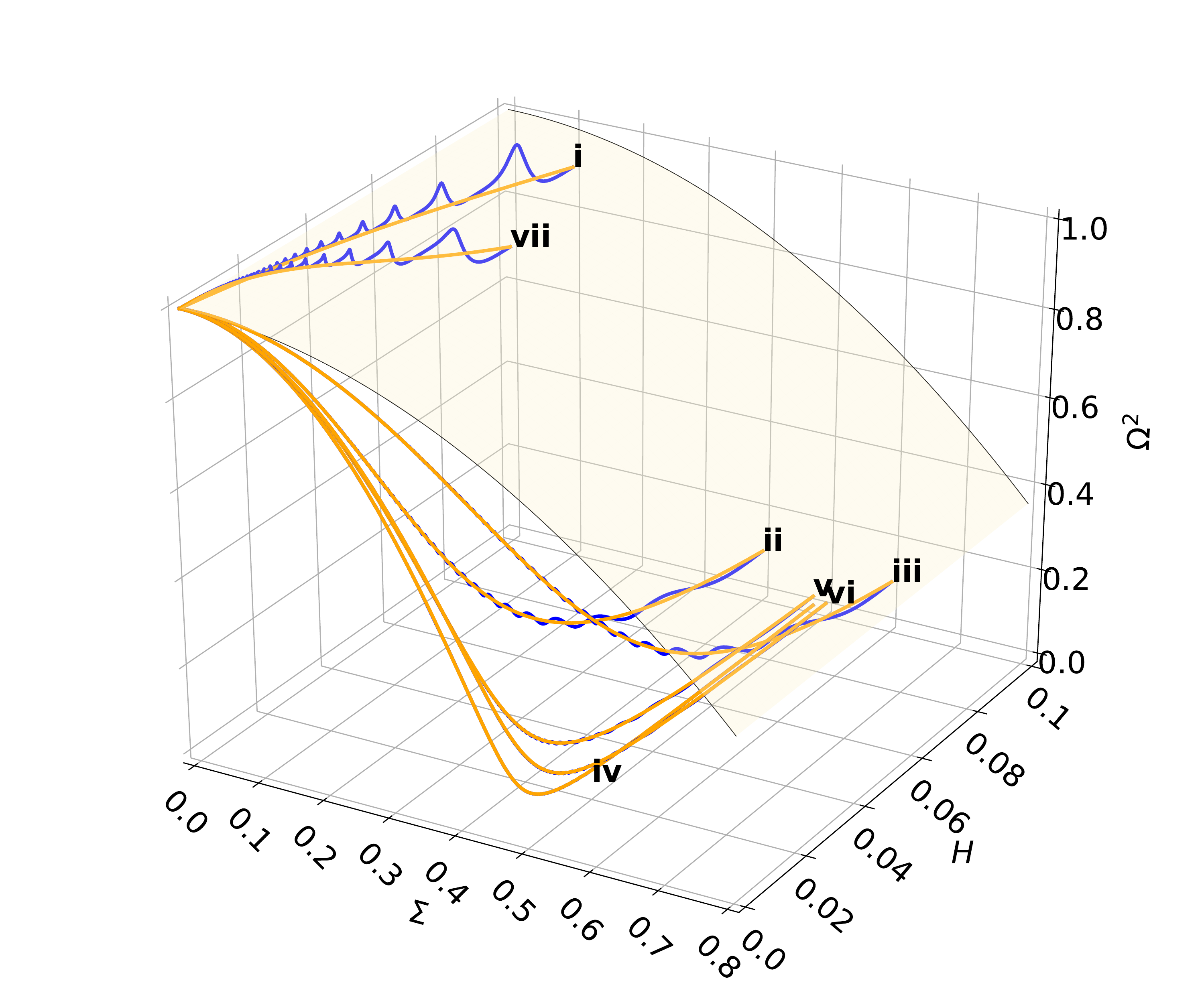}}
    \subfigure[\label{fig:BIStiff2D} Projection in the space $(\Sigma, \Omega^2)$. The black line represents the constraint $\Omega^{2}=1-\Sigma^{2}$.]{\includegraphics[scale = 0.51]{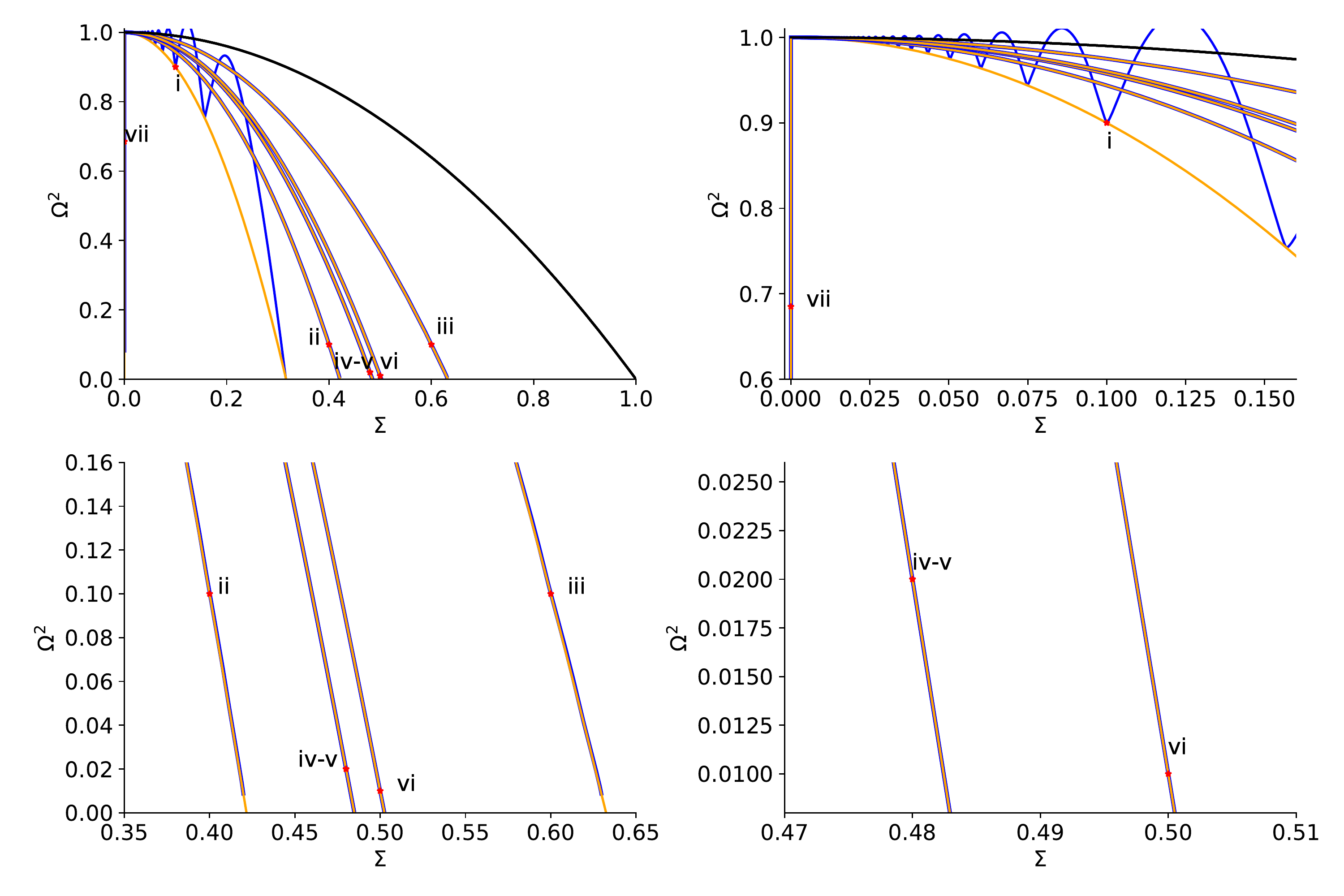}}
    \caption{Some solutions of the full system \eqref{unperturbed1} (blue) and time-averaged system \eqref{avrgsyst} (orange) for  Bianchi I metric when $\gamma=2$. We have used for both systems the initial data sets that are presented in Table \ref{Tab2}.}
\end{figure*}

\begin{figure*}
    \centering
    \subfigure[\label{fig:FlatFLRWCC} Projections in the space $(H, \Omega^{2})$ for $\gamma=0$.]{\includegraphics[scale = 0.52]{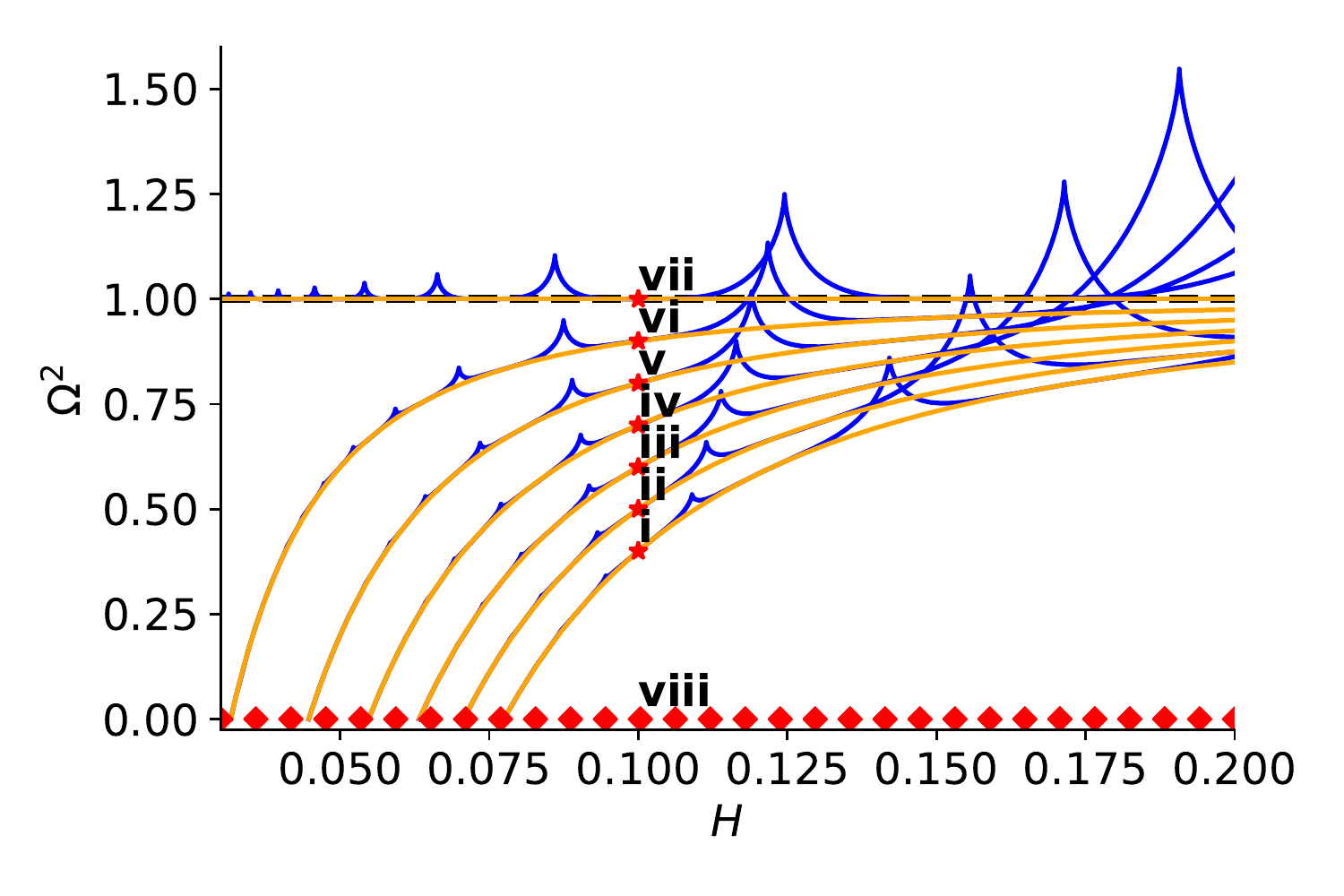}}
    \subfigure[\label{fig:FlatFLRWDust} Projections in the space $(H, \Omega^{2})$ for $\gamma=1$.]{\includegraphics[scale = 0.52]{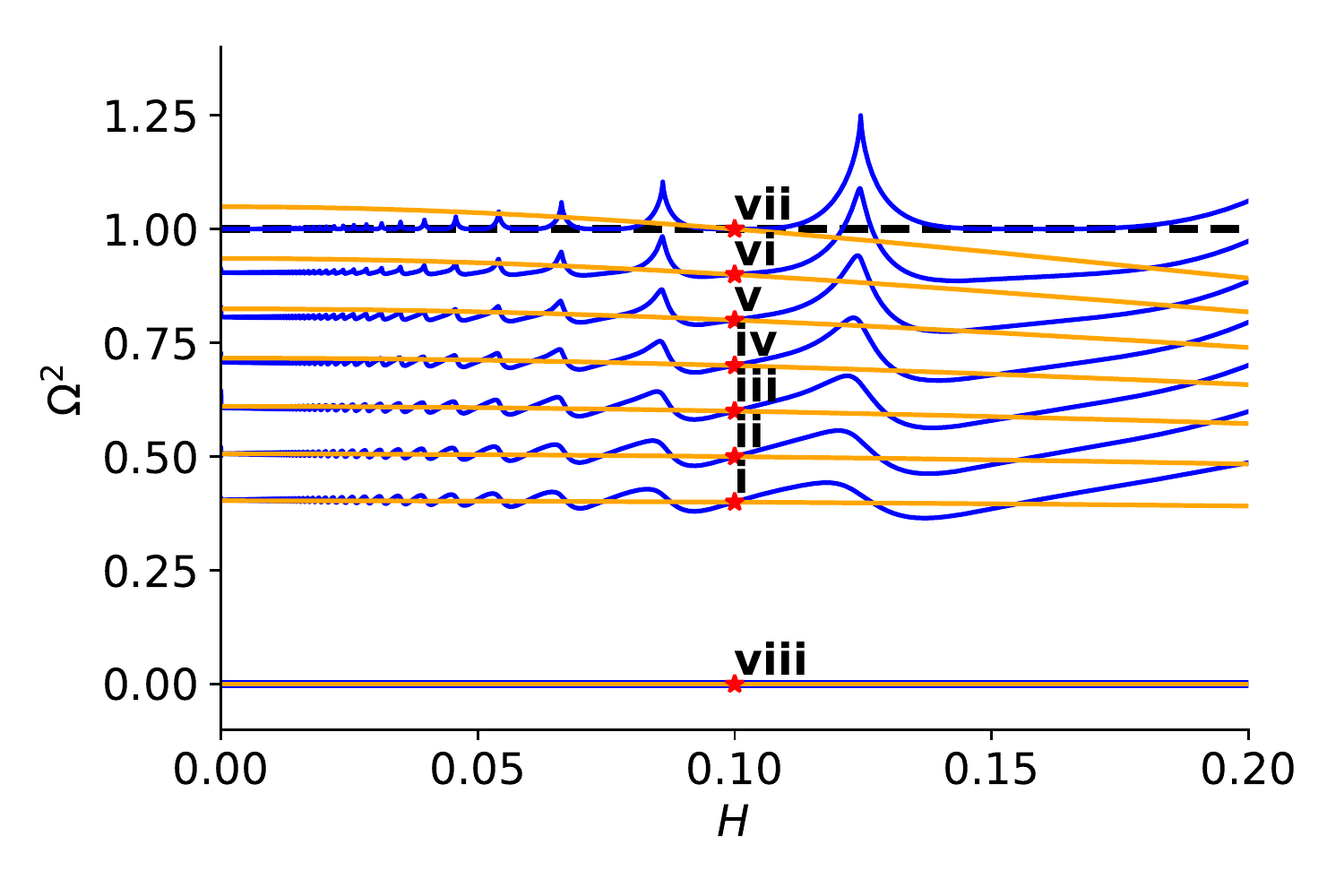}}
    \subfigure[\label{fig:FlatFLRWRad} Projections in the space $(H, \Omega^{2})$ for $\gamma=4/3$.]{\includegraphics[scale = 0.52]{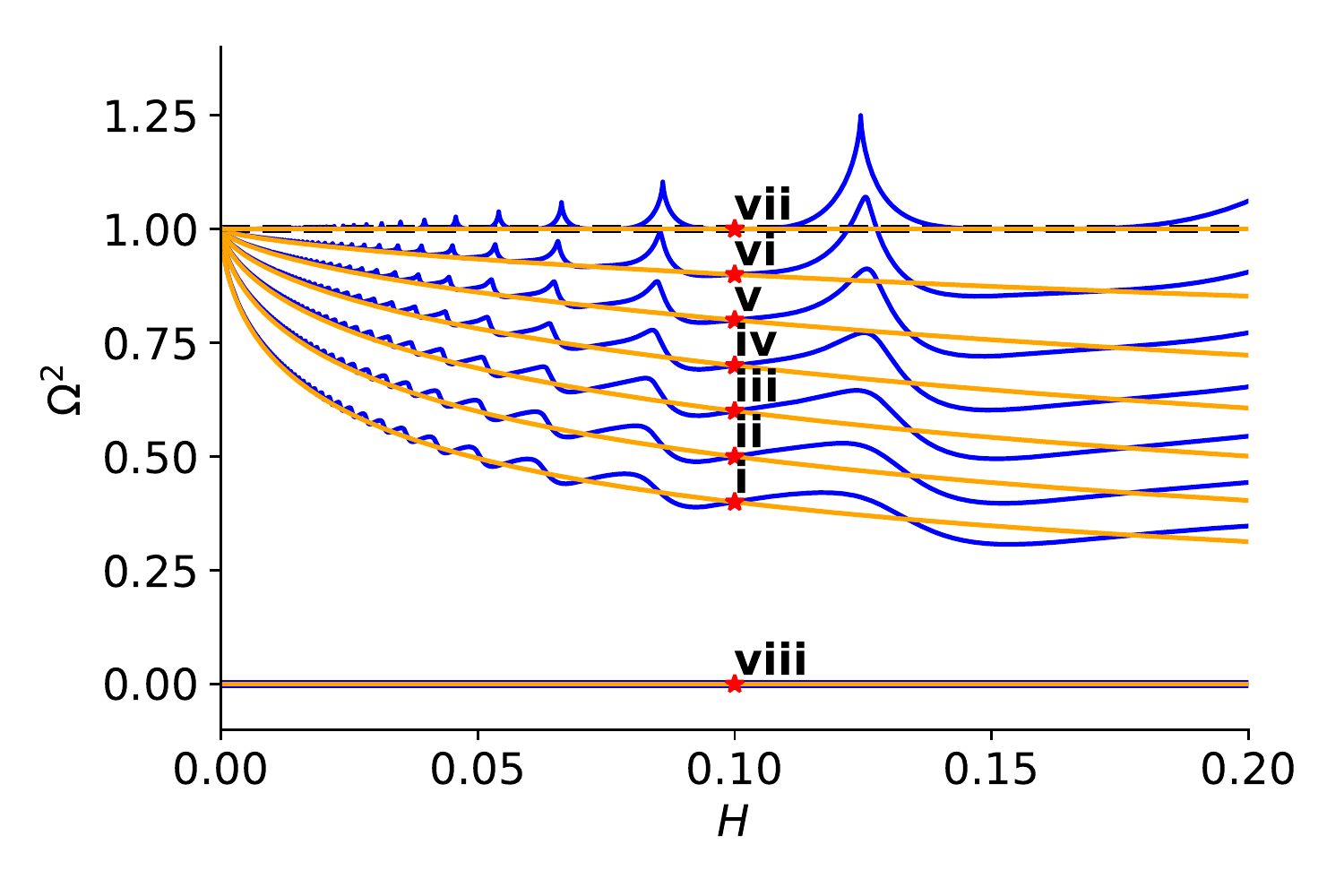}}
    \subfigure[\label{fig:FlatFLRWStiff} Projections in the space $(H, \Omega^{2})$ for $\gamma=2$.]{\includegraphics[scale = 0.52]{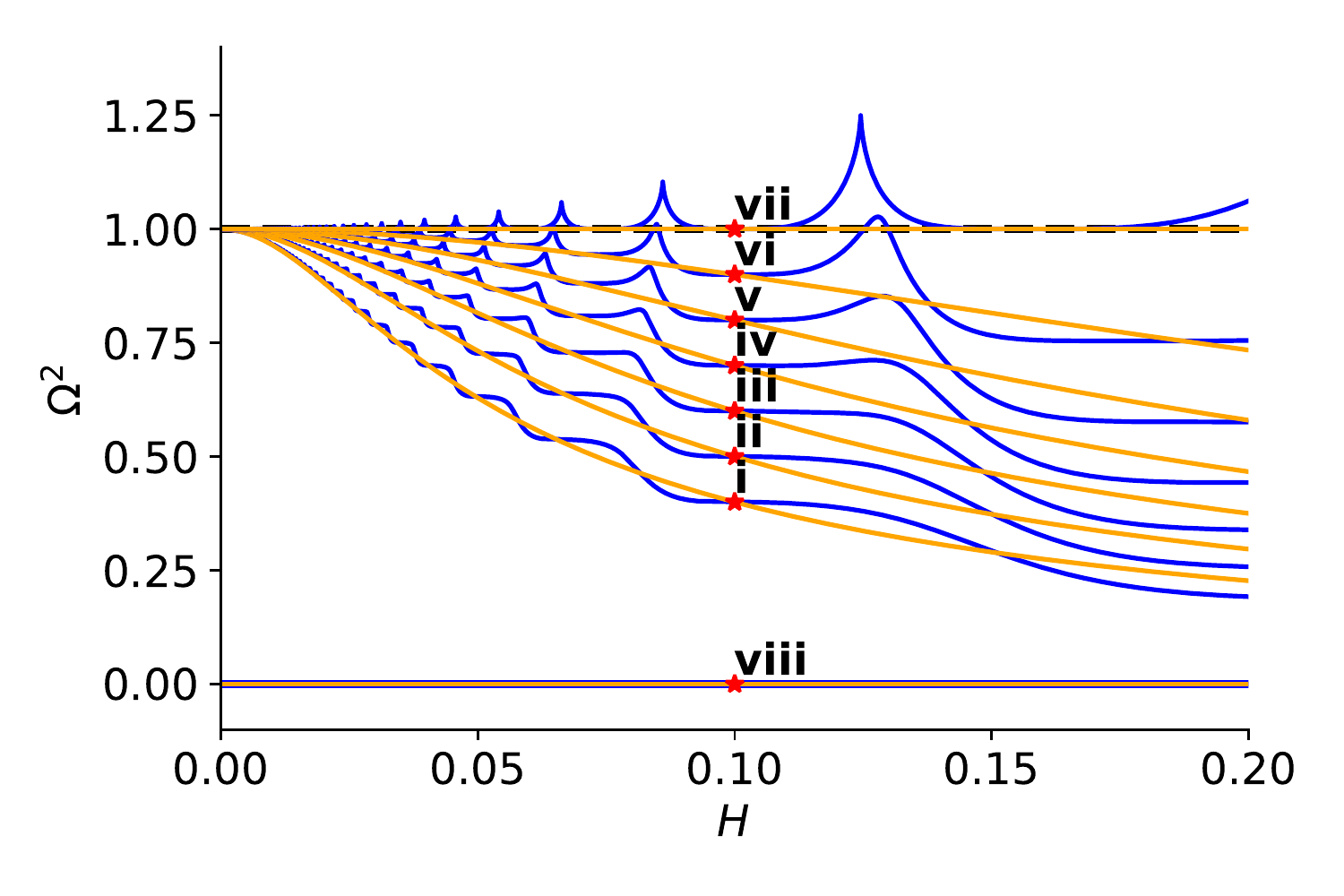}}
    \caption{Some solutions of   full system \eqref{unperturbed1} with $\Sigma=0$  (blue) and time-averaged system \eqref{avrgsystFLRW} or \eqref{avrgflatFLRWgamma1} if $\gamma=1$ (orange) for   flat FLRW metric ($k=0$) when the matter fluid correspond to: (a) cosmological constant, (b) dust, (c) radiation and (d) stiff fluid. The black line represents the constraint $\Omega^{2}=1$. In (a) the line denoted by red diamonds corresponds to the attracting line of equilibrium points (de Sitter solutions).  We have used for both systems  initial data sets that are presented in Table \ref{Tab3b}.}
\end{figure*}
\noindent
In figures \ref{fig:BICC3D}-\ref{fig:BIStiff2D}  projections of some solutions in the $(\Sigma, H, \Omega^{2})$ space of full system \eqref{unperturbed1} and time-averaged system \eqref{avrgsyst} along  with their respective projection in  the subspace $H=0$ are presented. Both systems were integrated  using the same initial data sets from Table \ref{Tab2}. Figures \ref{fig:BICC3D}-\ref{fig:BICC2D} show solutions for a matter fluid which corresponds to cosmological constant ($\gamma=0$). Figures \ref{fig:BIDust3D}-\ref{fig:BIDust2D} show solutions for a matter fluid which corresponds to dust ($\gamma=1$). Figures \ref{fig:BIRad3D}-\ref{fig:BIRad2D} show solutions for a matter fluid which corresponds to radiation ($\gamma=4/3$). Figures \ref{fig:BIStiff3D}-\ref{fig:BIStiff2D} show solutions for a matter fluid which corresponds to stiff fluid ($\gamma=2$). These figures numerically support the main theorem that is presented in Section \ref{SECT:II} for Bianchi I metric. As an interesting point, in this example, for $\gamma=0$ when the matter fluid corresponds to a cosmological constant,  $H$ tends asymptotically to a constant, which is consistent to de Sitter expansion.

\subsection{Flat FLRW model}
For flat FLRW model we integrate:
\begin{enumerate}
    \item The full system given by \eqref{unperturbed1} with $\Sigma=0$. 
\item The time-averaged system \eqref{avrgsystFLRW} if $\gamma\neq 1$ or the system \eqref{avrgflatFLRWgamma1} truncated at fourth order if $\gamma=1$.
\end{enumerate}
As initial conditions we use eight data sets that are presented in  Table \ref{Tab3b}. 

    \begin{table}[H]
    \centering   
\caption{\label{Tab3b} Eight initial data sets for the simulation of full system \eqref{unperturbed1} with $\bar{\Sigma}=0$ and time-averaged system \eqref{avrgsystFLRW} if $\gamma \neq 1$ or \eqref{avrgflatFLRWgamma1} if $\gamma=1$ for flat FLRW metric ($k=0$) are displayed. All initial conditions satisfy $\bar{\Omega}^{2}
(0)+\bar{\Omega}_{m}(0)=1$.}
\setlength{\tabcolsep}{5pt}
    \begin{tabular}{lccccccc}\hline
Sol.  & {$H(0)$}  &  {$\bar{\Omega}^2(0)$} &   {$\bar{\Omega}_m(0)$}  &  {$\bar{\Phi}(0)$}  &  {$t(0)$}  \\\hline
        i &  $0.1$ & $0.4$  & $0.6$ & $0$ & $0$\\
        ii &  $0.1$  & $0.5$  & $0.5$ & $0$ & $0$\\
        iii &  $0.1$  & $0.6$ & $0.4$ & $0$ & $0$\\
        iv &  $0.1$  & $0.7$  & $0.3$  & $0$ & $0$\\
        v &  $0.1$ &  $0.8$   & $0.2$ & $0$ & $0$\\
        vi &  $0.1$  & $0.9$  & $0.1$ & $0$ & $0$\\
        vii &  $0.1$  & $1.0$  & $0.0$ & $0$ & $0$\\
        viii &  $0.1$  & $0.0$  & $1.0$ & $0$ & $0$\\\hline
 \end{tabular}
\end{table}
In figures \ref{fig:FlatFLRWCC}-\ref{fig:FlatFLRWStiff}  projections  of some solutions in the $(H, \Omega^{2})$ space of full system \eqref{unperturbed1} with $\Sigma=0$  and time-averaged system \eqref{avrgsystFLRW} for  flat FLRW metric ($k=0$) are presented. Both systems were integrated  using the same initial data sets from Table  \ref{Tab3b}. Figure \ref{fig:FlatFLRWCC} shows solutions for a matter fluid which corresponds to cosmological constant ($\gamma=0$). Figure \ref{fig:FlatFLRWDust} shows solutions for a matter fluid which corresponds to dust ($\gamma=1$). Observe  that as $H\rightarrow 0$ the values of $\bar{\Omega}$ lying on the orange line (solution of  time-averaged system at fourth order) give an upper bound to the value of $\Omega$ of the original system. Figure \ref{fig:FlatFLRWRad} shows solutions for a matter fluid which corresponds to radiation ($\gamma=4/3$). Figure \ref{fig:FlatFLRWStiff} shows solutions for a matter fluid which corresponds to stiff fluid ($\gamma=2$). These figures support numerically   the main theorem that is presented in Section \ref{SECT:II}  for  flat FLRW metric. It is interesting to note that in the flat FLRW case when the matter fluid corresponds to a cosmological constant,  $H$ tends asymptotically to constant values depending on  initial conditions which is consistent to de Sitter expansion (see figure \ref{fig:FlatFLRWCC}).
With our approach, the oscillations which enter the system through the  KG equation can be controlled and smoothed out as the Hubble parameter $H$ tends monotonically to zero. This fact was analytically proved  in \ref{gBILFZ11} and we have  presented  numerical simulations as evidence of such behavior in \ref{numerics}.

\bibliographystyle{alpha}

\end{document}